\documentclass[aps,prb,twocolumn,superscriptaddress,showpacs]{revtex4-1}
\usepackage{amsmath,amssymb}
\usepackage{bm}

\usepackage{graphicx}
\usepackage{color}
\usepackage[colorlinks=true, linkcolor=magenta,citecolor=blue,urlcolor=magenta]{hyperref}
\DeclareGraphicsExtensions{.pdf}	%%% Extensions for pdflatex
\newcommand{\oH}[1]{\hat{H}_{\text{#1}}}
\newcommand{\oN}[1]{\hat{N}_{\text{#1}}}

\newcommand{\E}[1]{E_{\text{#1}}}

\newcommand{\Ex}[1]{\langle #1 \rangle}
\newcommand{\EX}[1]{\left\langle #1 \right\rangle}
\newcommand{\Tr}[1]{\text{Tr}[#1]}

\newcommand{\leftoperation}[1]{\overset{{}_\leftharpoonup}{#1}}
\newcommand{\rightoperation}[1]{\overset{{}_\rightharpoonup}{#1}}

\def\oa{\hat{a}}
\def\oad{\hat{a}^{\dagger}}
\def\oc{\hat{c}}
\def\ocd{\hat{c}^{\dagger}}
\def\ob{\hat{b}}
\def\obd{\hat{b}^{\dagger}}
\def\oPsi{\hat{\varPsi}} 
\def\oPsid{\hat{\varPsi}^{\dagger}}
\def\oO{\hat{O}}

\def\Hc{\mathrm{H.c.}}

\def\ii{\mathrm{i}}
\def\dd{\mathrm{d}}

\def\kB{k_{\text{B}}}

\def\k{\bm{k}}
\def\q{\bm{q}}

\def\R{\text{R}}
\def\A{\text{A}}
\def\K{\text{K}}
\def\B{\text{B}}
\def\C{\mathcal{C}}
\def\F{\text{F}}
\def\H{\text{H}}

\def\ph{\text{ph}}
\def\e{\text{e}}
\def\h{\text{h}}
\def\eh{\text{eh}}

\def\MF{\text{MF}}
\def\BGR{\text{BGR}}
\def\eff{\text{eff}}
\def\SS{\text{SS}}

\def\U{\mathcal{U}}
\def\T{\mathcal{T}}
\def\W{\mathcal{W}}

\def\Tc{\mathcal{T}_{\mathcal{C}}}
\def\Tcbar{\bar{\mathcal{T}}_{\mathcal{C}}}
\begin{document}
%===========================
%   Title, Author, Affiliation
%===========================
\title{Generating Functional Approach for Spontaneous Coherence
in Semiconductor Electron-Hole-Photon Systems}

\author{Makoto Yamaguchi}
\altaffiliation{E-mail: makoto.yamaguchi@riken.jp}
\affiliation{Department of Physics, Osaka University, 1-1 Machikaneyama, Toyonaka, Osaka 560-0043, Japan}
\affiliation{Center for Emergent Matter Science, RIKEN, Wakoshi, Saitama, 351-0198, Japan}
\author{Ryota Nii}
\affiliation{Department of Physics, Osaka University, 1-1 Machikaneyama, Toyonaka, Osaka 560-0043, Japan}
\author{Kenji Kamide}
\affiliation{Institute for Nano Quantum Information Electronics (NanoQuine), University of Tokyo, Tokyo 153-8505, Japan}
\author{Tetsuo Ogawa}
\affiliation{Department of Physics, Osaka University, 1-1 Machikaneyama, Toyonaka, Osaka 560-0043, Japan}
\affiliation{Photon Pioneers Center, Osaka University, 2-1 Yamada-oka, Suita, Osaka 565-0871, Japan}
\author{Yoshihisa Yamamoto}
\affiliation{E. L. Ginzton Laboratory, Stanford University, Stanford, California 94305, USA}
\affiliation{ImPACT Program, Japan Science and Technology Agency, 7 Gobancho, Chiyoda-ku, Tokyo 102-0076, Japan}

\date{\today}

%======================================================
%   Abstract
%======================================================
\begin{abstract}
Electrons, holes, and photons in semiconductors are interacting fermions and bosons.
In this system, a variety of ordered coherent phases can be formed through the spontaneous phase symmetry breaking because of their interactions.
The Bose-Einstein condensation (BEC) of excitons and polaritons is one of such coherent phases, which can potentially crossover into the Bardeen-Cooper-Schrieffer (BCS) type ordered phase at high densities under quasi-equilibrium conditions, known as the BCS-BEC crossover.
In contrast, one can find the semiconductor laser, superfluorescence (SF), and superradiance as relevant phenomena under nonequilibrium conditions. 
In this paper, we present a comprehensive generating functional theory that yields nonequilibrium Green's functions in a rigorous way. 
The theory gives us a starting point to discuss these phases in a unified view with a diagrammatic technique.
Comprehensible time-dependent equations are derived within the Hartree-Fock approximation, which generalize the Maxwell-Semiconductor-Bloch equations under the relaxation time approximation.
With the help of this formalism, we clarify the relationship among these cooperative phenomena and we show theoretically that the Fermi-edge SF is directly connected to the \mbox{e-h} BCS phase.
We also discuss the emission spectra as well as the gain-absorption spectra.
\end{abstract}

\maketitle

%======================================================
%   Introduction
%======================================================
\section{Introduction}\label{sec:intro}
Spontaneous development of macroscopic coherence is at the very heart of cooperative phenomena in condensed matter physics.
One major example is the superconductivity~\cite{Gennes66, Abrikosov75} in metals successfully explained by the Bardeen-Cooper-Schrieffer (BCS) theory.~\cite{BCS57}
In this case, weakly bound pairs of two electrons (Cooper pairs) are formed around the Fermi surface by their attractive many-body interaction and condensed by a similar mechanism to the Bose-Einstein condensation (BEC).~\cite{Pitaevskii03}
In the last decades, these cooperative phenomena have been intensively studied in several physical systems such as ultracold atomic systems~\cite{Anderson95, Regal04, Zwierlein04, Giorgini08, Barankov04-1, *Barankov04-2, *Barankov06, Andreev04, Szymanska05, Yuzbashyan06-1, *Yuzbashyan06-2} as well as superconductors.~\cite{Beck11,Bissbort11,Matsunaga12, *Matsunaga13}

In semiconductor systems, in a similar way, Cooper pairs of an electron and a hole can be considered through the Coulomb attractive interaction when the density is high enough to form the Fermi surface.~\cite{Keldysh64}
With decreasing the density, however, the electron-hole (\mbox{e-h}) Cooper pairs can smoothly change into excitons, that is, tightly bound \mbox{e-h} pairs through the Coulomb attraction.
As a result, the \mbox{e-h} BCS phase is expected to crossover into the exciton BEC.~\cite{Comte82, Leggett80, NSR85, Moskalenko00}
The BCS-BEC crossover recently highlighted in atomic Fermi gas systems,~\cite{Regal04, Zwierlein04, Giorgini08} in fact, arises partly from these considerations of the semiconductor \mbox{e-h} systems.~\cite{NSR85,Littlewood04}
In this sense, fundamental research on semiconductors is of great importance as it provides a stage to find concepts applicable to a wide range of fields.~\cite{Yoshioka11,*Yoshioka12,KYoshioka11,Versteegh12}

Open and dissipative nature of the system, however, should be taken care, particularly when electrons and holes have non-negligible interactions with photons because they are easily lost into free space even if confined in a cavity.~\cite{Kavokin07}
This is in stark contrast to the BCS and BEC phases---concepts basically for closed systems following equilibrium statistical physics.
Pictures and approaches in quantum optics,~\cite{Kavokin07,Scully97,Carmichael93} then, play a significant role to understand the appearance of macroscopic coherence in such nonequilibrium situations.
Striking examples are the super-radiance (SR) and the super-fluorescence (SF) as well as the laser.~\cite{Scully97,Gross82,Benedict96,*Andreev93,Vrehen80,Jho06,Noe12,Kim13}
Here, the SR is known as the cooperative radiation process where individual dipoles of emitters are synchronized with one another through their common radiation field.~\cite{Dicke54,Bonifacio71-1, *Bonifacio71-2}
The SF is a special case of the SR for the cooperative emission started from an initial state with no macroscopic coherence.~\cite{Bonifacio75-1, *Bonifacio75-2, Polder79}
These radiative processes are sometimes called mirror-less lasers~\cite{Gross82} because the cavity plays no essential role and is not necessarily required, in contrast to the standard lasers.
Although these cooperative phenomena can be found in quantum optics by using atomic discrete energy-level systems,~\cite{Schawlow58,Maiman60,Javan61,Skribanowitz73,Vrehen79,Scully09} semiconductor electron-hole-photon (\mbox{e-h-p}) systems are unique in relation to the pairing condensation, as described above, and provoke a non-trivial fundamental question about the {\em relationship among these cooperative phenomena}.

Vasil'ev and co-workers, for instance, studied the SF in an electrically pumped GaAs/AlGaAs heterostructure and suggested a hypothesis that the generation of super-fluorescent pulses is a result of the radiative recombination of the \mbox{e-h} BCS-like state.~\cite{Vasilev01,*Vasilev05,*Vasilev09}
Unfortunately, however, their discussions on this scenario remain largely speculative even though outstanding.
Nevertheless, the Fermi-edge SF~\cite{Kim13} recently demonstrated by Kim {\it et al.} is rather suggestive
where the macroscopic coherence is spontaneously developed near the Fermi edge due to the Coulomb-induced many-body effects; the physics seems closely related to the \mbox{e-h} BCS phase in our view, even though this similarity is not pointed out in the literature.~\cite{Kim13}
In an analogous sense, Dai and Monkman studied the SF in a highly excited bulk ZnTe crystal and claimed that the SF can be viewed as the exciton BEC developed on an ultrafast timescale.~\cite{Dai11}
These expectations might be plausible in terms of the spontaneous phase symmetry breaking and highly related to the above question.
However, such a question could not be previously addressed by any theoretical work because it is not trivial to incorporate both physics simultaneously.

Further intensive debate on this issue can be seen in the exciton-polariton systems;~\cite{Weisbuch92,Bloch98,Khitrova99,Deng02,Kasprzak06,Balili07,Deng10,Bajoni12,Horikiri13,Ishida14}
the relationship between the exciton-polariton BEC and the semiconductor laser.~\cite{Imamoglu96,Dang98,Deng03,Balili09,Nelsen09,Snoke12,Tempel12-1,*Tempel12-2,Tsotsis12,Kammann12}
Two distinct thresholds observed in several experiments were discussed in this context and the second threshold was interpreted in terms of a change from the exciton-polariton BEC into the standard lasing, the mechanism of which is attributed to a shift into the weak coupling regime due to dissociations of the excitons into the \mbox{e-h} plasma.~\cite{Balili09,Nelsen09,Tempel12-1,Tempel12-2,Tsotsis12,Kammann12} 
However,  there is no convincing discussion why such dissociations lead to nonequilibration of the system essential for lasing,~\cite{Yamaguchi13} while the laser is inherently a nonequilibrium phenomenon.~\cite{Bajoni08,Kasprzak08}
In a similar context, the distinction between the lasers and the photon BEC is also one of hot issues.~\cite{Klaers10,Kirton13,Leeuw13,Chiocchetta14,Schmitt14}

One difficulty to understand these phenomena results from a theoretical aspect;
in most cases in quantum optics, equations do not recover results expected from thermal equilibrium statistical physics even when equilibrium situations are considered.~\cite{Nakatani10,*Nakatani10E}
To overcome this difficulty, special care is required to use, for example, a quantum master equation (QME) approach.~\cite{Nakatani10,Nakatani10E,Breuer02}
{Szyma\ifmmode \acute{n}\else \'{n}\fi{}ska} and co-workers, in contrast, showed that a nonequilibrium Green's function (NEGF) approach is equally helpful to this problem even though the excitons are simply modeled by two-level systems with no internal \mbox{e-h} structures.~\cite{Szymanska06, *Szymanska07,Keeling10}

In previous papers,~\cite{Yamaguchi12,Yamaguchi13} motivated by their seminal work, we developed a steady-state framework based on the NEGF approach, which can treat the phases of the BEC, BCS and laser in a unified way with the \mbox{e-h} pairing mechanisms as well as an appropriate \mbox{e-h} picture.
This formalism results in the BCS theory~\cite{Kamide10, *Kamide11,Byrnes10} when the system can be regarded as in (quasi-)equilibrium, while it recovers the Maxwell--Semiconductor--Bloch equations~\cite{Chow02,Kamide11-MSBE} (\mbox{MSBEs}) of describing the laser when nonequilibrium features become important.
The mechanisms of the second threshold are, then, discussed and it is found that light-induced bound \mbox{e-h} pairs must remain alive even after the second threshold,~\cite{Yamaguchi13} in contrast to the above scenario.
At the same time, light-induced band renormalization causes the pairing gaps inside the conduction and valence bands.
In this paper, we elucidate several aspects of such a BEC-BCS-LASER crossover which we did not address in our previous papers.
In particular, we study the influence of the detuning and the pumping strength by showing the phase diagram and clearly reveal the possible types of the ordered phases, their individual mechanisms of appearance, and the criteria to distinguish these phases.
Spectral structures included in the emission spectra as well as the gain-absorption spectra are also clarified by introducing the energy- and momentum-resolved distribution functions.
One of our main purposes is thus to understand the nature lying between equilibrium and nonequilibrium steady states. 

A time-dependent formalism is, however, required to fully discuss the relationship of the cooperative phenomena because the SR and the SF are inherently transient phenomena.
In this context, another main purpose in this paper is to give a comprehensive generating functional theory~\cite{Martin59, Baym61,Baym62,Dominicis64,Hohenberg65} that yields NEGFs systematically in a time-dependent manner.~\cite{Kragler80,Henneberger88,Fujikawa05}
As a result, we show that the unknown variables in the \mbox{MSBEs} should evolve simultaneously with the time-dependent band renormalization, at least in principle.
This is quite natural for theorists because the NEGF approach originally describes the evolutions of the retarded, advanced, and Keldysh Green's functions (GFs);
the retarded and advanced GFs correspond to the band renormalization effects, while the Keldysh GF describes the distributions.
Nevertheless, we emphasize it because the band renormalization is critical for a unified view of the cooperative phenomena.
With the help of this formalism, we can directly tackle the problem of the relationship between the SF and the equilibrium phases.
As a result, we show that the Fermi-edge SF can be seen as a precursor of the \mbox{e-h} BCS phase in a sense that the Fermi-edge SF evolves toward the \mbox{e-h} BCS phase under the continuous pumping.
This is striking because the presence of the \mbox{e-h} BCS phase is a subject of long-time active interest not yet evidenced experimentally. 
Our result promisingly foresees the experimental observation of the \mbox{e-h} BCS phase in the context of the Fermi-edge SF.

Finally, the last purpose of this paper is to show the theoretical usefulness of the generating functional approach~\cite{Martin59,Baym61,Baym62,Dominicis64,Hohenberg65}  that can offer several advantages over the standard NEGF~\cite{Szymanska06,Szymanska07,Keeling10,Yamaguchi12,Yamaguchi13,Torre13,Kamenev11}
and QME~\cite{Nakatani10,Nakatani10E,Breuer02,Shirai14,Yuge14} approaches as follows;
(a) double counting problems of the Feynman diagrams are removed because dressed diagrams are directly obtained;
(b) at least in principle, equations can be closed when the hierarchy of the coupled GFs is truncated at certain level;
(c) except for initial states, the Born approximation is not required in contrast to the QME approach;
(d) two-particle GFs required for the calculations of the emission spectrum as well as the gain-absorption spectrum can be obtained in a convincing way.
These features seem somewhat technical but become significant if one extends our theory or develops a framework in similar open-dissipative systems.
However, there are few theoretical reports pointing out these features and no reports taking such an approach to address the relationship of the cooperative phenomena ranging from equilibrium to nonequilibrium in the semiconductor \mbox{e-h-p} systems.
We therefore describe our detailed theoretical treatment of the generating functional approach, which gives a starting point to study the above-described cooperative phenomena in a unified view.

As we now know, this paper covers cross-sectoral issues ranging from condensed matter physics to quantum optics. 
In order to make the paper accessible to experimentalists as well as theorists in both fields, therefore, we try to provide sufficient explanations and reinterpretations of the formalism and physics as far as possible even if these are well-known to some specialists.

The remainder of the paper is organized as follows.
In Section~\ref{sec:framework}, as a typical example of the semiconductor \mbox{e-h-p} systems, we consider the exciton-polariton system and introduce our Hamiltonians.
We then show our key results of the formalism after briefly reviewing the BCS theory and the MSBEs under the relaxation time approximation (RTA).
Our theoretical formulation is not shown here but will be presented in later sections for clarity (Sec.~\ref{sec:Generating Functional} and \ref{sec:real time}). 
In Section~\ref{sec:Relationship}, we study the relationship between the cooperative phenomena.
We first show that our formalism is appropriate to study the cooperative phenomena in a unified way, and then, study the steady-state phase diagrams.
Here, we will give detailed insights to the BEC-BCS-LASER crossover.~\cite{Yamaguchi12,Yamaguchi13,Yamaguchi14}
We then discuss the connections between the Fermi-edge SF and the \mbox{e-h} BCS phases, the theoretical study of which has been impossible before.
In Section~\ref{sec:Spectral}, we shortly explain our formalism to calculate the emission spectrum and the gain-absorption spectrum, and then, present several numerical results.
With the help of the energy- and momentum-resolved distribution functions, we will clarify that the underlying physics can basically be understood from the picture of the Mollow triplet in quantum optics.~\cite{Scully97,Mollow69}
In addition, it is further found that the gain-absorption spectra can be affected by the phase difference between the external probe field and the spontaneous coherence developed in the system.
In Section~\ref{sec:Generating Functional}, we present a general formalism based on the generating functional approach.
We define the relevant NEGFs and explain their equations of motion on the closed-time contour, together with their diagrammatic representations.
In Section~\ref{sec:real time}, we transform the NEGFs into the real-time formulation.
Within the Hartree-Fock (HF) approximation, we derive a time-dependent framework that generalizes the \mbox{MSBEs} under the RTA.
Readers who are not familiar with the NEGFs can, however, skip Section~\ref{sec:Generating Functional} and \ref{sec:real time} because these sections are mainly devoted to the theoretical explication of our formalism.
Finally, in Section~\ref{sec:Conclusions}, our main results are summarized with some final remarks and the paper is concluded.

%======================================================
%   Theoretical Framework
%======================================================
\section{Theoretical Framework}\label{sec:framework}
As a typical model of the semiconductor \mbox{e-h-p} systems, we consider the exction-polariton systems where electrons and holes are in quantum wells while photons are confined in a microcavity;~\cite{Kavokin07,Weisbuch92,Bloch98,Khitrova99,Deng02,Kasprzak06,Balili07,Deng10} see also Ref.~\onlinecite{Byrnes14} for a recent review.
In this section, we introduce our Hamiltonians and describe key results of our formalism.
For simplicity, we set $\hbar = \kB = 1$ throughout this paper.

%=======================================
% Hamiltonians
%=======================================
\subsection{Hamiltonians}\label{subsec:hamiltonians}
\begin{figure}[!tb] 
\centering
\includegraphics[width=.45\textwidth]{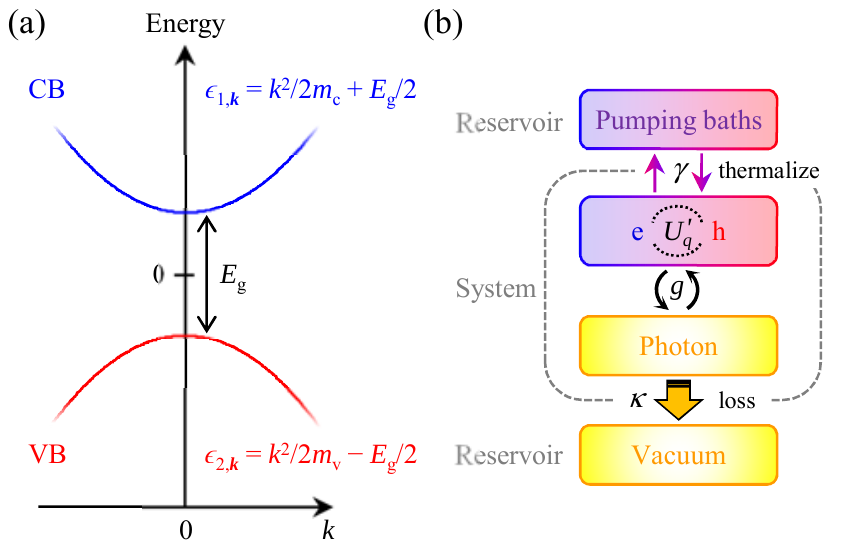} 
\caption{(Color online) Schematic illustration of the model.
(a) The structure of the conduction band (CB) and the valence band (VB).
(b) Relationship between the system and reservoirs.~\cite{Yamaguchi14}
}
\label{fig01:Model}
\end{figure}

Open and dissipative nature of a certain {\em system} is commonly described by its interactions with {\em reservoirs} in quantum optics.~\cite{Kavokin07,Scully97,Carmichael93}
Our Hamiltonian for the exciton-polariton system can, therefore, be described as
%----------
% Equation
%----------
\begin{align}
&\hat{H} =  \oH{S} + \oH{R} + \oH{SR}, 
\label{eq:H}
\end{align}
%----------
in the Schr\"odinger picture, where $\oH{S}$, $\oH{R}$ and $\oH{SR}$ are the system, reservoir and their interaction Hamiltonians, respectively.
Here, the system Hamiltonian $\oH{S}$ is given by
%----------
% Equation
%----------
\begin{align}
\oH{S} = \oH{0} + \oH{e-e} + \oH{e-ph},
\label{eq:HS}
\end{align}
%----------
where
%----------
% Equation
%----------
\begin{align}
\oH{0} & = \sum_{\alpha,\k} \epsilon_{\alpha,\k} \ocd_{\alpha,\k} \oc_{\alpha,\k}
       + \sum_{\k} \epsilon_{\ph,\k} \oad_{\k} \oa_{\k},
\label{eq:H0}
\end{align}
%----------
describes the free-particle Hamiltonian with $\alpha \in \{1, 2\}$.
$\oc_{1,\k}$ ($\oc_{2,\k}$) is the fermionic annihilation operator of electrons in the conduction (valence) band, while $\oa_{\k}$ is the bosonic annihilation operator of photons inside the cavity, with in-plane wave number $\k$.
$\epsilon_{1(2),\k} \equiv k^2/2m_{\text{c(v)}} \pm \E{g} /2$ denotes the energy dispersion of the conduction (valence) band with the effective mass $m_{\text{c(v)}}$ and the band gap energy $\E{g}$ (Figure~\ref{fig01:Model}(a)).
Similarly, $\epsilon_{\ph,\k} \equiv k^2/2m_{\text{cav}} + E_{\text{cav}}$ denotes the energy dispersion of photons with the effective mass $m_{\text{cav}}$ and the cavity mode energy $E_{\text{cav}}$ for $\k=0$.~\cite{Deng10}
We note that, instead of holes, electrons in the valence band are treated in our model and the \mbox{e-h} picture will be introduced after the formulation.

%----------
% Figure
%----------
%----------

$\oH{e-e}$ and $\oH{e-ph}$ in Eq.~\eqref{eq:HS} are the interactions between the particles and described as~\cite{Kamide10,Kamide11,Byrnes10}
%----------
% Equation
%----------
\begin{align}
&\oH{e-e} = \frac{1}{2}\sum_{\k,\k',\q}\sum_{\alpha,\alpha'}U'_{\q}\ocd_{\alpha,\k+\q}
		\ocd_{\alpha',\k'-\q}\oc_{\alpha',\k'}\oc_{\alpha,\k},
\label{eq:He-e}\\
&\oH{e-ph} = -\sum_{\k,\q}(g^*\oa_{\q}\ocd_{1,\k+\q}\oc_{2,\k} + \Hc).
\label{eq:He-ph}
\end{align}
%----------
Here, $g$ is the light-matter coupling constant under the dipole approximation and
%----------
% Equation
%----------
\begin{align}
U'_{\q} & \equiv \left \{
\begin{array}{ll}
U_{\q} & \text{for}~\q \neq 0 \\
0 & \text{for}~\q = 0 \\
\end{array}
\right.,
\label{eq:Coulomb}
\end{align}
%----------
is the Coulomb interaction.
Our model, thus, treats electrons, holes (electrons in the valence band) and photons explicitly in contrast to the well-known approaches, such as the Gross-Pitaevskii equations in the exciton-polariton community,~\cite{Wouters07,Wouters08,Savenko13,Haug14} where the excitons are regarded as simple bosons.
This is because our interest includes, for example, the \mbox{e-h} BCS phase where the phase space filling of electrons and holes plays an important role.
In this way, the semiconductor \mbox{e-h-p} system can be described by the Hamiltonians in Eqs.~\eqref{eq:HS}--\eqref{eq:He-ph} if nonequilibrium effects are not taken into account.

$\oH{R}$ and $\oH{SR}$ in Eq.~\eqref{eq:H} are, however, required to consider the pumping and loss of the system, as schematically shown in Figure~\ref{fig01:Model}(b), and described as
%----------
% Equation
%----------
\begin{align}
\oH{R} = &\sum_{\alpha,\k} \epsilon_{\alpha,\k}^{\text{B}} \obd_{\alpha,\k}\ob_{\alpha,\k}
       + \sum_{\k} \epsilon_{\ph,\k}^{\text{B}} \oPsid_{\k}\oPsi_{\k},
\label{eq:HR} \\
\oH{SR}  = & \sum_{\alpha,\k,\q} \varsigma_{\alpha,\k} (\ocd_{\alpha,\k}\ob_{\alpha,\q} + \Hc) \nonumber \\
       & + \sum_{\k,\q} \zeta_{\k} (\oad_{\k} \oPsi_{\q} + \Hc).
\label{eq:HSR}
\end{align}
%----------
Here, $\ob_{1,\k}$ and $\ob_{2,\k}$ are fermionic annihilation operators of the pumping baths for the conduction and valence band, respectively, and $\oPsi_{\k}$ is a bosonic annihilation operator of the free-space vacuum fields.
$\varsigma_{\alpha,\k}$ and $\zeta_{\k}$ are coupling constants between the system and the respective reservoirs, which are assumed to satisfy the standard approximations in quantum optics~\cite{Yamaguchi12,Yamaguchi13}
%----------
% Equation
%----------
\begin{subequations}
\label{alleq:gamma-kappa}
\begin{align}
\gamma & \cong \gamma_{\alpha,\k} \cong \pi |\varsigma_{\alpha,\k}|^2D^{\text{B}}_{\alpha}(\epsilon),
\label{eq:gamma}\\
\kappa & \cong \kappa_{\k} \cong \pi |\zeta_{\k}|^2D^{\text{B}}_{\ph}(\epsilon),
\label{eq:kappa} 
\end{align}
\end{subequations}
%----------
with the following definitions of the density of states:
%----------
% Equation
%----------
\begin{subequations}
\label{alleq:DOS}
\begin{align}
D^{\text{B}}_{\alpha}(\epsilon) \equiv \sum_{\k'} \delta(\epsilon_{\alpha,\k'}^{\text{B}} - \epsilon),
\label{eq:DOS_el}\\
D^{\text{B}}_{\ph}(\epsilon) \equiv \sum_{\k'} \delta(\epsilon_{\ph,\k'}^{\text{B}} - \epsilon).
\label{eq:DOS_ph}
\end{align}
\end{subequations}
%----------
We note that the thermalization rate of the \mbox{e-h} system and the cavity photon loss rate will be described by $2\gamma$ and $2\kappa$, as seen in later discussion, while their dependence on the wave number is neglected in Eq.~\eqref{alleq:gamma-kappa} for simplicity.~\cite{Yamaguchi12}

Finally, we note that $[\hat{H}, \hat{N}]=0$ can be found when a total excitation number is defined as 
%----------
% Equation
%----------
\begin{align}
\hat{N} \equiv \oN{S} + \oN{R},
\label{eq:total ex num} 
\end{align}
%----------
where
%----------
% Equation
%----------
\begin{align}
\oN{S} & \equiv \sum_{\k}  \left\{ \frac{1}{2}(\ocd_{1, \k} \oc_{1, \k} - \ocd_{2, \k} \oc_{2, \k}) + \oad_{\k} \oa_{\k} \right\},
\label{eq:ex num S} \\
\oN{R} & \equiv \sum_{\k}  \left\{ \frac{1}{2}(\obd_{1, \k} \ob_{1, \k} - \obd_{2, \k} \ob_{2, \k}) + \oPsid_{\k} \oPsi_{\k} \right\}.
\label{eq:ex num R} 
\end{align}
%----------
In the followings, therefore, we redefine $\oH{x} - \mu \oN{x}$ as $\oH{x}$ with $\text{x} \in \{\text{S}, \text{R} \} $.
This means that the dynamics of certain physical quantities is captured on a rotating frame with a frequency $\mu$ for time-dependent problems, while a grand canonical ensemble can be considered with a chemical potential $\mu$ if the system is identified as being in \mbox{(quasi-)equilibrium phases}.~\cite{Yamaguchi14}
As a result, $\epsilon_{1(2), \k}$ and $\epsilon_{\ph,\k}$ in Eq.~\eqref{eq:H0} are replaced by $\xi_{1(2), \k} \equiv \epsilon_{1(2), \k} \mp \mu/2$ and  $\xi_{\ph, \k} \equiv \epsilon_{\ph, \k} - \mu$, respectively.
In the same manner, $\epsilon_{1(2), \k}^{\text{B}}$ and $\epsilon_{\ph,\k}^{\text{B}}$ in Eq.~\eqref{eq:HR} are replaced by $\xi_{1(2), \k}^{\text{B}} \equiv \epsilon_{1(2), \k}^{\text{B}} \mp \mu/2$ and $\xi_{\ph,\k}^{\text{B}} \equiv \epsilon_{\ph, \k}^{\text{B}} - \mu$, respectively.

%=======================================
%   BCS theory and the MSBEs
%=======================================
\subsection{BCS theory and the MSBEs}\label{subsec:the BCS theory and the MSBEs}
Based on the Hamiltonians presented above, physical quantities of our interest are the cavity photon amplitude $a_{0}(t) \equiv \Ex{\oa_{\k=0}(t)}$, the polarization function $p_{\k}(t) \equiv \Ex{\ocd_{2,\k}(t)\oc_{1,\k}(t)}$, and the distribution functions of electrons in the conduction band $n_{1,\k}(t) \equiv \Ex{\ocd_{1,\k}(t) \oc_{1,\k}(t) }$ and in the valence band $n_{2,\k}(t) \equiv \Ex{\ocd_{2,\k}(t) \oc_{2,\k}(t) }$.
Here, $\Ex{\oO (t)}$ denotes the expectation value and is equivalent to $\Tr{\oO \hat\rho(t)}$ in the Schr\"odinger picture.
To study these physical quantities, one of the well-known approaches is the mean-field approximation that reduces the many-body problems to the single-particle one.
Here, let us shortly review such an approach~\cite{Yamaguchi14} to make our discussion and standpoint as clear as possible and to fix the notations.

In the mean-field approximation, certain operators $\oO_{i}$ $(i = 1, 2, \cdots)$ are described by $\oO_{i} = \Ex{\oO_{i}} + \delta\oO_{i}$ and the quadratic terms $\delta\oO_{i}\delta\oO_{j}$ are neglected in the Hamiltonians.
By taking $\oO_{i} \in \{\oa_{\k},  \ocd_{2,\k}\oc_{1,\k'},  \ocd_{1,\k}\oc_{1,\k'},  \ocd_{2,\k}\oc_{2,\k'} \}$, with definitions $\Ex{\oa_{\k}} \equiv \delta_{\k,0} a_{0}$, $\Ex{\ocd_{2,\k}\oc_{1,\k'}} \equiv \delta_{\k,\k'} p_{\k}$, and $\Ex{\ocd_{\alpha,\k}\oc_{\alpha,\k'}} \equiv \delta_{\k,\k'} n_{\alpha,\k}$, we obtain the mean-field Hamiltonian $\oH{S}^{\MF}$ for the system Hamiltonian $\oH{S}$ as
%----------
% Equation
%----------
\begin{align}
\oH{S}^{\MF} = & \sum_{\k} \left( \sum_{\alpha} \tilde{\xi}_{\alpha,\k} \ocd_{\alpha,\k}\oc_{\alpha,\k}
   - [\varDelta_{\k}\ocd_{1,\k}\oc_{2,\k} + \Hc] \right) \nonumber\\
&  + \sum_{\k} \left( \xi_{\ph, \k} \oad_{\k}\oa_{\k} - [g p_{\k} \oad_{0} + g^{*} p^{*}_{\k}\oa_{0}] \right).
\label{eq:HMF} 
\end{align}
%----------
Here,  $\varDelta_{\k} \equiv g^{*}a_{0} + \sum_{\k'} U'_{\k'-\k} p_{\k'}$ is the generalized Rabi frequency describing the effect of forming the \mbox{e-h} pairs~\cite{Schmitt-Rink88, Yamaguchi13, Yamaguchi14} and $\tilde {\xi}_{\alpha,\k} \equiv \xi_{\alpha,\k} + \varSigma^{\BGR}_{\alpha,\k}$ denotes the single-particle energy renormalized by the Coulomb interaction $\varSigma^{\BGR}_{\alpha,\k} \equiv -\sum_{\k'} U'_{\k'-\k} n_{\alpha,\k'}$, which includes the band-gap renormalization (BGR) in semiconductor physics.
In Eq.~\eqref{eq:HMF}, constants are ignored because the following discussion is not affected.

In the Schr\"odinger picture, therefore, the density operator of the system $\hat{\rho}^{\MF}$ is determined by the mean-field Hamiltonian $\oH{S}^{\MF}$ which includes $\hat{\rho}^{\MF}$ through the definition of the expectation values.
In this context, the self-consistency condition
%----------
% Equation
%----------
\begin{align}
\Ex{\oO_{i}} = \Tr{\oO_{i} \hat{\rho}^{\MF}(\Ex{\oO_1}, \Ex{\oO_2}, \cdots) },
\label{eq:self-consistency}
\end{align}
%----------
should be satisfied.
The BCS theory and the MSBEs for the exciton-polariton systems can be derived from this type of self-consistent equations, as described in the following.

%---------------------------------------
% the BCS theory
%---------------------------------------
\subsubsection{BCS theory}\label{subsubsec:BCS}
By assuming that the exciton-polariton system is in equilibrium at temperature $T$, the density operator $\hat{\rho}^{\MF}$ is given by
%----------
% Equation
%----------
\begin{align}
\hat{\rho}^{\MF} = \hat{\rho}^{\MF}_{\text{eq}} \equiv \frac{1}{Z}\exp(-\beta \oH{S}^{\MF}),
\label{eq:equilibrium}
\end{align}
%----------
where $Z \equiv \Tr{\exp(-\beta\oH{S}^{\MF})}$ and $\beta \equiv 1/T$.
We note that, in this case, $\mu$ is a given parameter corresponding to the chemical potential, as described above.
With the aid of the e-h picture in Table~\ref{tableI}, by assuming $\epsilon_{\e,\k} = \epsilon_{\h,\k}$ for simplicity, it is straightforward to obtain the following self-consistent equations from Eq.~\eqref{eq:self-consistency},
%----------
% Equation
%----------
\begin{subequations}
\label{alleq:BCS}
\begin{align}
&a_{0} = \sum_{\k'} \frac{g}{\xi_{\ph,0}} p_{\k'},
\label{eq:BCS--a0}\\
&p_{\k} = \frac{\varDelta_{\k}}{2E_{\k}}\tanh\left( \frac{\beta E_{\k}}{2} \right),
\label{eq:BCS--pk}\\
&n_{\e,\k} = n_{\h,\k} = \frac{1}{2} \left\{ 1 - \frac{\tilde{\xi}^{+}_{\eh, \k}}{E_{\k}} \tanh \left( \frac{\beta E_{\k}}{2} \right) \right\},
\label{eq:BCS--nehk}
\end{align}
\end{subequations}
%----------
by diagonalizing $\oH{S}^{\MF}$ through the Bogoliubov transformation for $\oc_{1,\k}$ and $\oc_{2,\k}$ and a displacement of $\oa_{0}$.
This can be performed because the Hilbert space of the first (second) line in Eq.~\eqref{eq:HMF} is spanned solely by the electron (photon) degrees of freedom.
Here, we have defined
%----------
% Equation
%----------
\begin{align*}
E_{\k} \equiv \sqrt{ ( \tilde{\xi}^{+}_{\eh,\k} )^2 + |\varDelta_{\k}|^2 },
\end{align*}
%----------
with $\tilde{\xi}^{\pm}_{\eh,\k} \equiv (\tilde{\xi}_{\e,\k} \pm \tilde{\xi}_{\h,\k})/2$ in the derivation.

%----------
% TABLE I
%----------
\begin{table}[!tb]
\caption{\label{tableI} Definitions of the variables in the e-h picture.}
\begin{ruledtabular}
\begin{tabular}{cc|cc}
Variable & Definition & Variable & Definition \\
\hline
$n_{\e,\k}$ & $ n_{1,\k} $ & $n_{\h,\k}$ & $ 1- n_{2,-\k} $ \\
$f^{\B}_{\e}(\nu)$ & $f^{\B}_{1}(\nu)$ & $f^{\B}_{\h}(\nu)$ & $1 - f^{\B}_{2}(-\nu)$\\
$m_{\e}$ & $m_{\text{c}}$ & $m_{\h}$ & $-m_{\text{v}}$ \\
$\epsilon_{\e,\k}$ & $ \epsilon_{1,\k} $ & $\epsilon_{\h,\k}$ & $ -\epsilon_{2,\k} +  \sum_{\k'} U'_{\k'}$\\
$\tilde{\epsilon}_{\e,\k}$ & $\epsilon_{\e,\k} + \varSigma^{\BGR}_{\e,\k}$ & $\tilde{\epsilon}_{\h,\k}$ & $\epsilon_{\h,\k} + \varSigma^{\BGR}_{\h,\k}$ \\
$\xi_{\e,\k} $ & $ \epsilon_{\e,\k} - \mu/2 $ & $\xi_{\h,\k} $ & $ \epsilon_{\h,\k} - \mu/2 $ \\
$\tilde{\xi}_{\e,\k}$ & $\xi_{\e,\k} + \varSigma^{\BGR}_{\e,\k}$ & $\tilde{\xi}_{\h,\k}$ & $\xi_{\h,\k} + \varSigma^{\BGR}_{\h,\k}$ \\
$\varSigma^{\BGR}_{\e,\k} $ & $- \sum_{\k'} U'_{\k' - \k} n_{\e,\k'}$ & $\varSigma^{\BGR}_{\h,\k} $ & $- \sum_{\k'} U'_{\k' - \k} n_{\h,\k'}$  \\
$\mu^{\B}_{\e}$ & $\mu^{\B}_{1}$ & $\mu^{\B}_{\h}$ & $-\mu^{\B}_{2}$ \\
\end{tabular}
\end{ruledtabular}
\end{table}

By putting Eqs.~\eqref{eq:BCS--a0} and \eqref{eq:BCS--pk} into the definition of $\varDelta_{\k}$, the equations for $a_{0}$ and $p_{\k}$ can be combined into one equation:
%----------
% Equation
%----------
\begin{align}
& \varDelta_{\k} = \sum_{\k'} U^{\eff}_{\k', \k} \frac{\varDelta_{\k'}}{2E_{\k'}} \tanh \left( \frac{\beta E_{\k'}}{2} \right),
\label{eq:BCSgap1}
\end{align}
%----------
which is formally equivalent to the gap equation of the BCS theory for superconductors. 
In this context, $\varDelta_{\k}$ describes an order parameter for the \mbox{e-h} pairing and $U^{\eff}_{\k', \k} \equiv |g|^2/\xi_{\ph,0} + U'_{\k' - \k}$ represents an effective attractive \mbox{e-h} interaction.
As a result, the equations are closed by Eqs.~\eqref{eq:BCS--nehk} and \eqref{eq:BCSgap1} with the unknown variables $\varDelta_{\k}$, $n_{\e,\k}$ and $n_{\h,\k}$.
Especially for $T = 0$, this treatment is known to cover the equilibrium phases from the BEC to the BCS states.~\cite{Comte82, Kamide10, Byrnes10}

%---------------------------------------
% the MSBEs
%---------------------------------------
\subsubsection{MSBEs}\label{subsubsec:BCS}
The BCS theory described above is, however, not appropriate to treat non-equilibrium cooperative phenomena, such as the SR, SF, and lasing, because of the excitation and thermalization of the \mbox{e-h} system and the loss of photons from the microcavity.
In this context, the effect of reservoirs cannot be neglected.
For this reason, it is convenient to discuss the dynamics of the total density operator $\hat{\rho}^{\MF}$ with the total mean-field Hamiltonian $\hat{H}^{\MF} \equiv \oH{S}^{\MF} + \oH{SR} + \oH{R}$.
Since $\ii \partial_t \hat{\rho}^{\MF} = [\hat{H}^{\MF}, \hat{\rho}^{\MF}]$ in the Schr\"odinger picture, a time derivative of Eq.~\eqref{eq:self-consistency} yields
%----------
% Equation
%----------
\begin{align}
\ii \partial_t \Ex{\oO_i} = \Tr{[\oO_i, \oH{S}^{\MF}]\hat{\rho}^{\MF}} + \Tr{[\oO_i, \oH{SR}]\hat{\rho}^{\MF}},
\label{eq:self-consistency2}
\end{align}
%----------
where $[\oO_i, \oH{R}] = 0$ and $\Tr{\hat{A}\hat{B}} = \Tr{\hat{B}\hat{A}}$ are used.
Substitution of Eq.~\eqref{eq:HMF} into the first term, then, reads the MSBEs
%----------
% Equation
%----------
\begin{subequations}
\label{alleq:MSBEs}
\begin{align}
&\partial_{t} a_0 = -\ii \xi_{\ph,0}a_0 + \ii g \textstyle{\sum_{\k}} p_{\k} - \kappa a_0, 
\label{eq:MSBEs--a0}\\
&\partial_{t} p_{\k} = - 2\ii \tilde{\xi}^{+}_{\eh,\k} p_{\k}  - \ii \varDelta_{\k} N_{\k} - 2 \gamma ( p_{\k} - p^0_{\k} ),
\label{eq:MSBEs--pk}\\
&\partial_{t} n_{\e/\h, \k} = -2 \Im [\varDelta_{\k} p^{*}_{\k}] - 2\gamma (n_{\e/\h,\k}- n^{0}_{\e/\h,\k} ),
\label{eq:MSBEs--nehk}
\end{align}
\end{subequations}
%----------
where $N_{\k} = n_{\e, \k} + n_{\h,\k} - 1$ denotes the population inversion.
In the derivation, the second term in Eq.~\eqref{eq:self-consistency2} has been replaced by phenomenological relaxation terms proportional to $\gamma$ and $\kappa$ and we have introduced
%----------
% Equation
%----------
\begin{align}
& p^{0}_{\k} \equiv 0, \quad
n^{0}_{\e/\h, \k} \equiv f_{\e/\h, \k}.
\label{eq:RTA}
\end{align}
%----------
Here, $f_{\e/\h, \k} \equiv [1 + \exp \{ \beta (\tilde{\epsilon}_{\e/\h, \k} - \mu^{\B}_{\e/\h} ) \} ]^{-1}$ denotes the Fermi distribution with the chemical potential $\mu^{\B}_{\e/\h}$ of the electron (hole) pumping bath, the approximation of which is called the RTA.~\cite{Henneberger92}
Each relaxation term suggests that the photon field $a_{0}$ decays with a rate of $\kappa$, the distribution function $n_{\e/\h,\k}$ is driven to approach the Fermi distribution $f_{\e/\h, \k}$, namely, the thermalization, and $p_{\k}$ is reduced due to the thermalization-induced dephasing.

Under the steady-state condition, $\partial_t \Ex{\oO_i} = 0$, for example, the lasing solution can be obtained by determining the unknown variable $a_0$, $p_{\k}$, $n_{\e/\h,\k}$, and $\mu$ in Eqs.~\eqref{alleq:MSBEs} and \eqref{eq:RTA}.
We note that, in contrast to the BCS theory, $\mu$ is {\em not} a given parameter but an unknown variable corresponding to the laser frequency.
This is equivalent to find an appropriate frequency with which the lasing oscillation of $a_0$ and $p_{\k}$ seems to remain stationary on the rotating frame.

%=======================================
% Key results of our formalism
%=======================================
\subsection{Key results of our formalism} \label{subsec:Key results}
As seen in Subsection~\ref{subsec:the BCS theory and the MSBEs}, the BCS theory and the MSBEs are based on the common Hamiltonian with the same mean-field approximation.
However, the way of deriving the self-consistent equations are different from each other.
In the case of the BCS theory, the density operator $\hat{\rho}^{\MF}$ is directly described by $\oH{S}^{\MF}$ [Eq.~\eqref{eq:equilibrium}].
In contrast, in the case of the MSBEs, Eq.~\eqref{eq:self-consistency2} is alternatively used to introduce the phenomenological relaxation terms.
Here, we should notice that any assumption is not used for $\hat{\rho}^{\MF}$ in Eq.~\eqref{eq:self-consistency2}, which indicates that the MSBEs under the RTA may incorporate the BCS theory at least in principle.

It would therefore be instructive to discuss an approach to derive the BCS theory from the MSBEs under the RTA.
It is, however, apparent that the BCS theory cannot be reproduced by the MSBEs when the relaxation term is completely neglected ($\kappa = \gamma = 0$) because there is no term to drive the system into equilibrium in the MSBEs.~\cite{Note7}
In this context, we should consider a physically natural limit of $\gamma \to 0^{+}$ after $\kappa \to 0$ in order to thermalize the system into equilibrium.
Unfortunately, however, the MSBEs under the RTA cannot recover the BCS theory even by taking this limit.
Obviously, the phenomenological RTA is the cause of this failure.

Based on the generating functional approach (see Sections~\ref{sec:Generating Functional} and \ref{sec:real time}), our key result to this problem is to simply replace Eq.~\eqref{eq:RTA} by 
%----------
% Equation
%----------
\begin{subequations}
\label{alleq:p0k2-n0k2}
\begin{align}
&p^{0}_{\k}(t) = \ii \int^{\infty}_{-\infty} \frac{\dd \nu}{2 \pi } 
\left[ \{ 1-f^{\B}_{\h}(-\nu) \} G^{\R}_{12,\k}(t;\nu) \right. \nonumber \\
& \hspace{4.0cm} \left. - f^{\B}_{\e}(\nu) G^{\R*}_{21,\k}(t;\nu) \right], 
\label{eq:p0k2} \\
&n^{0}_{\e/\h,\k}(t) = \int^{\infty}_{-\infty} \frac{\dd \nu}{2 \pi} f_{\e/\h}^{\B}(\nu) A_{11/22, \k}(t; \pm\nu),
\label{eq:n0k2} 
\end{align}
\end{subequations}
%----------
where $f^{\B}_{\e/\h}(\nu) = [1+ \exp\{ \beta (\nu - \mu^{\B}_{\e/\h} + \mu/2) \} ]^{-1}$ denotes the Fermi distribution of the electron (hole) pumping bath and $G^{\R}_{\alpha \alpha' , \k}(t;\nu)$ is an element of the $2\times2$ matrix which evolves according to 
%----------
% Equation
%----------
\begin{align}
& G^{-1}_{0, \k} G^{\R}_{\k}(t;\nu) - G^{\R}_{\k}(t;\nu) [ G^{-1}_{0, \k} ]^{\dagger} \nonumber \\
& = \varSigma^{\R}_{\k}(t) G^{\R}_{\k}(t;\nu) - G^{\R}_{\k}(t;\nu)\varSigma^{\R}_{\k}(t) \nonumber \\
& \quad - \frac{\ii}{2} \left\{ \partial_{t}\varSigma^{\R}_{\k}(t) \partial_{\nu}G^{\R}_{\k}(t;\nu) + \partial_{\nu}G^{\R}_{\k}(t;\nu) \partial_{t}\varSigma^{\R}_{\k}(t)  \right\},
\label{eq:EOM_GR}
\end{align}
%----------
where
%----------
% Equation
%----------
\begin{subequations}
\label{alleq: GR_Inv2-Sigma2}
\begin{align}
&G^{-1}_{0, \k} \equiv
\begin{pmatrix}
\frac{\ii}{2} \partial_{t} + \nu - \xi_{\e,\k} & 0 \\
0 & \frac{\ii}{2} \partial_{t} + \nu + \xi_{\h,\k} 
\end{pmatrix},
\label{eq:EOM_GR_Inverse2}\\
&\varSigma^{\R}_{\k}(t) \equiv
\begin{pmatrix}
\varSigma^{\BGR}_{\e, \k} - \ii \gamma & - \varDelta_{\k} \\
- \varDelta^{*}_{\k}  & - \varSigma^{\BGR}_{\h, \k} - \ii \gamma
\end{pmatrix}.
\end{align}
\end{subequations}
%----------
The time-dependent single-particle spectral function $A_{\alpha \alpha', \k}$ is then given by 
%----------
% Equation
%----------
\begin{align}
A_{\alpha \alpha', \k}(t;\nu) \equiv \ii (G^{\R}_{\alpha \alpha' , \k}(t;\nu) - G^{\R*}_{\alpha' \alpha, \k}(t;\nu)),
\label{eq:A}
\end{align}
%----------
which couples to Eq.~\eqref{alleq:MSBEs} through Eq.~\eqref{alleq:p0k2-n0k2}.

Although the equations still keep the form of MSBEs [Eq.~\eqref{alleq:MSBEs}], the important difference is that the renormalization of the electronic band structures, caused by the \mbox{e-h} pairing $\varDelta_{\k}$, for example, is taken into account through $A_{\alpha \alpha', \k}$, or equivalently $G^{\R}_{\alpha \alpha', \k}$.
In this sense, the formalism generalizes the \mbox{MSBEs} under the RTA.
The frequency $\nu$-dependence in Eq.~\eqref{alleq:p0k2-n0k2} means that the correlations with the pumping baths, or the past history of the system-bath interactions, influence on the dynamics in the non-Markovian way.
This is important to describe the redistributions of the carriers in the renormalized bands because the particle energies cannot be measured instantaneously due to the uncertainty principle.
In the next section, we also see that the band renormalization and the correlations are essential to study the cooperative phenomena ranging form (quasi-)equilibrium to nonequilibrium in a unified way.

%======================================================
% Relationship of the cooperative phenomena
%======================================================
\section{Relationship of the cooperative phenomena}\label{sec:Relationship}
In the previous section, we have introduced our model Hamiltonians and described our key results based on the generating functional approach. 
Although we will postpone our theoretical treatment and derivation to Sections~\ref{sec:Generating Functional} and \ref{sec:real time} for clarity, we alternatively show here that our formalism is appropriate to discuss the cooperative phenomena, such as the BEC, BCS, LASER, SR and SF, in a unified way.
Then, as important examples, we study the BEC-BCS-LASER crossover in the exciton-polariton systems~\cite{Imamoglu96,Dang98,Deng03,Balili09,Nelsen09,Snoke12,Tempel12-1,Tempel12-2,Tsotsis12,Kammann12,Yamaguchi12,Yamaguchi13,Yamaguchi14} and the connections between the Fermi-edge SF~\cite{Kim13} and the \mbox{e-h} BCS phase with several numerical calculations.

%=======================================
% Connections to the BCS theory and the MSBEs
%=======================================
\subsection{Connections to the BCS theory and the MSBEs} \label{subsec:Connections}
One of the fastest ways to understand our formalism is to find the conditions to recover the BCS theory and the MSBEs under the RTA.
For this purpose, we first consider the situation where the band renormalization caused by the \mbox{e-h} pairing $\varDelta_{\k}$ is neglected. 
In this case, by considering only the electron-electron (\mbox{e-e}) and hole-hole (\mbox{h-h}) Coulomb interactions, the single-particle spectral function can be approximated as
%----------
% Equation
%----------
\begin{align}
&A_{11/22,\k}(t;\nu) \simeq 2 \pi \delta(\nu \mp \tilde{\xi}_{\e/\h,\k}),
\label{eq:QuasiParticleApprox1}
\end{align}
%----------
and, in the same accuracy, the off-diagonal element of $G^{\R}_{\alpha \alpha', \k}$ becomes
%----------
% Equation
%----------
\begin{align}
G^{\R}_{12/21,\k}(t;\nu) \simeq 0,
\label{eq:QuasiParticleApprox2}
\end{align}
%----------
which are essentially the same approximation known as the quasi-particle approximation.~\cite{Rammer07}
As a result, we do not have to solve Eq.~\eqref{eq:EOM_GR} any more.
By substituting Eqs.~\eqref{eq:QuasiParticleApprox1} and \eqref{eq:QuasiParticleApprox2} into Eq.~\eqref{alleq:p0k2-n0k2}, we obtain
%----------
% Equation
%----------
\begin{align}
&p^{0}_{\k} = 0, 
&n^{0}_{\e/\h, \k} = \frac{1}{1 + \exp \{ \beta (\tilde{\epsilon}_{\e/\h, \k} - \mu^{\B}_{\e/\h} ) \}},
\label{eq:ConditionMSBE}
\end{align}
%----------
which is now exactly identical to Eq.~\eqref{eq:RTA} and the standard \mbox{MSBEs} under the RTA are recovered.
In this context, in the standard \mbox{MSBEs} describing the semiconductor lasers, the effects of the \mbox{e-h} pairing are {\em not} taken into account in the band renormalization.
This would be one of the major reasons why many authors believe that the \mbox{e-h} pairs are dissociated under the standard lasing condition, based on the knowledge under the non-lasing conditions.~\cite{Houdre95}
This is, however, only an approximation to simply describe the lasing physics in semiconductors; Eqs.~\eqref{eq:QuasiParticleApprox1} and \eqref{eq:QuasiParticleApprox2} are validated, for example, when $|\varDelta_{\k}| \ll \gamma \ll T$ if the time-dependence of the band renormalization can be adiabatically eliminated,
the situation of which is similar to the gapless superconductor.~\cite{Szymanska03}
At least in principle, therefore, there should be bound \mbox{e-h} pairs whenever lasing,~\cite{Yamaguchi13} or more generally, whenever the phase symmetry is broken, as discussed later.~\cite{Note1}

For the description of the SR, we note that the standard time-dependent \mbox{MSBEs} can be used in the limit of large $\kappa$ in an analogous way to the two-level systems interacting with a single-mode photon field, called the Dicke model.~\cite{Dicke54, Bonifacio71-1,Bonifacio71-2, Bonifacio75-1, Bonifacio75-2, Polder79} 
In the case of the SF, however, the initial condition~\cite{Note2} should be determined by the quantum fluctuations, or equivalently the spontaneous emission to the photon field, which triggers the spontaneous development of the macroscopic coherence; see also Ref.~\onlinecite{Gross82} for a review.
In this context, our formalism is also available to discuss the SF if the initial condition is determined correctly.

In contrast to the standard \mbox{MSBEs}, however, our treatment can drive the system toward the quasi-equilibrium state as well as the nonequilibrium steady state (NESS), after a certain period of time.
To see this, we next consider the steady state condition $t \rightarrow \infty$ by taking $\partial_{t} = 0$ in Eqs.~\eqref{alleq:MSBEs} and \eqref{eq:EOM_GR_Inverse2}.
In this situation, we can find the solution for Eq.~\eqref{eq:EOM_GR} as 
%----------
% Equation
%----------
\begin{align}
&G^{\R}_{\k}(\nu) =
\begin{pmatrix}
\nu - \tilde{\xi}_{\e, \k} + \ii \gamma & \varDelta_{\k} \\
\varDelta^{*}_{\k} & \nu + \tilde{\xi}_{\h, \k} + \ii \gamma
\end{pmatrix}^{-1}.
\label{eq:GR_SS}
\end{align}
%----------
As a result, the single-particle spectral function becomes 
%----------
% Equation
%----------
\begin{multline}
A_{11/22,\k}(\nu) = 2|u_{\k}|^2 \frac{\gamma}{(\nu - \tilde{\xi}^{-}_{\eh,\k} \mp E_{\k})^2 + \gamma^2} \\
 + 2|v_{\k}|^2 \frac{\gamma}{(\nu - \tilde{\xi}^{-}_{\eh,\k} \pm E_{\k})^2 + \gamma^2},
\label{eq:A_SS}
\end{multline}
%----------
where $u_{\k}$ and $v_{\k}$ are the Bogoliubov coefficients
%----------
% Equation
%----------
\begin{align*}
u_{\k} \equiv \sqrt{\frac{1}{2} + \frac{\tilde{\xi}^{+}_{\eh,\k}}{2 E_{\k}}}, \quad
v_{\k} \equiv e^{\ii \theta_{\k}} \sqrt{\frac{1}{2} - \frac{\tilde{\xi}^{+}_{\eh,\k}}{2 E_{\k}}},
\end{align*}
%----------
with $\theta_{\k} \equiv \arg (\varDelta_{\k})$.
Now, Eqs.~\eqref{alleq:MSBEs} and \eqref{alleq:p0k2-n0k2} with Eqs.~\eqref{eq:GR_SS} and \eqref{eq:A_SS} [Eq.~\eqref{eq:A}] are the very equations shown in our previous work.~\cite{Yamaguchi12, Yamaguchi13}
Note that, under the steady-state condition $\partial_{t} = 0$, $\mu$ becomes one of the unknown variables with which the temporal oscillation of the photon amplitude $a_{0}$ and polarization function $p_{\k}$ seems to remain stationary on the rotating frame, as described above.
Hence, $\mu$ corresponds to the frequency of the cavity photon amplitude $a_0$, at which the photoluminescence has a main peak (Section~\ref{sec:Spectral}).
At the same time, $a_0$ can be set to be real without loss of generality.

Under the steady-state condition, the formalism now allows us to clearly understand the standpoint of the BCS theory. 
For this purpose, let us discuss the limit of equilibrium, namely, $\gamma \to 0^{+}$ after $\kappa \to 0$.
By assuming $\epsilon_{\e, \k} = \epsilon_{\h, \k}$ ($m_{\e} = m_{\h} $) with the charge neutrality $\mu^{\B}_{\e} = \mu^{\B}_{\h}$ for simplicity, $\mu = \mu_{\B} \equiv \mu^{\B}_{\e} + \mu^{\B}_{\h}$ can be obtained in the vanishing limit of $\kappa$.
This means that the system reaches in chemical equilibrium with the pumping baths because photons are not lost any more from the microcavity.
As a result, $\mu$ becomes a given parameter equivalent to $\mu_{\B}$.
Then, by taking the limit of $\gamma \to 0^{+}$, the integrals in Eq.~\eqref{alleq:p0k2-n0k2} can be performed analytically; the BCS theory (Subsubsec.~\ref{subsubsec:BCS}) is then successfully recovered.
In this derivation, $\gamma \ne 0$ is required to be canceled down even though $\gamma$ does not appear in the final form.
This means that thermalization is essential to recover the equilibrium theory.

%----------
% Figure
%----------
\begin{figure*}[!tb] 
\centering
\includegraphics[width=.90\textwidth, clip]{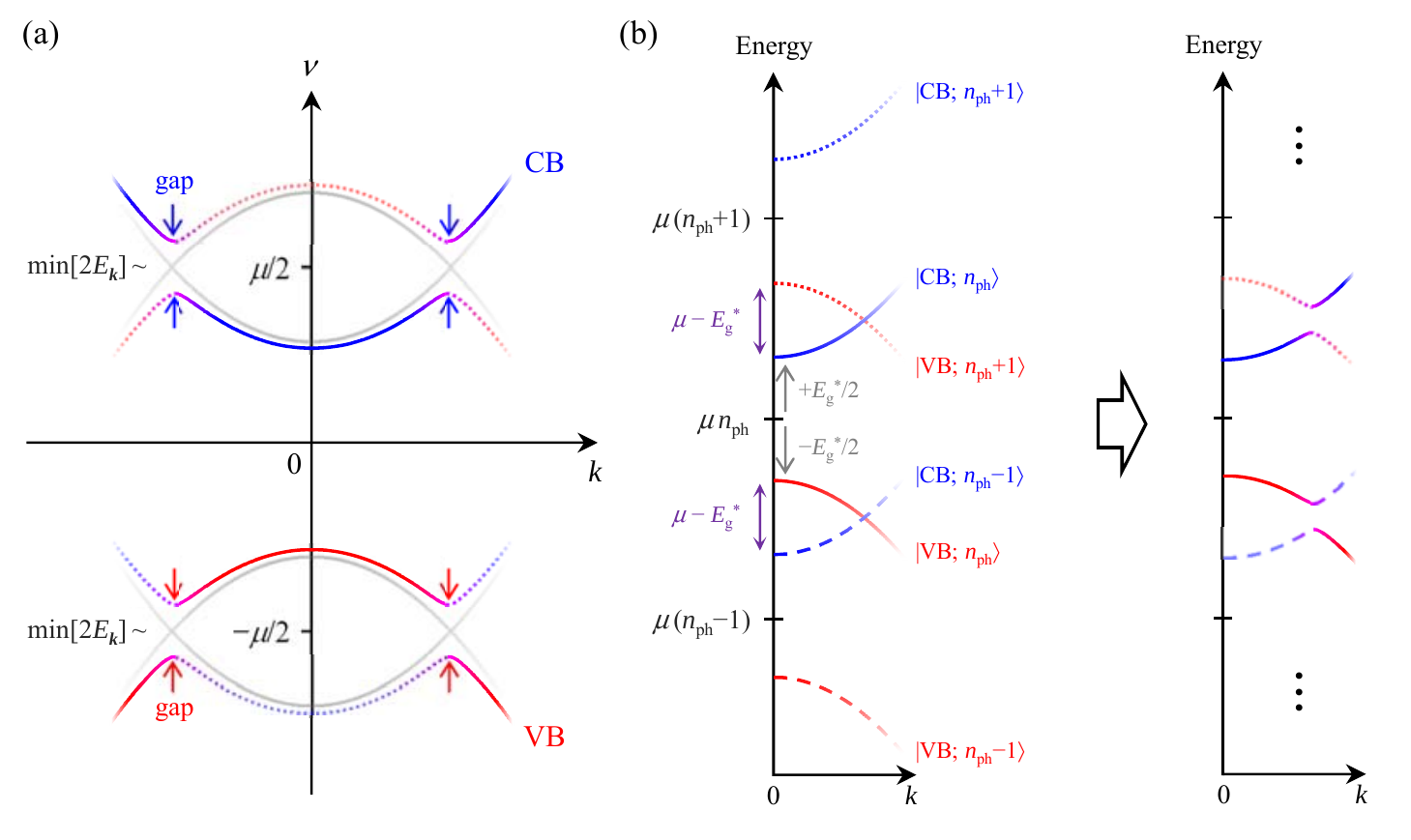} 
\caption{(Color online) Schematic illustration of the typical band renormalization.
(a)~Typical renormalized CB and VB structures obtained from the single-particle spectral functions $A_{11,\k}(\nu)$ and $A_{22,\k}(\nu)$ in Eq.~\eqref{eq:A_SS}, respectively.
The energies are shifted by $\pm \mu/2$ to recover the laboratory frame.
The gray solid lines show the band structures without the \mbox{e-h} pairing effect $\varDelta_{\k}$.
One can find remarkable similarities to the excitation spectrum in the BCS theory for the metal superconductors.
(b)~The mechanism of opening the gaps. 
Without the \mbox{e-h} paring, the CB and VB structures with $n_{\ph}$ photons are shown by the solid lines (left).
These bands are further shifted by $\pm \mu$ in energy when the number of photons are incremented (decremented) by one, as indicated by the dotted and dashed lines.
The gaps are then opened in the same manner as the Rabi splitting or the Mollow triplet in resonance fluorescence~\cite{Scully97,Schmitt-Rink88,Henneberger92,Mollow69,Horikiri14} when the phase symmetry is broken (right).
Thus, the BCS theory has close relationship to the Mollow triplet in semiconductors.
For simplicity, $\epsilon_{\e,\k} = \epsilon_{\h,\k}$ is assumed with the charge neutrality $\mu^{\B}_{\e} = \mu^{\B}_{\h}$.
$\E{g}^{*} \equiv \E{g} + \varSigma^{\BGR}_{\e, \k = 0} + \varSigma^{\BGR}_{\h, \k = 0}$ is the Coulomb-renormalized band gap energy.
}
\label{fig02:BandRenormalization}
\end{figure*}
%----------

In a physical sense, however, this limit is trivial because the nonequilibrium theory should recover the equilibrium theory by taking the negligible limit of reservoirs.
More important situation is that the system can be identified as being in equilibrium (quasi-equilibrium) as long as the \mbox{e-h} system is excited and thermalized even though photons are continuously lost.
In this context, we have revealed~\cite{Yamaguchi12, Yamaguchi13} that the system can indeed be in quasi-equilibrium if the condition
\begin{center}
\hypertarget{CI}{(I)~$\min [2E_{\k}] \gtrsim \mu_{\B} - \mu + 2\gamma + 2T$},
\end{center}
is satisfied, where $\min [2E_{\k}]$ denotes the minimum of $2E_{\k}$ when the wave number $\k$ is changed.
From Eqs.~\eqref{alleq:MSBEs} and \eqref{alleq:p0k2-n0k2} with Eqs.~\eqref{eq:GR_SS} and \eqref{eq:A_SS}, then, we can obtain the gap equation  [Eq.~\eqref{eq:BCSgap1}] and the number equation [Eq.~\eqref{eq:BCS--nehk}].
In this case, the effective \mbox{e-h} attractive potential $U^{\eff}_{\k', \k}$ is replaced by $U^{\eff, \kappa}_{\k', \k} \equiv |g|^2/(\xi_{\ph, 0} - \ii \kappa) + U'_{\k' - \k}$ where the effect of $\kappa$ is included.~\cite{Note3}
We remark that, in this situation, $\beta$ and $\mu$ can be regarded as the inverse temperature and the chemical potential of the {\em system}, respectively, even though $\beta$ and $\mu$ are originally introduced as the inverse temperature of the pumping baths and the frequency of the rotating frame, respectively; see below Eq.~\eqref{alleq:p0k2-n0k2}.
Also, $\min [2E_{\k}]$ is equivalent to the minimum energy required to break the \mbox{e-h} pairs in a similar context to the metal superconductors.
The equilibrium phases from the BEC to the BCS states can then be covered by our formalism at least for $T$ = 0.~\cite{Comte82,Kamide10,Kamide11,Byrnes10}

However, the system can no longer be in quasi-equilibrium when the condition \hyperlink{CI}{(I)} is violated and nonequilibrium effect becomes significant.
The standard steady-state MSBEs are then recovered in $\k$-regions satisfying 
\begin{center}
(II$'$)~$\mu_{\B} - \mu \gtrsim 2E_{\k} + 2\gamma + 2T$,
\end{center}
which can be found whenever 
\begin{center}
\hypertarget{CII}{(II)~$\mu_{\B} - \mu \gtrsim \min [2E_{\k}] + 2\gamma + 2T$},
\end{center}
is fulfilled.~\cite{Note3}
The system thus enters into the lasing regime in the NESS~\cite{Note4}
 and the physical meaning of $\mu$ changes into the oscillating frequency of the laser action.
In this context, the formalism can naturally describe the change from the quasi-equilibrium (the BEC and BCS phases) to nonequilibrium phenomena (lasing) without {\it a priori} assuming the quasi-equilibrium and nonequilibrium situations.~\cite{Note5}

It is now instructive to note that the single-particle spectral function $A_{11/22,\k}(\nu)$ in Eq.~\eqref{eq:A_SS} 
 has remarkable similarities to the superconductivities in the equilibrium statistical theory.~\cite{Abrikosov75}
However, if the unknown variables are determined by Eqs.~\eqref{alleq:MSBEs} and \eqref{alleq:p0k2-n0k2}, Eq.~\eqref{eq:A_SS} can be used even in the lasing regime because only the steady-state assumption ($\partial_t = 0$) is required in the derivation.
It is then obvious that the pairing gaps of $\min [2E_{\k}]$ are opened around $\pm\mu/2$ in the renormalized CB and VB structures in a very similar way to the superconductivities.
In the case of the conduction band, for example, $A_{11,\k}(\nu)$ has peaks around $\nu = \tilde{\xi}^{-}_{\eh,\k} \pm E_{\k}$ and the difference of the two peaks becomes 2$E_{\k}$ in energy at fixed $\k$; the gap therefore corresponds to $\min [2E_{\k}]$, as typically shown in Figure~\ref{fig02:BandRenormalization}(a).

The mechanism of opening the gaps is closely related to the Rabi splitting or the Mollow triplet in resonance fluorescence~\cite{Scully97,Mollow69,Schmitt-Rink88,Henneberger92,Horikiri14} although the phase symmetry is spontaneously broken in our case; see also Refs.~\onlinecite{,Valle09,*Valle10} and \onlinecite{,Valle11} for the Mollow triplet under the incoherent pumping.
Without the \mbox{e-h} pairing effect, the CB and VB structures are represented by the solid lines in the left side of Figure~\ref{fig02:BandRenormalization}(b) when $n_{\ph}$ photons in the cavity.
Here, the CB and VB are renormalized by the \mbox{e-e} and \mbox{h-h} Coulomb interactions, as already seen in Eq.~\eqref{eq:QuasiParticleApprox1}, and the total energy is shifted by $\mu n_{\ph}$ from Figure~\ref{fig01:Model}.
The total energy can further be shifted by $\pm \mu$ when the number of photons is changed by one, as illustrated by the dotted and dashed lines.
These energy bands are not mixed with each other because the total Hamiltonian $\hat{H}$ commutes with the total excitation number $\hat{N}$.
However, this is {\em not} the case once the phase symmetry is broken, or equivalently, the photon amplitude is developed.
As a result, the pairing gaps are inherently opened [the right side of Figure~\ref{fig02:BandRenormalization}(b)] whenever the photon amplitude $a_0$ has a non-zero value, regardless of whether the system is in quasi-equilibrium.
In other words, there must be \mbox{e-h} pairs whenever the symmetry is broken, at least in principle.
This is an important result because this means that (light-induced) bound \mbox{e-h} pairs should exist even in the standard lasing regime in contrast to earlier expectations.~\cite{Yamaguchi13}

For later convenience, we further point out that, under the steady-state assumption, Eqs.~\eqref{alleq:MSBEs} and \eqref{alleq:p0k2-n0k2} with Eqs.~\eqref{eq:GR_SS} and \eqref{eq:A_SS} can be rewritten as 
%----------
% Equation
%----------
\begin{subequations}
\label{alleq:BCSform}
\begin{align}
& \varDelta_{\k} = \sum_{\k'} U^{\eff, \kappa}_{\k', \k} \int^{\infty}_{-\infty} \frac{\dd \nu}{2 \pi} f^{\SS}_{\eh, \k'}(\nu) A_{12, \k'}(\nu),
\label{eq:BCSform1}\\
& n_{\e/\h, \k} = \int^{\infty}_{-\infty} \frac{\dd \nu}{2 \pi} f^{\SS}_{\e/\h, \k}(\nu) A_{11/22, \k}(\pm \nu), 
\label{eq:BCSform2}
\end{align}
\end{subequations}
%----------
where $f^{\SS}_{\eh, \k}(\nu)$ and $f^{\SS}_{\e/\h, \k}(\nu)$ are defined as
%----------
% Equation
%----------
\begin{subequations}
\label{alleq:f_SS}
\begin{align}
& f^{\SS}_{\eh, \k}(\nu) \equiv \frac{1}{2} \{ f^{\B}_{\e}(\nu) - f^{\B}_{\h}(-\nu) \} \nonumber \\
& \qquad\qquad + \frac{1}{2} \{ f^{\B}_{\e}(\nu) + f^{\B}_{\h}(-\nu) - 1 \} \frac{\tilde{\xi}^{+}_{\eh, \k} + \ii \gamma}{\nu - \tilde{\xi}^{-}_{\eh,\k}}, \\
& f^{\SS}_{\e/\h, \k}(\nu) \equiv f^{\B}_{\e/\h}(\nu) \eta_{\e/\h, \k} (\nu) \nonumber \\
& \qquad\qquad\qquad + \{ 1-f^{\B}_{\h/\e}(-\nu) \} \{ 1 - \eta_{\e/\h,\k}(\nu) \},
\label{eq:f_SS-2}
\end{align}
\end{subequations}
%----------
respectively; see also Appendix~\ref{app:Effective distribution} for the derivation.
Eq.~\eqref{alleq:BCSform} is formally analogous to the BCS gap equation and the number equation [Eqs.~\eqref{eq:BCSgap1} and \eqref{eq:BCS--nehk}] rather than the MSBEs under the RTA.
In particular, $f^{\SS}_{\e/\h, \k}(\nu)$ corresponds to the effective steady-state distribution function in which
%----------
% Equation
%----------
\begin{align}
\eta_{\e/\h, \k}(\nu) \equiv \frac{ (\nu+\tilde{\xi}_{\h/\e,\k})^2 + \gamma^2}{(\nu+\tilde{\xi}_{\h/\e,\k})^2 + \gamma^2 + |\varDelta_{\k}|^2},
\label{eq:eta}
\end{align}
%----------
denotes a weighting factor ($0 < \eta_{\e/\h, \k}(\nu) \le 1$) due to the mixture of the conduction and valence bands.
Note that $\eta_{\e/\h, \k}(\nu) = 1$ and $f^{\SS}_{\e/\h, \k}(\nu) = f^{\B}_{\e/\h}(\nu)$ when there is no \mbox{e-h} pairing effect ($\varDelta_{\k}=0$).
However, Eqs.~\eqref{eq:BCSform2}, \eqref{eq:f_SS-2} and \eqref{eq:eta} mean that $n_{\e, \k}$ and $n_{\h, \k}$ are also influenced by the hole and electron bath distributions [$f^{\B}_{\h}(-\nu)$ and $f^{\B}_{\e}(-\nu)$], respectively, when the band mixing occurs ($\varDelta_{\k} \ne 0$).

%=======================================
% BEC-BCS-LASER crossover
%=======================================
\subsection{BEC-BCS-LASER crossover} \label{subsec:BEC-BCS-LASER}
To gain further insight into the relationship among the cooperative phenomena, we now discuss numerical results calculated under the steady-state condition.
In the calculations, the $\k$-dependence of $\varDelta_{\k}$ is eliminated by using a contact potential $U'_{\q} = U$ with the replacement of $\sum_{\k} \rightarrow \frac{S}{2\pi} \int^{k_{\text{c}}}_{0}\dd k k$.~\cite{Haug09}
Here, $S$ is the area of the system and $k_{\text{c}}$ is the cut-off wave number.
For simplicity, $m_{\e} = m_{\h} $ is also assumed with the charge neutrality $\mu^{\B}_{\e} = \mu^{\B}_{\h}$.
In this context, our calculation is not quantitative but qualitative even though the parameters are taken as realistic as possible;
unless otherwise stated, we use the parameters shown in Table~\ref{tableII}.
In this situation, the exciton level $\E{ex}$ is formed at 10~meV below $\E{g}$ ($\E{ex} = \E{g} - 10$~meV) and the lower polariton (LP) level $\E{LP}$ is formed at 20~meV below $\E{g}$ ($\E{LP} = \E{g} - 20$~meV $= \E{ex} - 10$~meV) under the resonant condition $\E{cav} = \E{ex}$.~\cite{Note3}
The Rabi splitting ($\equiv \E{Rabi}$) is therefore 20~meV. 
To see the nonequilibrium effects, $\kappa = $ 0.1~$\mu$eV, 100~$\mu$eV and 100~meV are used for comparison but we note that $\kappa = $ 100~$\mu$eV is a reasonable value in current experiments.~\cite{Yamaguchi13,Byrnes14}

\begin{figure}[!tb] 
\centering
\includegraphics[width=.48\textwidth, clip]{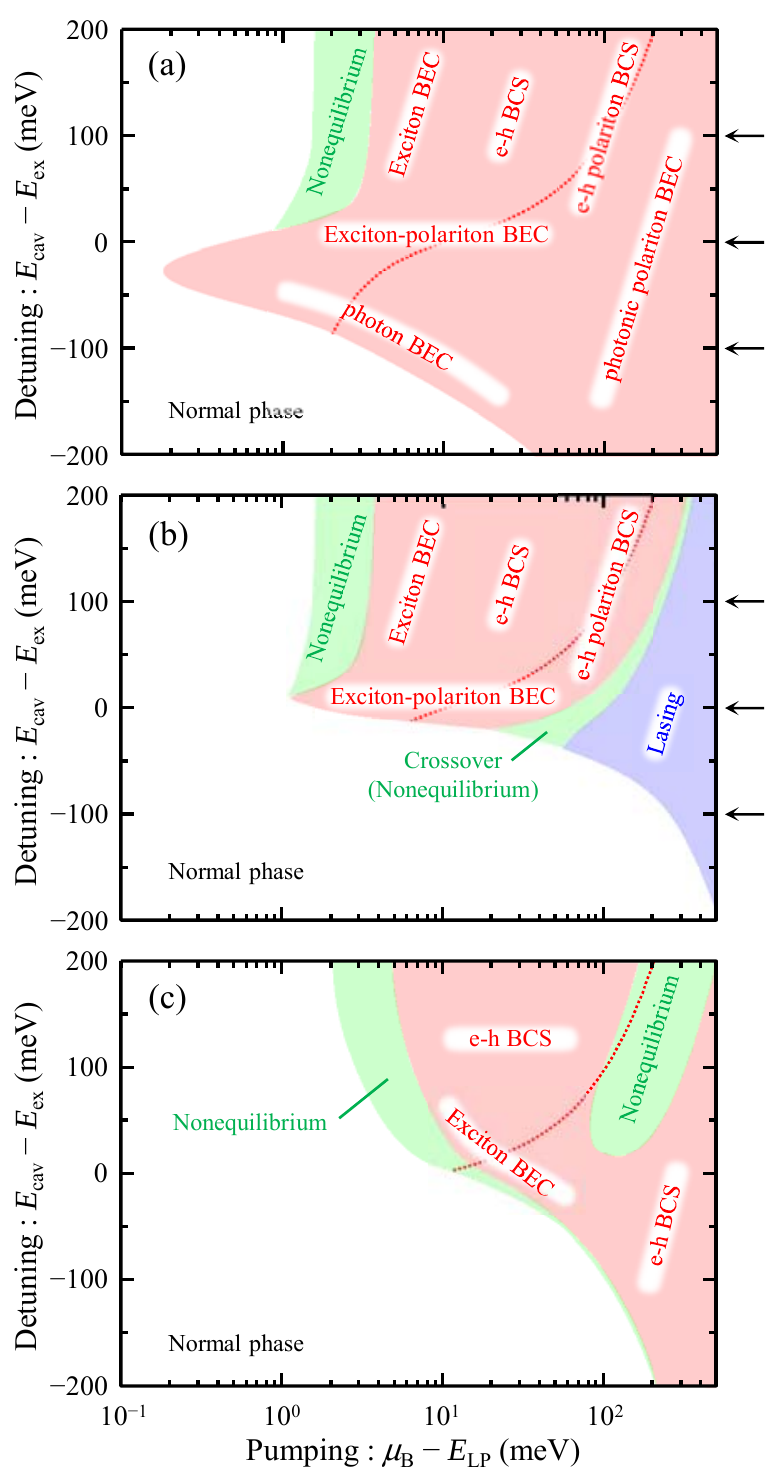} 
\caption{(Color online) Phase diagrams for (a) $\kappa = $ 0.1~$\mu$eV, (b) $\kappa = $ 100~$\mu$eV and (c) $\kappa = $ 100~meV.
Red and blue colors indicate that the quasi-equilibrium condition (I) and the lasing condition (II) are satisfied, respectively, while green colors indicate that neither of them is satisfied. 
A condition of $\mu_{\B} = \E{cav}$ is indicated by the dotted line;
the influence of the cavity becomes large when $\mu_{\B}$ goes over the line.
We note that $\E{LP}$ in the horizontal axis depends on the cavity resonance $\E{cav}$.
}
\label{fig03:PhaseDiagram}
\end{figure}

Figure~\ref{fig03:PhaseDiagram} shows the phase diagrams calculated by changing the detuning $\E{cav} - \E{ex}$ and the chemical potential of the pumping baths $\mu_{\B}$, the pumping parameter.
The landscape of the phase diagram is significantly modified by the rate of the cavity photon loss $\kappa$.
For the case with $\kappa = $ 0.1~$\mu$eV in Figure~\ref{fig03:PhaseDiagram}(a), most of the area is dominated by quasi-equilibrium phases satisfying the condition \hyperlink{CI}{(I)} due to the low rate of the cavity photon loss and, as a result, one finds a variety of distinct BEC and BCS phases smoothly connected with each other.
In contrast, for the case with $\kappa = $ 100~$\mu$eV in Figure~\ref{fig03:PhaseDiagram}(b), there arises the lasing phase satisfying the condition \hyperlink{CII}{(II)} and the whole area of the ordered phases is decreased.
We note that $\E{Rabi} \gg \kappa$ holds in Figures~\ref{fig03:PhaseDiagram}(a) and \ref{fig03:PhaseDiagram}(b).
However, when $\kappa$ becomes sufficiently large ($\kappa \gg \E{Rabi}$) in Figure~\ref{fig03:PhaseDiagram}(c), the quasi-equilibrium phases again spread over the large area despite the increased photon loss.
The emergence of the quasi-equilibrium phases is seemingly counterintuitive but the situation is quite similar to the Purcell effect~\cite{Purcell46,Goy83,Scully97} known for a two-level emitter inside a single-mode cavity; the emission rate of the two-level emitter is {\em decreased} when the cavity photon loss is {\em increased} in the weak coupling regime, the physics of which is intuitively the same as the impedance matching.
Hence, there exists an optimal $\kappa$ to maximize the decay rate~\cite{Yamaguchi08-2} and, in the ultimate limit of $\kappa = \infty$, the effect of the cavity loss inversely becomes negligible.
In other words, $\kappa = \infty$ is identical to the situation that the cavity is practically non-existent.
As a result, the quasi-equilibrium phases dominate the phase diagram when $\kappa$ becomes sufficiently large.
This situation is, in turn, appropriate to study the SF under the continuous pumping, as we will see later, because the cavity plays no essential role.

%----------
% TABLE II
%----------
\begin{table}[!tb]
\caption{\label{tableII} Parameters in the calculation. $m_{0}$ denotes the electron mass in vacuum.}
\begin{ruledtabular}
\begin{tabular}{clc}
Quantity & Value & Unit \\
\hline
$m_{\e}$ & 0.068 & $m_{0}$ \\
$\gamma$ & $4.0 \times 10^{-3}$ & eV \\
$U$ & $2.66 \times 10^{-10}$ & eV \\
$g$ & $6.29 \times 10^{-7}$ & eV \\
$S$ & $100 \times 100$ & $(\mu m)^2$ \\
$k_{\text{c}}$ & $1.36 \times 10^{9}$ & $m^{-1}$ \\
$T$ & 10 & K \\
\end{tabular}
\end{ruledtabular}
\end{table}

%----------
% Figure
%----------
%----------

%----------
% Figure
%----------
\begin{figure*}[!tb] 
\centering
\includegraphics[width=.90\textwidth, clip]{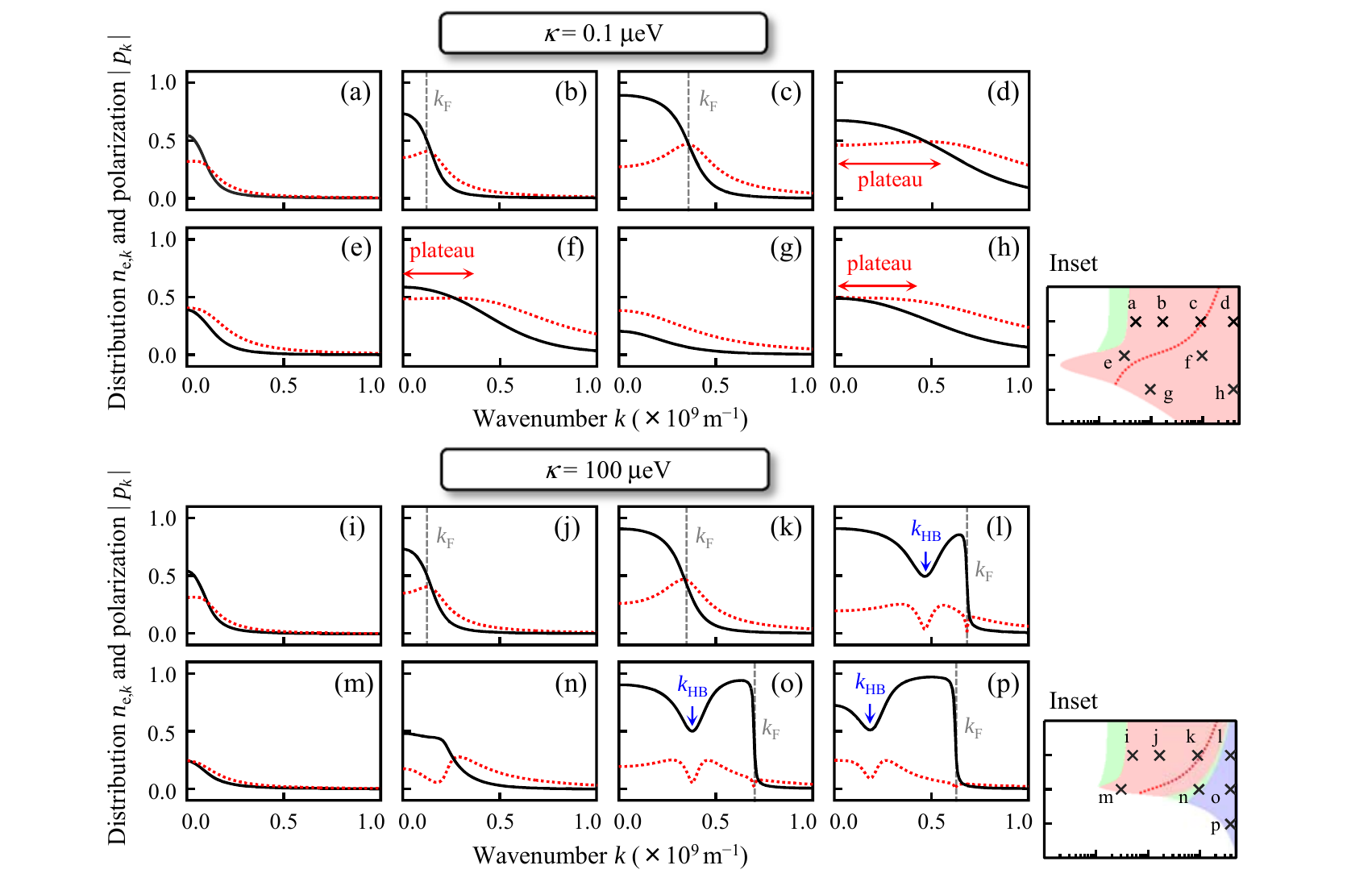} 
\caption{(Color online) The distribution functions $n_{\e, k}$ (solid lines) and the polarization functions $|p_{k}|$ (dotted lines) for various values of $\mu_{\B}$ at the detuning of $+$100~meV, 0~meV and $-$100~meV, indicated by the arrows in Figure~\ref{fig03:PhaseDiagram}.
At the detuning of $+$100~meV, $\mu_{\B} - \E{LP}$ = 5~meV for panels (a) and (i), 15~meV for panels (b) and (j), 90~meV for panels (c) and (k), and 400~meV for panels (d) and (l).
For the zero detuning case, $\mu_{\B} - \E{LP}$ = 3~meV for panels (e) and (m), 100~meV for panels (f) and (n), and 400~meV for panel (o).
Finally, when the detuning is $-$100~meV, $\mu_{\B} - \E{LP}$ = 10 meV for panel (g) and 400~meV for panels (h) and (p). 
Insets show the parameters corresponding to the panels (a)--(p) in the phase diagrams.
$k_{\F}$ and $k_{\text{HB}}$ denote the momentum of the Fermi edge and the kinetic hole burning, respectively.
In our calculations, $k_{\F}$ is introduced by $k_{\F} = \sqrt{2m_{\text{r}}(\mu_{\B}-\E{g}^{*})}$ with $m_{\text{r}}^{-1} \equiv m_{\e}^{-1} + m_{\h}^{-1}$.
}
\label{fig04:Distribution-Polarization}
\end{figure*}
%----------

In the low density regime with $\E{Rabi} \gg \kappa$ [Figures~\ref{fig03:PhaseDiagram}(a) and \ref{fig03:PhaseDiagram}(b)], the behaviors are roughly understood from the photonic and excitonic component of the LP state.~\cite{Kasprzak08, Deng10}
In the positively detuned regime, the excitonic component is increased in the LP state and the photonic component becomes negligible in the limit of $\E{cav} - \E{ex} \gg \E{Rabi}$.
Around the area labeled by the exciton BEC in Figures~\ref{fig03:PhaseDiagram}(a) and \ref{fig03:PhaseDiagram}(b), therefore, the ordered phase is insensitive to the value of the detuning and the cavity photon loss.
In the negatively detuned regime, in contrast, the LP state is dominated by the photonic component in the limit of $- \E{cav} + \E{ex} \gg \E{Rabi}$.
As a result, the system is susceptible to the photonic effect around the area labeled by the photon BEC~\cite{Klaers10} in Figure~\ref{fig03:PhaseDiagram}(a) and the ordered phase disappears in the corresponding area in Figure~\ref{fig03:PhaseDiagram}(b) due to the increased cavity photon loss.
For the case with $\kappa \gg \E{Rabi}$, on the other hand, the normal-mode splitting does not take place.
However, the system still experiences the photonic effect weakly when $|\E{cav} - \E{ex}| \lesssim \kappa$ in the low density regime.
As a result, in the positively detuned regime, the boundary with the normal phase depends on the detuning in Figure~\ref{fig03:PhaseDiagram}(c), while it is almost constant in Figures~\ref{fig03:PhaseDiagram}(a) and \ref{fig03:PhaseDiagram}(b).

In the high density regime, however, the pictures of the excitons are not available any more because the phase space filling of the \mbox{e-h} system becomes non-negligible.
In this situation, the relation between $\mu_{\B}$ and the bare cavity resonance $\E{cav}$ is important to discuss the photonic effect in the case with the positive detuning.
For $\mu_{\B} \lesssim \E{cav}$, there is no carrier around the cavity resonance because the cavity is far above the Fermi edge of the \mbox{e-h} system if the broadening effect of $\kappa$ is neglected.
The system is therefore still insensitive to the photonic effect, as seen in the area of the \mbox{e-h} BCS in Figures~\ref{fig03:PhaseDiagram}(a) and \ref{fig03:PhaseDiagram}(b).
As the pumping is further increased, however, the photonic effect would be discernible for $\mu_{\B} \simeq \E{cav}$ (indicated by the dotted lines) and become prominent for $\mu_{\B} \gtrsim \E{cav}$.
These are given by the change from the \mbox{e-h} polariton BCS to the photonic polariton BEC~\cite{Kamide10,Note6} [Figure~\ref{fig03:PhaseDiagram}(a)] and to the lasing phase [Figure~\ref{fig03:PhaseDiagram}(b)].
However, when $\kappa$ is sufficiently large, the photonic effect becomes week due to the Purcell-like effect and the high-density regime is basically in the \mbox{e-h} BCS phase, as in Figure~\ref{fig03:PhaseDiagram}(c).

We can thus give rough explanations for the phase diagrams even without studying the details of the variables, such as $a_{0}$, $p_{\k}$, $n_{\e/\h, \k}$ and $\mu$.
However, to identify the variety of the BEC and BCS phases, we need more careful discussions with clear criteria to distinguish the respective phases, for example, the photon BEC and the photonic polariton BEC.
For this purpose, in the followings, we focus on the phases with $\kappa = $ 0.1~$\mu$eV and 100~$\mu$eV for the moment to keep our discussion as simple as possible. 

Figure~\ref{fig04:Distribution-Polarization} shows the distribution function $n_{\e, k}$ and the polarization function $|p_{k}|$ obtained for various values of $\mu_{\B}$ at the detuning of $+$100~meV, 0 meV and $-$100~meV, indicated by the arrows in Figures~\ref{fig03:PhaseDiagram}(a) and \ref{fig03:PhaseDiagram}(b).
The lasing phase is then easily distinguished from the other phases when the kinetic hole burning (the dip in the distribution function) is seen [Figures~\ref{fig04:Distribution-Polarization}(l), (o) and (p)].
In contrast, the BCS phase can be distinguished by the presence of a peak in $p_{k}$ around the Fermi momentum $k_{\F}$ resulting from the phase space filling effect [Figures~\ref{fig04:Distribution-Polarization}(b), (c), (j) and (k)], whereas $p_{k}$ and $n_{\e, k}$ are slowly decreased as a function of $k \equiv |\k|$ in the BEC phase [Figures~\ref{fig04:Distribution-Polarization}(a), (d), (e)--(h), (i) and (m)].
In particular, in Figures~\ref{fig04:Distribution-Polarization}(d), (f) and (h), the plateau of $p_{k} \simeq 0.5$ is known as a signature for the photonic polariton BEC.~\cite{Kamide10}

%----------
% Figure
%----------
\begin{figure}[!tb] 
\centering
\includegraphics[width=.45\textwidth, clip]{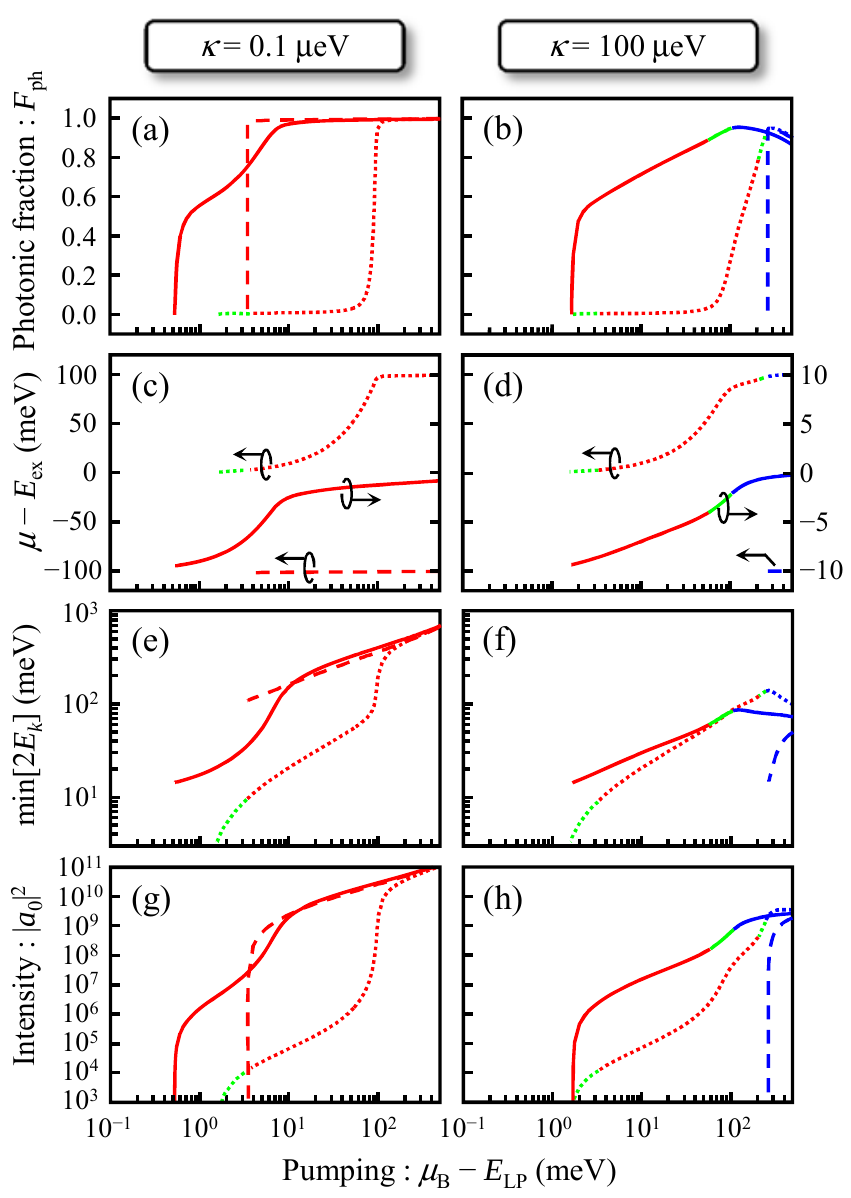} 
\caption{(Color online) Numerical results for the photonic fraction $F_{\ph}$, the frequency $\mu$, the gap energy $\min[2E_{\k}]$, and the coherent number of photons in the cavity $|a_{0}|^2$.
The detuning is $+$100~meV (dotted lines), 0~meV (solid lines) and $-$100~meV (dashed lines), as indicated by the arrows in Figure~\ref{fig03:PhaseDiagram}.
Left panels are for $\kappa = $ 0.1~$\mu$eV and right panels are for $\kappa = $ 100~$\mu$eV.
}
\label{fig05:Results}
\end{figure}
%----------

%----------
% Figure
%----------
\begin{figure*}[!tb] 
\centering
\includegraphics[width=.90\textwidth, clip]{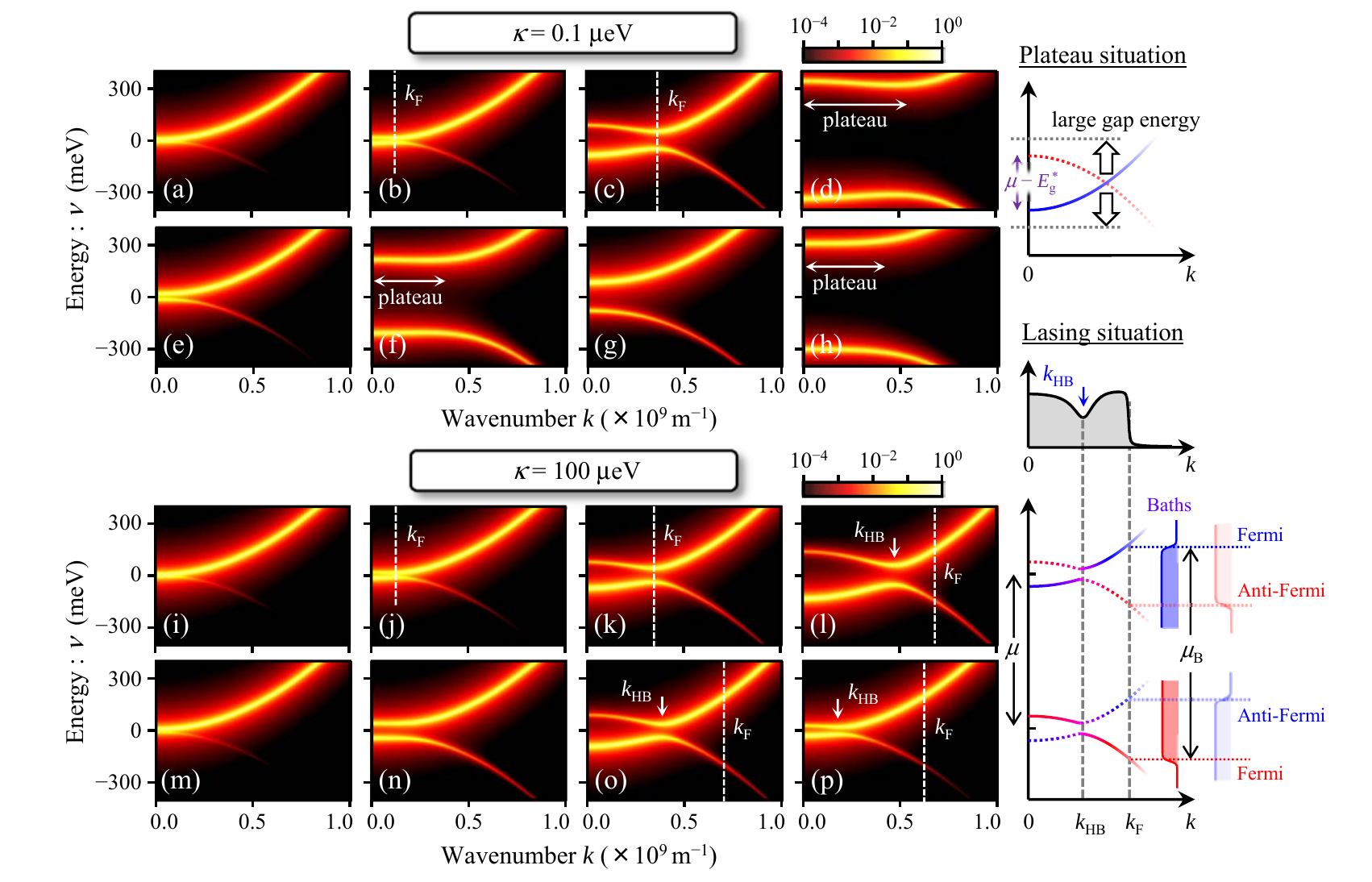} 
\caption{(Color online) The single-particle spectral function $A_{11,k}(\nu)$ for various values of $\mu_{\B}$ at the detuning of $+$100~meV, 0~meV and $-$100~meV (the same as in Figure~\ref{fig04:Distribution-Polarization}).
The origin of the vertical axis is identical to the energy $\mu/2$ in the laboratory frame; see also Figure~\ref{fig02:BandRenormalization}(a).
The right upper inset shows the mechanism of forming the plateau; the gap energy $\min[2E_{\k}]$ is greater than $|\mu - \E{g}^*|$.
The right lower inset shows the relationship of the distribution function $n_{\e,k}$, the renormalized band structures, and the pumping baths under the lasing condition.
We remark that the distribution of electrons (holes) is also influenced by the hole (electron) bath distribution due to the band mixing; see also Eqs.~\eqref{alleq:BCSform}--\eqref{eq:eta}.
We refer to the Fermi level formed by this mechanism as the anti-Fermi level to distinguish from the ordinary one.
}
\label{fig06:Single-Particle}
\end{figure*}
%----------

The photonic effect can then be estimated from the photonic fraction $F_{\ph}$, the ratio of photons to the effective excitation density contributing to the ordered phase; see also Appendix~\ref{app:Photonic fraction} for details.
At the detuning of $+$100~meV with $\kappa = $ 0.1~$\mu$eV, $F_{\ph}$ [the dotted line in Figure~\ref{fig05:Results}(a)] is nearly zero for $\mu_{\B} - \E{LP} \lesssim$ 70~meV but grows rapidly around $\mu_{\B} - \E{LP} \simeq $ 100~meV, and then, $F_{\ph} \simeq 1.0$ for larger pumping.
We note that $\mu_{\B} - \E{LP} \simeq $ 100~meV is almost identical to the dotted line in Figure~\ref{fig03:PhaseDiagram}(a) at $+$100~meV detuning.
These results indeed reveal that the photonic effect is negligible in the low density regime but is discernible for $\mu_{\B} \simeq \E{cav}$ and finally becomes dominant for $\mu_{\B} \gtrsim \E{cav}$ with increasing the pumping, as described above.
As a result, Figures~\ref{fig04:Distribution-Polarization}(a)--\ref{fig04:Distribution-Polarization}(d) are identified as the exciton BEC [$F_{\ph}\simeq0.0$], \mbox{e-h} BCS [$F_{\ph}\simeq 0.0$], \mbox{e-h} polariton BCS [$F_{\ph} \simeq 0.5$] and photonic polariton BEC [$F_{\ph}\simeq1.0$], respectively.
The \mbox{e-h} polariton BCS state has been explicitly distinguished from the \mbox{e-h} BCS phase because the \mbox{e-h} attraction is enhanced by the cavity photons to form the \mbox{e-h} Cooper pairs.~\cite{Kamide10, Byrnes10}
This identification is also evidenced by the behavior of $\mu$ (dotted line) in Figure~\ref{fig05:Results}(c), where $\mu \simeq \mu_{\B} \simeq \E{ex}$ for the exciton BEC, $\mu \simeq \mu_{\B}$ for the \mbox{e-h} BCS and \mbox{e-h} polariton BCS phases (not shown), and $\mu \simeq \E{cav}$ for the photonic polariton BEC.

It is here interesting to notice that $\mu$ is not necessarily in the vicinity of the cavity resonance even though $\mu$ can be regarded as the frequency of the cavity photon amplitude under the steady-state assumption.
This is intuitively equivalent to the classical forced oscillation of the cavity mode;
in the case of the exciton BEC (\mbox{e-h} BCS state), for example, the coherence is developed at the exciton resonance (at the Fermi level) which in turn drives the cavity photon amplitude forcibly, resulting in $\mu \simeq \E{ex}$ ($\mu \simeq \mu_{\B}$).

For $\kappa = $ 100~$\mu$eV, essentially the same identification can be performed for Figures~\ref{fig04:Distribution-Polarization}(i)--\ref{fig04:Distribution-Polarization}(k) with the results shown in Figure~\ref{fig05:Results}(b) and \ref{fig05:Results}(d).
The similarities between the panels (i)--(k) and (a)--(c) in Figure~\ref{fig04:Distribution-Polarization} directly shows that the ordered phases become insensitive to the photonic effect when $\mu_{\B} \lesssim \E{cav}$ in the positively detuned regime.

In the case of zero detuning $\E{cav} = \E{ex}$, the situation is slightly different in particular in the low density regime; $F_{\ph} \sim 0.5$ can be seen immediately after the ordered phase is developed [Figures~\ref{fig05:Results}(a) and \ref{fig05:Results}(b); solid lines].
At the same time, $\mu \simeq \E{LP}$ ($\mu - \E{ex}  \simeq $ $-$10~meV) is observed [Figures~\ref{fig05:Results}(c) and \ref{fig05:Results}(d); solid lines].
In this context, it is reasonable to identify the ordered phase [Figures~\ref{fig04:Distribution-Polarization}(e) and \ref{fig04:Distribution-Polarization}(m)] as the exciton-polariton BEC.
With the increased pumping for $\kappa = $ 0.1~$\mu$eV, however, $F_{\ph} \simeq 1.0$ [Figure~\ref{fig05:Results}(a)] is again found with $\mu \simeq \E{cav}$ [Figure~\ref{fig05:Results}(c)].
Figure~\ref{fig04:Distribution-Polarization}(f) is thus identical to the photonic polariton BEC.
For $\kappa = $ 100~$\mu$eV, in contrast, the system enters into the lasing phase as already revealed by the kinetic hole burning [Figure~\ref{fig04:Distribution-Polarization}(o)] through the crossover regime [Figure~\ref{fig04:Distribution-Polarization}(n)].

Finally, when the detuning is $-$100~meV, for $\kappa = $ 0.1~$\mu$eV, $F_{\ph} \simeq 1.0$ is almost always maintained with $\mu \simeq \E{cav}$ [Figures~\ref{fig05:Results}(a) and \ref{fig05:Results}(c); dashed lines].
The panel~(h) in Figure~\ref{fig04:Distribution-Polarization} is therefore the photonic polariton BEC as also evidenced by the plateau structure of $p_{k} \simeq 0.5$.
However, in Figure~\ref{fig04:Distribution-Polarization}(g), the plateau cannot be found even with $F_{\ph} \simeq 1.0$ and $\mu \simeq \E{cav}$.
We can therefore understand this phase as a kind of the photon BEC because the LP state is dominated by the photonic component.
We note that the photon BEC in the present case is given by the quasi-equilibrium for the {\em whole} \mbox{e-h-p} system in the negligible limit of the \mbox{e-h} system and indeed covered by the original sense of the photon BEC,~\cite{Klaers10} in which only the photon system is in quasi-equilibrium and the state of the medium is not taken care.
For $\kappa =$ 100~$\mu$eV, in contrast, the system directly goes into the lasing phase when the pumping is increased, as evidenced by Figure~\ref{fig04:Distribution-Polarization}(p). 
This is because, in the negatively detuned regime, the thermalization speed can easily become insufficient to recover the equilibrium phase due to the increased photonic component.~\cite{Kasprzak08}

In this way, all of the distinct phases can be identified definitely in Figures~\ref{fig03:PhaseDiagram}(a) and \ref{fig03:PhaseDiagram}(b).
The same identification procedure is applicable to the case with $\kappa =$ 100~meV [Figure~\ref{fig03:PhaseDiagram}(c)].
However, the gap energy $\min[2E_{\k}]$, the coherent number of photons inside the cavity $|a_0|^2$, and the renormalized band structure $A_{11/22,\k}(\nu)$ provide further insight into the underlying physics.

To see this, let us first focus on the case with $\kappa = $ 0.1~$\mu$eV.
For the detuning of $+$100~meV, $\min[2E_{\k}]$ is less than 10~meV when $\mu_{\B} - \E{LP} \lesssim $ 4~meV [Figure~\ref{fig05:Results}(e); dotted line].
The thermalization rate 2$\gamma$ (= 8~meV; Table~\ref{tableII}), therefore, becomes the same order as the gap energy $\min[2E_{\k}]$ in this regime.
This means that the thermalization-induced dephasing is significant and, as a result, the nonequilibrium phase appears in Figure~\ref{fig03:PhaseDiagram}(a).
By increasing the pumping, however, $\min[2E_{\k}]$ increases gradually and grows rapidly at $\mu_{\B} - \E{LP} \simeq $ 100~meV.
At the same time, $|a_0|^2$ shows the two threshold behavior [Figure~\ref{fig05:Results}(g); dotted line].
These results indicate that the change into the photonic phase $F_{\ph} \simeq 1.0$ indeed enhances the \mbox{e-h} attraction notably and can cause the two threshold behavior.~\cite{Yamaguchi13}

In the renormalized band structures [Figures~\ref{fig06:Single-Particle}(a)--\ref{fig06:Single-Particle}(d)], on the other hand, the gap is opened around $k \simeq 0$ for the exciton BEC when the pumping is small [Figure~\ref{fig06:Single-Particle}(a)] but it moves to $k_{\F}$ for the \mbox{e-h} BCS and \mbox{e-h} polariton BCS phases by increasing the pumping [Figures~\ref{fig06:Single-Particle}(b) and \ref{fig06:Single-Particle}(c)].
These results are consistent with the standard picture of the BCS-BEC crossover.~\cite{Comte82}
With increasing the pumping further, the plateau is formed when the photonic polariton BEC is achieved [Figure~\ref{fig06:Single-Particle}(d)].
Note that the $k$ region of the plateau corresponds to that in Figure~\ref{fig04:Distribution-Polarization}(d).
In this context, we can now understand that the plateau of $p_{k} \simeq 0.5$ originates from the enhanced gap energy due to $F_{\ph} \simeq 1.0$, namely, the large gap energy compared with $|\mu - \E{g}^{*}|$, as schematically illustrated in the right upper inset of Figure~\ref{fig06:Single-Particle}.

We remark that qualitatively similar features, i.e. the rapid enhancement of $\min[2E_{\k}]$, the two threshold behavior of $|a_0|^2$ and the plateau structure, can still be found in the resonant case in Figures~\ref{fig05:Results}(e) and \ref{fig05:Results}(g) (solid lines) and in Figures~\ref{fig06:Single-Particle}(e) and \ref{fig06:Single-Particle}(f) even though the clarity of the threshold behavior is reduced.
However, in the case of $-$100~meV detuning, only the monotonic increase of $\min[2E_{\k}]$ and $|a_0|^2$ can be seen in Figures~\ref{fig05:Results}(e) and \ref{fig05:Results}(g); as a consequence, there is only a single threshold for $|a_0|^2$.
This is because $F_{\ph} \simeq 1.0$ is satisfied almost from the beginning of the ordered phase.
These results also support the above-described interpretation that the increase of $F_{\ph}$ can cause the two threshold behavior.

Focusing next on the case with $\kappa = $ 100~$\mu$eV,  at the detuning of $+$100~meV, Figures~\ref{fig06:Single-Particle}(i)--\ref{fig06:Single-Particle}(k) are quite similar to  Figures~\ref{fig06:Single-Particle}(a)--\ref{fig06:Single-Particle}(c) because the ordered phases are insensitive to the change of $\kappa$ for $\mu_{\B} \lesssim \E{cav}$ in the positively detuned regime.
By increasing $\mu_{\B}$, however, the system enters into the lasing phase with the multiple threshold-like behavior of $|a_0|^2$ (dotted line) in Figure~\ref{fig05:Results}(h).
In this situation, the renormalized CB structure [Figure~\ref{fig06:Single-Particle}(l)] is quite different from Figure~\ref{fig06:Single-Particle}(d) but has a formal similarity to the BCS phases, for example, Figure~\ref{fig06:Single-Particle}(k).
The major difference is, however, that the paring gap is opened around the momentum of the kinetic hole burning $k_{\text{HB}}$ in the lasing phase, whereas the gap is around $k_{\F}$ in the BCS phase; see also the right lower inset of Figure~\ref{fig06:Single-Particle}.
This means that the (light-induced) \mbox{e-h} pairs are formed around the laser frequency under the lasing condition, while the \mbox{e-h} Cooper pairs are formed around the Fermi energy.
Even at different detuning, the same picture holds for Figures~\ref{fig06:Single-Particle}(o) and \ref{fig06:Single-Particle}(p) though $k_{\text{HB}}$ is located at different position.
Our theory thus predicts the existence of the bound \mbox{e-h} pairs even in the lasing phase in contrast to earlier expectations.~\cite{Balili09, Dang98, Tempel12-1, Tempel12-2,Tsotsis12,Kammann12}
However, we note that the \mbox{e-h} pair breaking energy is reduced by the crossover into the lasing phase, as shown in Figure~\ref{fig05:Results}(f); dotted and solid lines.
Such a ``lasing gap" has not been observed experimentally but, at least in principle, can be measured in the optical gain spectrum because it is strongly affected by the renormalized band structure in general;~\cite{Yamaguchi13, Yamaguchi14} the details will be discussed later (Section~\ref{sec:Spectral}).

It is now important to notice that the two or multiple threshold behavior found in Figure~\ref{fig05:Results}(h) (solid and dotted lines) cannot be explained solely by the increase of $F_{\ph}$ because $F_{\ph}$ is decreased after the crossover into the lasing phase [Figure~\ref{fig05:Results}(b)].
This means that there is another mechanism to cause the threshold-like behavior, explained as follows.
In the quasi-equilibrium phases, the quasi-equilibrium condition \hyperlink{CI}{(I)}, $\min [2 E_{\k}] \gtrsim \mu_{\B} - \mu$ is satisfied when $\gamma$ and $T$ is neglected for simplicity.
This condition is equivalent to the situation in which $\mu_{\B}$ stays inside the energy gaps $\min [2 E_{\k}]$ located at $\pm \mu/2$ [cf.~Figure~\ref{fig02:BandRenormalization}(a)].
As a result, the pumping is inherently blocked by the gaps.
In this sense, the system is protected by the gaps from the chemical nonequilibrium effect.
In contrast, the lasing condition \hyperlink{CII}{(II)}, $\mu_{\B} - \mu \gtrsim \min [2 E_{\k}]$, indicates that $\mu_{\B}$ goes beyond the energy gaps, as shown in the right lower inset of Figure~\ref{fig06:Single-Particle}.
This means that electrons and holes above the gaps are supplied suddenly when the system changes into the lasing phase and the rapid increase of photons is expected.
This mechanism can also cause the threshold-like behavior even without the increase of the photonic fraction $F_{\ph}$ and the combination of the two mechanisms can successfully explain the two or multiple thresholds in Figure~\ref{fig05:Results}(h).
In the case of zero detuning, in particular, the two threshold behavior as well as the blue shift of $\mu$ [Figures~\ref{fig05:Results}(h) and \ref{fig05:Results}(d); solid lines] are in good qualitative agreement with experiments.~\cite{Yamaguchi13}

We have thus described the fundamental relationship of the cooperative phenomena under the steady-state condition, that is, the BEC-BCS-LASER crossover.
We have shown that the phase diagram on the detuning and the pumping strength plane exhibits a variety of distinct ordered phase depending on the cavity photon loss.
The individual mechanisms of developing such phases and the criteria to distinguish them are clearly addressed by studying the physical quantities of $|a_0|^2$, $p_{\k}$, $n_{\e,\k}$ and $\mu$ as well as the renormalized band structure through the single-particle spectral function $A_{\alpha \alpha,\k}(\nu)$.
As another application of our theory, in the next subsection, we will discuss the dynamics of the system under the continuous pumping to study the connections between the Fermi-edge SF~\cite{Kim13} and the other phases, in particular.

%=======================================
% Fermi-edge SF and the e-h BCS phase
%=======================================
\subsection{Fermi-edge SF and the e-h BCS phase} \label{subsec:Fermi-edge SF}
We now turn to the study of the Fermi-edge SF recently found in the quantum-degenerate high-density \mbox{e-h} system,~\cite{Kim13} in which the macroscopic coherence is spontaneously developed near the Fermi edge due to the Coulomb-induced many-body effects.
As a result, the experiment showed the coherent pulsed radiation of photons, or equivalently the SF, at the Fermi level.
However, in our view, the physics seems closely related to the \mbox{e-h} BCS phase even though the Fermi-edge SF is a time-dependent phenomenon.
It is also worth noting that, to the best of our knowledge, there is no conclusive evidence for the presence of the \mbox{e-h} BCS phase in the past experiments.
In this context, it is of great importance to understand the relationship between the Fermi-edge SF and the the \mbox{e-h} BCS phase.

%----------
% Figure
%----------
\begin{figure}[!tb] 
\centering
\includegraphics[width=.45\textwidth, clip]{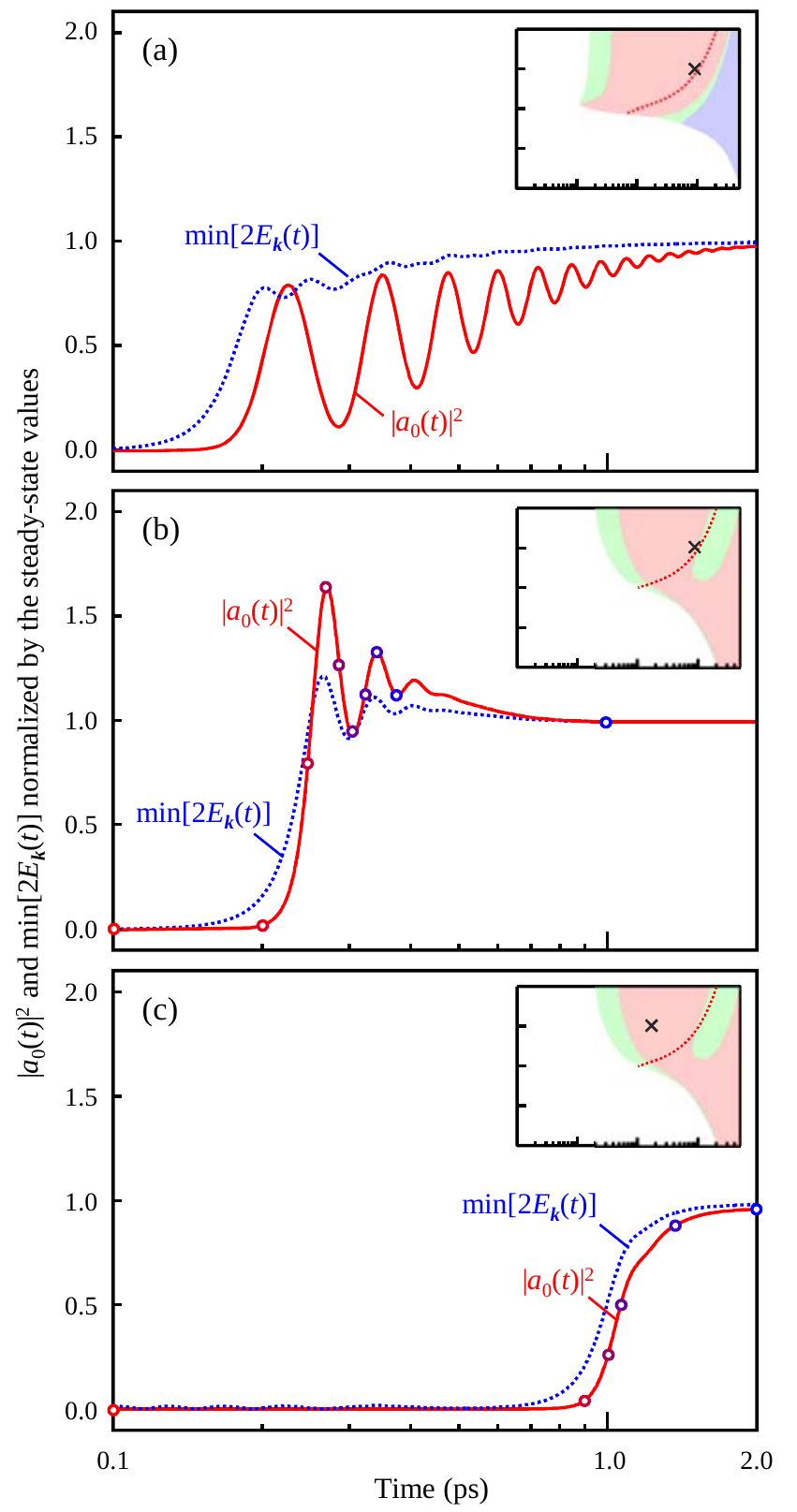} 
\caption{(Color online) Evolution of the coherent photon number $|a_0(t)|^2$ and the gap energy $\min[2E_{\k}]$ under the continuous pumping for (a) $\kappa =$ 100~$\mu$eV and (b),~(c) 100~meV.
Each value is normalized by the steady-state value.
$\mu_{\B} - \E{LP} =$ 90~meV in panels~(a) and (b), while 15~meV in panel~(c).
The detuning is $+$100~meV in all panels. 
Insets show the parameters in the phase diagram [see Figures~\ref{fig03:PhaseDiagram}(b) and \ref{fig03:PhaseDiagram}(c)].
In panel~(a), the oscillatory behavior is categorized as a kind of the relaxation oscillation because $\kappa =$ 100~$\mu$eV is much smaller than $\gamma =$ 4~meV.
In panels~(b) and (c), in contrast, $\kappa =$ 100~meV is much greater than $\gamma =$ 4~meV, which satisfies the condition for the SF, $\kappa \gg \gamma$.
}
\label{fig07:SF}
\end{figure}
%----------

For this purpose, we have to directly solve the time dependent equations [Eq.~\eqref{alleq:MSBEs} with Eqs.~\eqref{alleq:p0k2-n0k2}--\eqref{eq:A}], in principle.
For the reduction of numerical cost, however, we assume that the dynamics of the band renormalization [Eqs.~\eqref{eq:EOM_GR} and \eqref{alleq: GR_Inv2-Sigma2}] can be eliminated adiabatically by Eqs.~\eqref{eq:GR_SS} and \eqref{eq:A_SS} with the steady-state value of $\mu$.
We note that $\mu$ can be set to any value in principle because, for the time dependent problems, $\mu$ is merely the frequency of the rotating frame.
However, for the adiabatic elimination, it is reasonable to use the steady-state value of $\mu$, if exists, to recover the steady state.
At the same time, the cavity photon amplitude $a_0(t)$ now have to be treated as a complex variable.

In order to discuss the Fermi-edge SF, we further assume that, at $t = 0$, the distribution function is described by the Fermi distribution $n_{\e/\h, \k} = 1/[1 + \exp \{ \beta (\tilde{\epsilon}_{\e/\h, \k} - \mu^{\B}_{\e/\h} ) \}]$ with no polarization function $p_{\k} = 0$ [cf.~Eq.~\eqref{eq:ConditionMSBE}].
However, instead of $a_0 = 0$, the photon amplitude is initially set to $a_0 = 1$ to {\it ad hoc} trigger the development of the macroscopic coherence.
This indicates that the SF starts from the photon number of the order of vacuum fluctuation,~\cite{Gross82} or equivalently the spontaneous emission event.
However, we remark that the statistical feature of the initial condition is still a non-trivial problem in semiconductor systems, in contrast to the two-level systems.

With these assumptions, the evolution of the system is calculated under the continuous pumping.
Figure~\ref{fig07:SF}(a) shows the time dependence of the coherent photon number $|a_0(t)|^2$ and the gap energy $\min[2E_{\k}(t)]$ normalized by their steady-state values for $\kappa =$ 100~$\mu$eV.
The parameters are the same as in Figure~\ref{fig04:Distribution-Polarization}(k), and therefore, the evolution finally recover the \mbox{e-h} polariton BCS phase in the steady state.
We can see that $|a_0(t)|^2$ and $\min[2E_{\k}(t)]$ exponentially grow at early times, and then, show oscillatory behaviors with approaching their steady-state values.
In this situation, however,  $\kappa =$ 100~$\mu$eV is much smaller than $\gamma =$ 4~meV, namely $\kappa \ll \gamma$, which is in the opposite limit of $\kappa \gg \gamma$ for the SF.~\cite{Bonifacio71-1, Bonifacio71-2}
This means that the cavity has non-negligible effect on the dynamics and, as a result, the (Fermi-edge) SF is not allowed in Figure~\ref{fig07:SF}(a).
For this reason, we categorize the oscillation as a kind of the relaxation oscillation.

\begin{figure}[!tb] 
\centering
\includegraphics[width=.45\textwidth, clip]{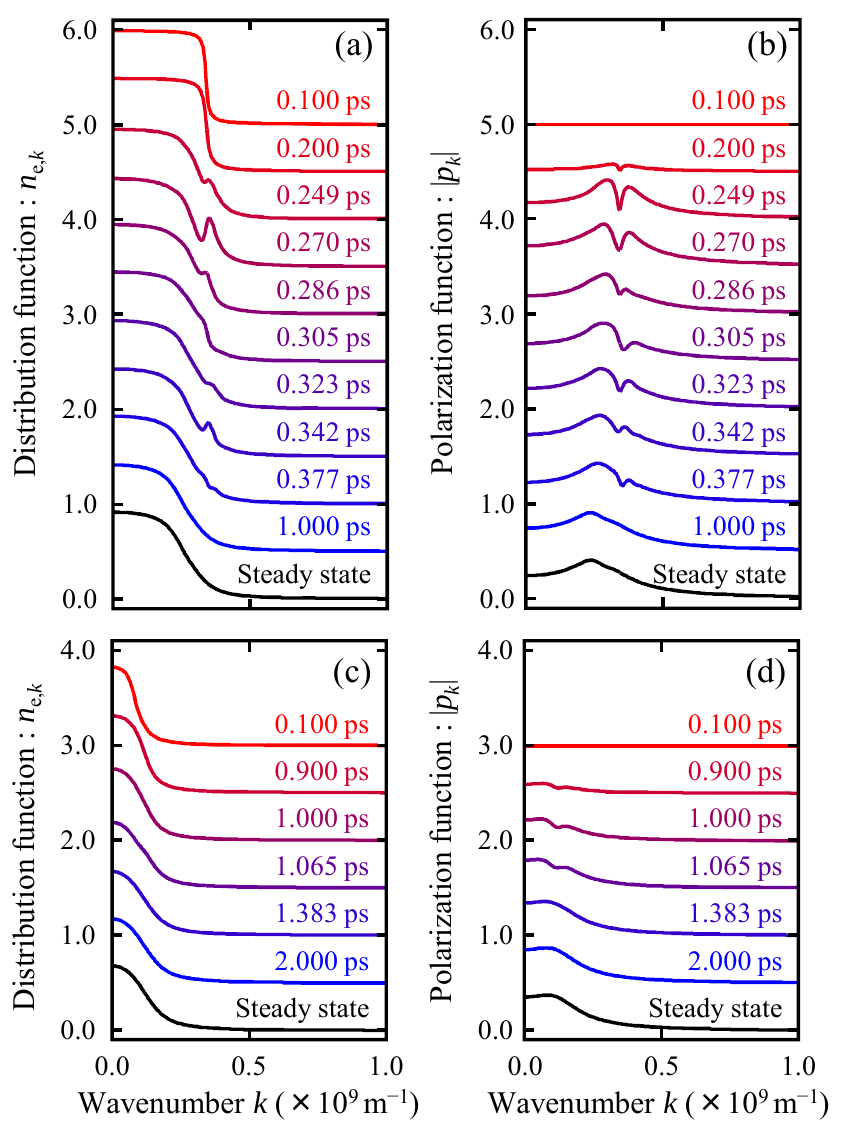} 
\caption{(Color online) The distribution function $n_{\e, \k}$ and the polarization function $p_{\k}$ at the times indicated by open circles in Figures~\ref{fig07:SF}(b) and \ref{fig07:SF}(c).
Panels~(a) and (b) are for Figure~\ref{fig07:SF}(b), while panels~(c) and (d) are for Figure~\ref{fig07:SF}(c).
For easy comparison, the curves have been shifted vertically.
}
\label{fig08:SF_distribution}
\end{figure}

To satisfy the necessary condition $\kappa \gg \gamma$,  in Figure~\ref{fig07:SF}(b), $\kappa$ is increased up to 100~meV with the other parameters unchanged.
In this situation, the cavity effect becomes sufficiently weak or negligible indeed in the corresponding steady-state phase diagram [Figure~\ref{fig03:PhaseDiagram}(c)] and the parameters are now appropriate to discuss the super-fluorescent emission.
Compared with Figure~\ref{fig07:SF}(a), in Figure~\ref{fig07:SF}(b), the visibility of the oscillation is reduced for $|a_0(t)|^2$ but the (normalized) peak value is increased with the exponential growth.
As a result, the behavior becomes similar to the ringing of the SF known for the two-level systems~\cite{Gross82} under the continuous pumping.~\cite{Bolda95}
Analogous qualitative behavior can also be seen for $\min[2E_{\k}(t)]$.
In the distribution function [Figure~\ref{fig08:SF_distribution}(a)], the major modification can be found around the Fermi momentum, whereas in the polarization function [Figure~\ref{fig08:SF_distribution}(b)], a peaked structure is developed around the same momentum with a dip.
$n_{\e,\k}$ and $p_{\k}$ then approach the profiles of the \mbox{e-h} BCS phase as the steady state.
The signature of the kinetic hole burning is also found around the Fermi momentum at 0.270~ps.
This means that the carriers are excessively expended at the Fermi-edge even without the cavity effect; the signature of the SF.
The Fermi-edge SF thus appears in our calculation and converges toward the \mbox{e-h} BCS phase.
The Fermi-edge SF can therefore be seen as a precursor of the \mbox{e-h} BCS phase.
This result is striking by considering the current situation of experiments; the e-h BCS phase is not yet evidenced but the Fermi-edge SF is recently demonstrated by Kim {\it et al}.~\cite{Kim13}
Our theory clearly predicts that the e-h BCS phase can be observed after the Fermi-edge SF, the result of which could not be obtained by the other past theories.
We remark that the Fermi-edge SF already has the macroscopic order through the spontaneous symmetry breaking, and therefore, the pairing gaps are opened, as evidenced by the non-zero value of $\min[2E_{\k}]$.
In this context, the Fermi-edge SF should not be confused with the preformed \mbox{e-h} Cooper pairs,~\cite{Versteegh12} in which such an order is not developed.

%----------
% Figure
%----------
%----------

However, we note that the ringing behavior is not necessarily observed in the evolution toward the \mbox{e-h} BCS phase when the pumping is reduced, as shown in Figure~\ref{fig07:SF}(c).
The delay time is also increased because it takes more time to form macroscopic coherence as the Fermi edge is decreased from $\E{cav}$.
In this case, $n_{\e, k}$ and $p_{k}$ smoothly turn into the profiles of the \mbox{e-h} BCS phase in Figures~\ref{fig08:SF_distribution}(c) and \ref{fig08:SF_distribution}(d).
In particular, the absence of the temporal kinetic hole burning indicates that the thermalization speed becomes relatively large to compensate the lost carriers instantaneously.
In a narrow sense, therefore, the evolution in Figure~\ref{fig07:SF}(c) would not be the SF because the above description means that the photon loss rate becomes effectively smaller than the thermalization rate, the very opposite limit of the ordinary SF.
We remark, however, that the evolution largely shares essential physics; the spontaneous process of developing macroscopic coherence as a result of quantum fluctuation.
The emission property naturally fluctuates from shot to shot also in this case.

The important point here is that, in both cases [Figure~\ref{fig07:SF}(b) and \ref{fig07:SF}(c)], the system eventually evolves toward the \mbox{e-h} BCS state after the spontaneous phase symmetry breaking.
In this context, we do not rule out the Fermi-edge SF experimentally demonstrated in Ref.~\onlinecite{Kim13} is already in the \mbox{e-h} BCS phase because the time scale of the pulsed emission is one order of magnitude larger than the presented results even though our calculations do not purpose quantitative discussions.
These results strongly encourage the experimental discovery of the \mbox{e-h} BCS phase in the context of the Fermi-edge SF.
In a theoretical viewpoint, we also stress that the results presented above are the physics elucidated only by considering the macroscopic coherence in a unified way.
To our knowledge, there has been no theoretical framework that has the ability to address the relationship between the SF and the BCS phase in the past.

%======================================================
% Formalism for calculating Spectral properties
%======================================================
\section{Spectral properties}\label{sec:Spectral}
We have thus explained the relationship of the cooperative phenomena.
However, the properties on the emission spectrum and the gain-absorption spectrum are still unclear.
In this section, by assuming the steady state for simplicity, we first explain the formalism briefly to calculate the spectral properties (Subsections~\ref{subsec:Emission spectrum}--\ref{subsec:Gain-absorption spectrum}).
We then show several numerical results for the BEC-BCS-LASER crossover in Subsection~\ref{subsec:Numerical results}.

%=======================================
% Emission spectrum
%=======================================
\subsection{Emission spectrum} \label{subsec:Emission spectrum}
According to the standard quantum optics,~\cite{Kavokin07,Scully97,Carmichael93} the steady-state emission spectrum observed outside is defined by the Fourier transformation of the correlation function
%----------
% Equation
%----------
\begin{align}
S_{\SS}(\omega, \q) & \equiv \frac{\kappa}{\pi} \lim_{t \to \infty} \int \dd \tau  e^{\ii \omega \tau} \Ex{\oad_{\q}(t) \oa_{\q}(t+\tau)},
\label{eq:DefEmissionSpectrum}
\end{align}
%----------
the definition of which can be rewritten as 
%----------
% Equation
%----------
\begin{align}
S_{\SS}(\omega, \q) = S_{\SS}^{\text{coh}}(\omega, \q) + S_{\SS}^{\text{inc}}(\omega, \q),
\label{eq:EmissionSpectrum}
\end{align}
%----------
where $S_{\SS}^{\text{coh}}(\omega, \q)$ and $S_{\SS}^{\text{inc}}(\omega, \q)$ denote, respectively, the coherent and incoherent parts of the spectrum;
%----------
% Equation
%----------
\begin{subequations}
\label{alleq:Scoh-Sinc}
\begin{align}
S_{\SS}^{\text{coh}}(\omega, \q) &\equiv 2\kappa \lim_{t \to \infty} |a_0(t)|^2 \delta_{\q, 0} \delta(\omega), 
\label{eq:Scoh}\\
S_{\SS}^{\text{inc}}(\omega, \q) &\equiv \frac{\kappa}{\pi} \lim_{t \to \infty} \int \dd \tau  e^{\ii \omega \tau} \Ex{\varDelta\oad_{\q}(t) \varDelta\oa_{\q}(t+\tau)} \nonumber \\
&= \ii \frac{\kappa}{\pi}  \lim_{t \to \infty} D^{<}_{11,\q}(t; \omega),
\label{eq:Sinc}
\end{align}
\end{subequations}
%----------
Here, we have used $\Ex{\oa_{\q}(t)} = \delta_{\q,0}a_0$ and $\varDelta \oa_{\q}(t) \equiv \oa_{\q}(t) - \Ex{\oa_{\q}(t)}$.
$D^{<}_{\alpha_1 \alpha_2, \q}(t; \nu)$ is the Wigner representation of the lesser photon GF, which will be introduced in Section~\ref{sec:Generating Functional}.
$\alpha_i \in \{1, 2\}$ denotes the index for the Nambu space of the photon GF.
In the RAK basis, $S_{\SS}^{\text{inc}}(\omega, \q)$ becomes
%----------
% Equation
%----------
\begin{align}
S_{\SS}^{\text{inc}}(\omega, \q) &= \ii \frac{\kappa}{2\pi} 
[D^{\K}_{11,\q}(\omega) - D^{\R}_{11,\q}(\omega) + D^{\A}_{11,\q}(\omega)],
\label{eq:Sinc2}
\end{align}
%----------
where we have dropped the argument $t$ because we assume the steady state throughout this section.
Eq.~\eqref{eq:Scoh} means that a delta function peak is formed at $\omega = 0$ when $|a_0|^2 \neq 0$ in the same manner as the Mollow triplet.~\cite{Scully97,Carmichael93}
Note that the origin of $\omega$ corresponds to $\mu$ because we are on the rotating frame.
In contrast, Eq.~\eqref{eq:Sinc2} means that the incoherent part of the emission spectrum can be calculated if the photon GFs are obtained.
We do not show the way to estimate the photon GFs here.
However, we emphasize that the {\em partially dressed} two-particle GF $\tilde{K}_{0,\C}$ plays an essential role to calculate the Dyson equation for the photon GFs (Section~\ref{sec:Generating Functional}); see also Appendix~\ref{app:Formalism for Spectra} for the estimation of the partially dressed two-particle GFs.

%----------
% Figure
%----------
\begin{figure*}[!htb] 
\centering
\includegraphics[width=.90\textwidth, clip]{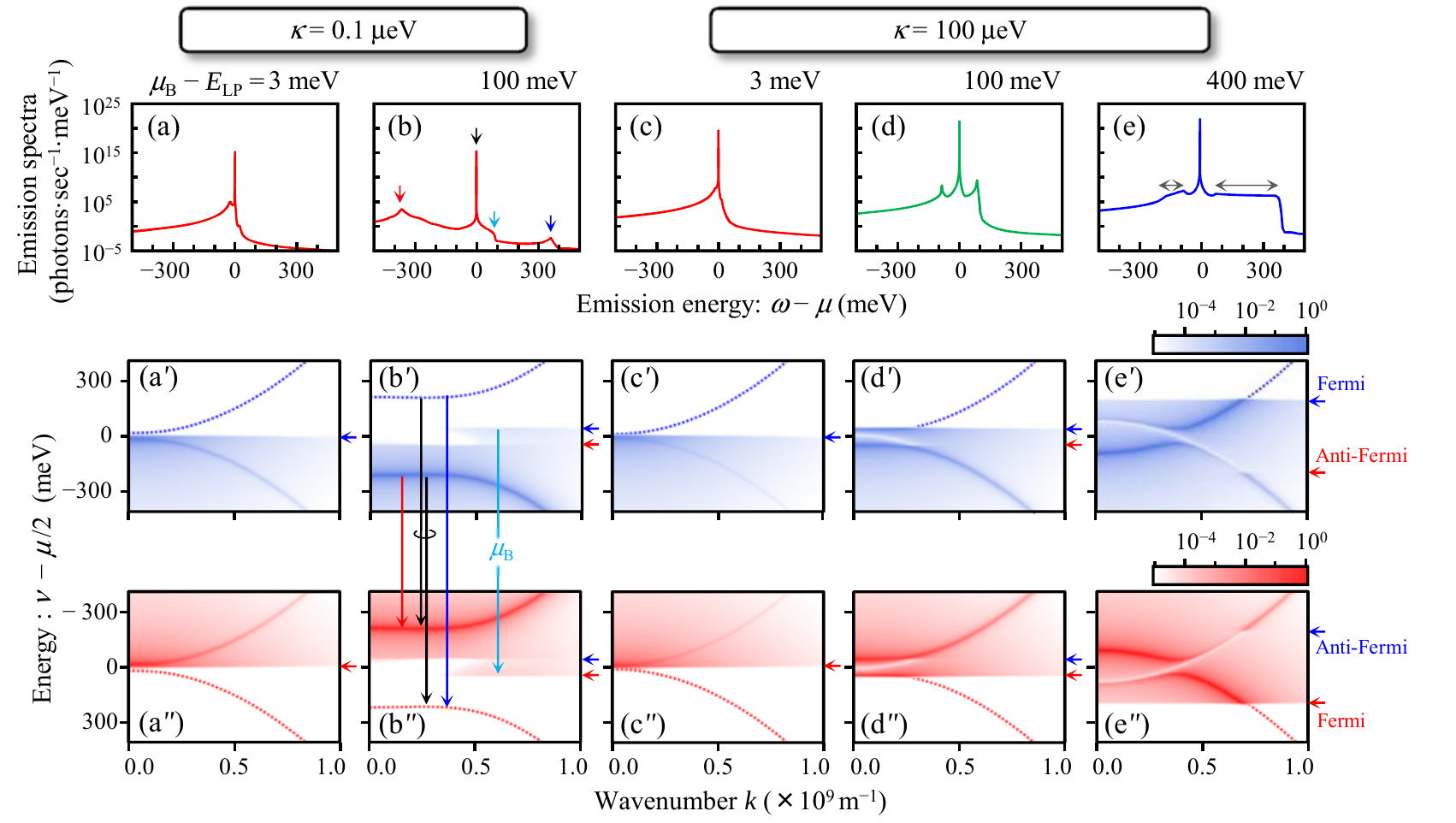} 
\caption{(Color online) 
Emission spectra $S^{\text{inc}}_{\SS} (\omega, \q=0)$ [panels~(a)--(e)] and the corresponding distributions of electrons $n_{\e,\k}(\nu)$ [panels~(a$'$)--(e$'$)] and holes $n_{\h,\k}(\nu)$ [panels~(a$''$)--(e$''$)] under the resonance condition ($\E{cav} = \E{ex}$).
The parameters for the panels~(a)--(e) are the same for the panels~(e), (f), (m)--(o) in Figures~\ref{fig04:Distribution-Polarization} and \ref{fig06:Single-Particle}, respectively.
In the panels~(a$'$)--(e$'$) and (a$''$)--(e$''$), the dotted lines indicate the peak position of $A_{11,\k}(\nu)$ and $A_{22,\k}(-\nu)$; see also the panels~(e), (f) and (m)--(o) in Figure~\ref{fig06:Single-Particle}.
The downward arrows between the panels (b$'$) and (b$''$) typically explain the recombinations of electrons and holes that form the peaks of the emission spectrum in the panel (b) in an analogous way to the Mollow triplet.
Basically, the same picture holds for understanding the emission spectra in the other panels.
The left arrows signify the Fermi level and anti-Fermi level formed by the pumping baths.
The two-headed arrows in the panel~(e) are the guides for the eye that indicate the continuum structures.
}
\label{fig09:Spectra}
\end{figure*}
%----------

%=======================================
% Gain-absorption spectrum
%=======================================
\subsection{Gain-absorption spectrum} \label{subsec:Gain-absorption spectrum}
Before proceeding further, let us turn to the gain-absorption spectrum.
For this purpose, we consider a situation where a weak probe field $F(t)$ is applied to the system in the steady state and interacts with the \mbox{e-h} system through the Hamiltonian,~\cite{Yamaguchi13}
%----------
% Equation
%----------
\begin{align}
\hat{H}'(t) = - F(t) \sum_{\k} [d_{12} \ocd_{1, \k} \oc_{2, \k} + d^*_{12} \ocd_{2, \k} \oc_{1, \k}].
\label{eq:Hprobe}
\end{align}
%----------
Here, $F(t)$ is real and $d_{12} = |d_{12}|\exp(\ii \phi)$ is the dipole matrix element.
Within the linear response theory for the NESS,~\cite{Shimizu10} the microscopic response function $\chi_{\k}(\tau)$ for the polarization $p_{\k}$ becomes
%----------
% Equation
%----------
\begin{align}
\chi_{\k}(\tau) = \ii \theta(\tau) \sum_{\k'} \Ex{[\hat{p}_{\k}(\tau), d^{*}_{12} \hat{p}_{\k'} + d_{12} \hat{p}^{\dagger}_{\k'}]}_{\SS},
\end{align}
%----------
where $\hat{p}_{\k} \equiv \ocd_{2, \k} \oc_{1, \k}$ and $\Ex{[\hat{O}_1(\tau), \hat{O}_2]}_{\SS}$ corresponds to $\lim_{t \to \infty}\Ex{[\hat{O}_{1}(t+\tau), \hat{O}_2(t)]}$ for certain operators $\hat{O}_1$ and $\hat{O}_2$.
The optical susceptibility~\cite{Haug09} is then given by $\chi(\omega) = d^{*}_{12} \sum_{\k} \chi_{\k}(\omega)$ in the Fourier domain.
It is then straightforward to rewrite $\chi(\omega)$ as 
%----------
% Equation
%----------
\begin{multline}
\chi(\omega) = - \ii |d_{12}|^2 \sum_{\k_1, \k_2} K^{\R}_{11, \q = 0}(\omega; \k_1 \k_2) \\
 - \ii (d^{*}_{12})^2 \sum_{\k_1, \k_2} K^{\R}_{12, \q = 0}(\omega; \k_1 \k_2),
\label{eq:Susceptibility}
\end{multline}
%----------
where $K^{\mathcal{Z}}_{\alpha_1 \alpha_2, \q}(t_1t_2; \k_1 \k_2)$ corresponds to the {\em fully dressed} two-particle GF $K_{\C}$, which will be introduced in Section~\ref{sec:Generating Functional}, and $\alpha_i$ denotes the index for the conduction band ($\alpha_i = 1$) and valence band ($\alpha_i = 2$), corresponding to the Nambu space in the matrix form of the GFs.
The gain-absorption spectrum $G(\omega)$ is then given by $G(\omega) = - \Im[\chi(\omega)]$.
Again, we do not go into the detail here but it is noteworthy that the required two-particle GF is not the partially dressed one but the fully dressed one.
Such a distinction is naturally obtained through the generating functional approach, as we shall see in Section~\ref{sec:Generating Functional}; see also Appendix~\ref{app:Formalism for Spectra} for the estimation of the fully dressed two-particle GFs.
% the same notation has been applied to $K^{\mathcal{Z}}_{\alpha_1 \alpha_2, \q}(t_1t_2; \k_1 \k_2)$ as introduced for $\tilde{K}^{\mathcal{Z}}_{0,\alpha_1 \alpha_2, \q}(t_1t_2; \k_1 \k_2)$ just below Eq.~\eqref{eq:Pi_RAK2}.

Here, we note that the second term in Eq.~\eqref{eq:Susceptibility} is non-zero only when the phase symmetry is broken because the off-diagonal elements in Nambu space naturally vanish in the normal phase.
In this context, $\chi(\omega)$ depends on the phase $\phi$ when the macroscopic coherence is developed, while 
it is independent of the phase $\phi$ in the non-ordered phase.
Eq.~\eqref{eq:Susceptibility}, therefore, suggests that the phase difference between the developed order in the system and the coherent probe field may affect the susceptibility even though $\phi = 0$ was implicitly assumed in our previous work.~\cite{Yamaguchi13}
The dependence will be discussed in the next subsection.

%=======================================
% Numerical results
%=======================================
\subsection{Numerical results} \label{subsec:Numerical results}
Based on the above formalism, Figures~\ref{fig09:Spectra}(a)--\ref{fig09:Spectra}(e) show the typical emission spectra for $\kappa =$ 0.1~$\mu$eV and 100~$\mu$eV under the resonant condition ($\E{cav} = \E{ex}$); the parameters are the same for the panels~(e), (f), (m)--(o) in Figures~\ref{fig04:Distribution-Polarization} and \ref{fig06:Single-Particle}, respectively.
We then find that the spectral profiles are significantly changed by increasing the pumping strength.
In the case with $\kappa =$ 0.1~$\mu$eV, for $\mu_{\B} - \E{LP} = $ 3~meV [Figure~\ref{fig09:Spectra}(a)], two side-band peaks can be found on each side of the main peak at $\omega = \mu$.
We also find that the intensity on the lower energy side is brighter than the higher energy side.
These properties become more prominent when the pumping is increased up to $\mu_{\B} - \E{LP} = $ 100~meV [Figure~\ref{fig09:Spectra}(b)].
Furthermore, there also appears a steep reduction of the intensity around $\omega - \mu \simeq $ 100~meV.

In the case for $\kappa =$ 100~$\mu$eV, the side peaks become brighter and more conspicuous when the system is in the crossover regime [Figure~\ref{fig09:Spectra}(d)] even though Figure~\ref{fig09:Spectra}(c) is quite similar to Figure~\ref{fig09:Spectra}(a) because the exciton-polariton BEC is the relevant phase in both cases [Figures~\ref{fig04:Distribution-Polarization}(e) and \ref{fig04:Distribution-Polarization}(m)].
In this situation, the relative peak intensity on the higher energy side is greater than the lower energy side.~\cite{Horikiri14}
However, the continuum structures are developed instead when the system enters deeply inside the lasing regime as seen in Figure~\ref{fig09:Spectra}(e).

To clearly explain these spectral properties, based on Eq.~\eqref{eq:BCSform2}, we here introduce energy- and momentum-resolved distributions of electrons and holes as 
%----------
% Equation
%----------
\begin{align*}
&n_{\e,\k}(\nu) \equiv f^{\SS}_{\e,\k}(\nu)A_{11,\k}(\nu), 
&n_{\h,\k}(\nu) \equiv f^{\SS}_{\h,\k}(\nu)A_{22,\k}(-\nu),
\end{align*}
%----------
respectively.
Notice that $n_{\e/\h,\k} = \int \frac{\dd \nu}{2 \pi} n_{\e/\h,\k}(\nu)$ by definition and $n_{\e/\h,\k}(\nu)$ allows us to estimate how electrons and holes are distributed in the renormalized band structures, as shown in Figures~\ref{fig09:Spectra}(a$'$)--\ref{fig09:Spectra}(e$'$) and \ref{fig09:Spectra}(a$''$)--\ref{fig09:Spectra}(e$''$).
This, in turn, enables us to discuss the \mbox{e-h} recombination process that illustrate the respective emission peaks.

The peak positions indicated by the arrows in Figure~\ref{fig09:Spectra}(b), for example, can be explained through the \mbox{e-h} recombination expressed by the downward arrows between Figures~\ref{fig09:Spectra}(b$'$) and \ref{fig09:Spectra}(b$''$).
The mechanism is again similar to the Mollow triplet in quantum optics.
In our case, however, the greater amount of distributions in the low energy side of the renormalized bands makes the lower sideband peak brighter than the higher one.
The steep reduction of the intensity is then attributed to the Fermi levels lying inside the gaps because there is small but non-zero density of states even inside the gaps due to $\gamma$ [see Eq.~\eqref{eq:A_SS}].
Essentially the same interpretation is possible for the spectra seen in Figures~\ref{fig09:Spectra}(a) and \ref{fig09:Spectra}(c) even though the energy difference between the Fermi levels approximately overlaps with the main peak.

In contrast, in Figures~\ref{fig09:Spectra}(d$'$) and \ref{fig09:Spectra}(d$''$), the Fermi and anti-Fermi levels reach the upper and lower edges of the gaps.
The increased distributions on the upper edges, then, brighten the higher energy sideband peak than the lower one.
In this situation, furthermore, the density of states is enhanced around the edges of the pairing gaps. 
As a result, the Fermi-edge enhancement~\cite{Horikiri14, Skolnick87, Schmitt-Rink86} is emphasized notably on these edges, which makes the side peaks more pronounced in Figure~\ref{fig09:Spectra}(d).

%----------
% Figure
%----------
\begin{figure*}[!htb] 
\centering
\includegraphics[width=.90\textwidth, clip]{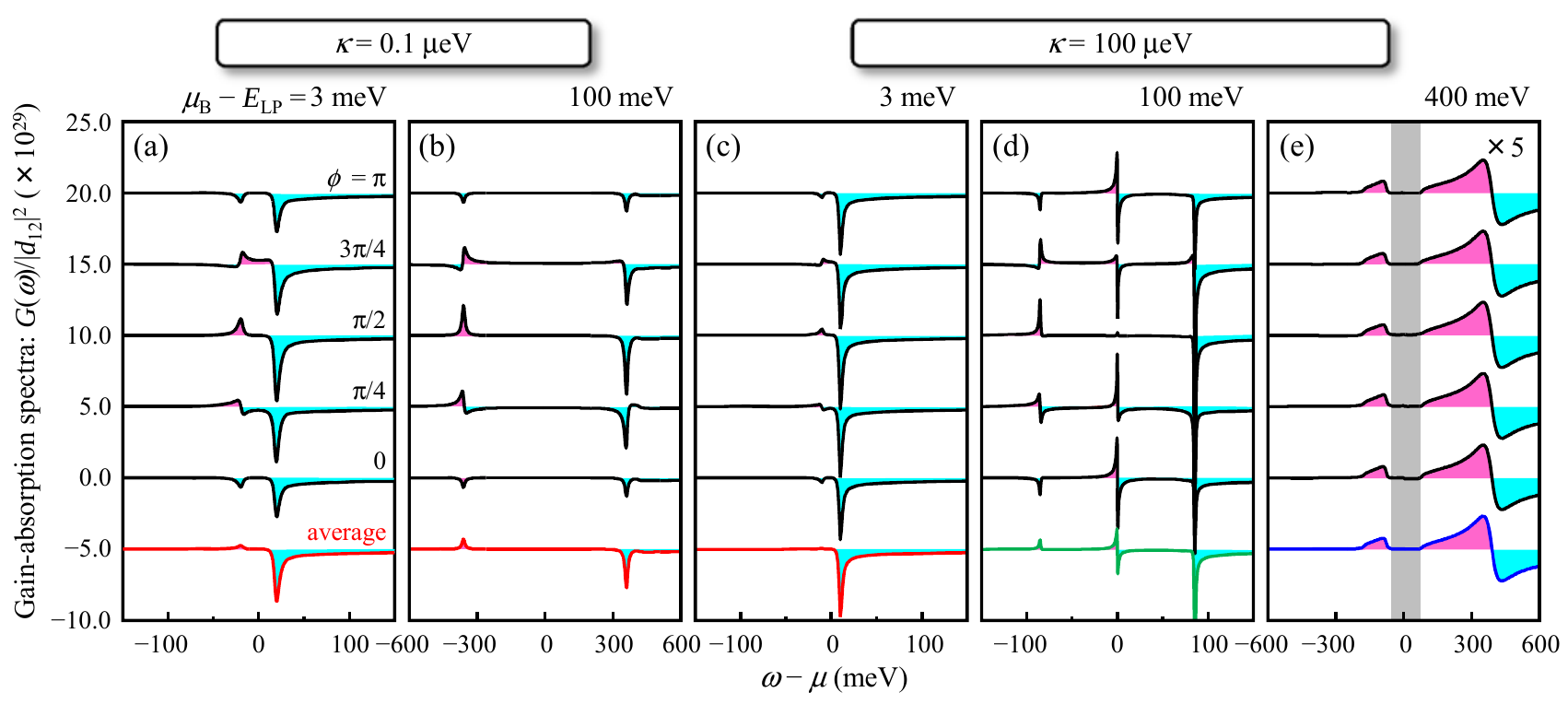} 
\caption{(Color online) 
Gain-absorption spectra corresponding to Figure~\ref{fig09:Spectra}.
In each panel, the phase difference $\phi$ is changed from zero to $\pi$ and the curves have been shifted vertically for easy comparison; the lowest one is the averaged profile over the phase.
In the panel~(e), the values have been multiplied by a factor of~5 for clarity and the gray shaded region denotes the lasing gap. 
}
\label{fig10:Gain}
\end{figure*}
%----------

In the lasing regime [Figures~\ref{fig09:Spectra}(e$'$) and \ref{fig09:Spectra}(e$''$)], the renormalized bands above the gaps are filled with electrons and holes because the Fermi levels are present sufficiently above the gaps.
As a result, the distributions are spread in energy, which forms the continuum structures in Figure~\ref{fig09:Spectra}(e).
The end of the continuum around $\omega - \mu =$ 400~meV in Figure~\ref{fig09:Spectra}(e) is again attributed to the Fermi levels of electrons and holes.
We remark that, in some cases, the anti-Fermi levels can also cause an additional weak structure in the emission spectra even though not shown in the figures.
The opposite end of the continuum around $\omega - \mu = -$200~meV, in contrast, is due to the renormalized band gap energy determined by the optical Stark effect as well as the Coulomb-induced BGR.

These results reveal that the distributions of carriers in the renormalized band structures are reflected mainly in the side peaks of the emission spectra.
Compared with the main peak, however, the intensity is fairly small in our calculations.
Therefore, we finally study the gain-absorption spectra $G(\omega)$, as shown in Figures~\ref{fig10:Gain}(a)--\ref{fig10:Gain}(e).
As discussed in Subsection~\ref{subsec:Gain-absorption spectrum}, the phase difference between the weak probe field and the spontaneously developed order of the system may change the gain-absorption spectra at least in principle.
In this context, the averaged result as well as the dependence on the phase $\phi$ are shown in each panel.
We note that the averaged one is equivalent to taking only the first term in Eq.~\eqref{eq:Susceptibility} into account.

By focusing on the averaged results in Figures~\ref{fig10:Gain}(a)--\ref{fig10:Gain}(c), the gain-absorption spectra are mainly dominated by the absorption.
This is roughly because there is no or little population inversion ($N_{\k} > 0$, or $n_{\e,\k} > 0.5$) in Figures~\ref{fig04:Distribution-Polarization}(e), \ref{fig04:Distribution-Polarization}(f) and \ref{fig04:Distribution-Polarization}(m).
However, the intensity of the gain peak is modified and even enhanced when $\phi$ is changed.
At $\phi = \pi/2$ in Figure~\ref{fig10:Gain}(b), for example, the gain peak becomes comparable to the absorption peak.
The positions of the two peaks are again understood from Figures~\ref{fig09:Spectra}(b$'$) and \ref{fig09:Spectra}(b$''$) and the separation between the peaks is determined by the sum of the gap energies $\min[4E_{\k}]$.
However, the dependence on $\phi$ cannot be understood from $n_{\e/\h,\k}(\nu)$.
To our knowledge, there has been no theoretical work pointing out that the gain-absorption spectrum is changed by the relative phase of the probe field.
However, our result is not surprising because there are two relevant phases as a result of the broken symmetry.

We here note that any structures cannot be found around $\omega \simeq \mu$ in Figure~\ref{fig10:Gain}(b) because such optical transitions for the external probe light are vanishingly low in Figures~\ref{fig09:Spectra}(b$'$) and (b$''$).
In the crossover regime, however, the Fermi and anti-Fermi levels are located at the edges of the pairing gaps, as described above [Figures~\ref{fig09:Spectra}(d$'$) and \ref{fig09:Spectra}(d$''$)]. 
As a result, the gain and absorption peaks corresponding to the relevant transitions become apparent, leading to the characteristic structures around $\omega \simeq \mu$ in Figure~\ref{fig10:Gain}(d).
The peaks are further pronounced because the Fermi-edge enhancement is emphasized by the increased density of states around the edges of the pairing gaps.
In fact, the other peaks around $\omega \simeq \mu \pm 85$~meV also arise from such transitions and indeed become prominent, compared with the other situations.

In the lasing regime [Figure~\ref{fig10:Gain}(e)], the structures around $\omega \simeq \mu$ still exist but become almost invisible, which is consistent with the above scenario.
However, in contrast to Figures~\ref{fig10:Gain}(a)--(d), the numerical results become independent of $\phi$ as a consequence of $K^{\R}_{11,\q=0} \gg K^{\R}_{12,\q=0}$ in Eq.~\eqref{eq:Susceptibility}.
We expect that this is because the distributions far from the gaps dominate the gain-absorption spectra in this situation; see Figure~\ref{fig09:Spectra}(e).
As a result, regardless of the phase $\phi$, the lasing gap appears around $\omega = \mu$ as a nearly transparent frequency window, or a kind of the spectral hole burning.~\cite{Henneberger92, Schmitt-Rink88}
The gain-absorption spectrum is thus one of important ways for the verification of the lasing gap, or equivalently the \mbox{e-h} pairing.

%======================================================
% Generating Functional Approach
%======================================================
\section{Generating Functional Approach}\label{sec:Generating Functional}
In the preceding sections, we have highlighted the relationship of the cooperative phenomena and their spectral properties, the study of which is enabled by our formalism based on the generating functional approach.
In particular, the relationship between the Fermi-edge SF and the \mbox{e-h} BCS phase is one of the most prominent results, which has not been reported previously.
However, until now, we did not show the detailed formalism of the generating functional.
In Section~\ref{sec:Generating Functional} and \ref{sec:real time}, therefore, we finally present our general framework to treat the semiconductor \mbox{e-h-p} systems and derive the key equations  [Eqs.~\eqref{alleq:MSBEs} with Eqs.~\eqref{alleq:p0k2-n0k2}--\eqref{eq:A}] shown in Subsection~\ref{subsec:Key results}.

\begin{figure*}[!tb] 
\centering
\includegraphics[width=.90\textwidth, clip]{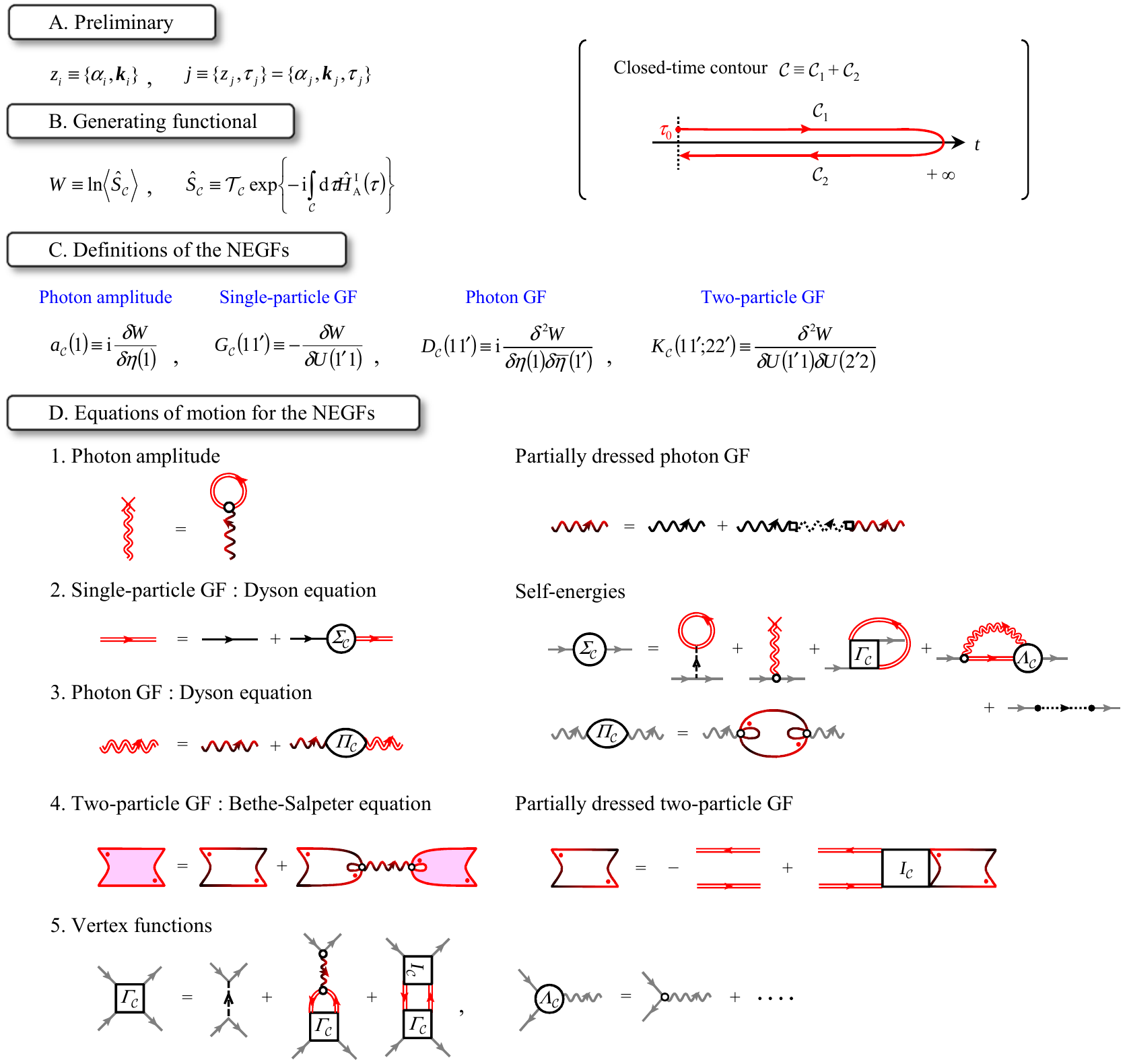} 
\caption{(Color) 
An overview of our generating functional approach depicted in the same structure as Section~\ref{sec:Generating Functional}.
The diagrammatic representations of the individual NEGFs are defined in Figure~\ref{fig12:Diagram_NEGFs}.
Inset shows the closed-time contour $\C \equiv \C_{1} + \C_{2}$.
The contour time $\tau$ goes forward from $\tau_{0}$ to $+\infty$ along the contour $\C_{1}$ and then backward from $+\infty$ to $\tau_{0}$ along the contour $\C_{2}$.
} 
\label{fig11:Overview}
\end{figure*}

An overview of our approach is schematically shown in Figure~\ref{fig11:Overview}, which is depicted in the same structure of this section.
We first explain our preliminary definitions and notations with the closed-time contour $\C \equiv \C_{1} + \C_{2}$ (inset of Figure~\ref{fig11:Overview}) in Subsection~\ref{sec:preliminary} and introduce the generating functional $W$ in Subsection~\ref{subsec:Generating functional}.
The relevant NEGFs are defined in Subsection~\ref{subsec:NEGFs} and their equations of motion are derived in relation to the Dyson equations and the Bethe-Salpeter equations (BSEs) in Subsection~\ref{subsec:Equations of motion for the NEGFs}.
As a result, we can naturally introduce a partially dressed photon GF and a partially dressed two-particle GF, as shown in Figure~\ref{fig11:Overview}.

As already mentioned in the introduction, this approach has several theoretical advantages to systematically study the relevant equations by the diagrammatic technique.
However, readers who are not interested in the formalism may go directly to Section~\ref{sec:Conclusions} because Sections~\ref{sec:Generating Functional} and \ref{sec:real time} involve long theoretical argument.

%----------
% TABLE III
%----------
\begin{table}[!tb]
\caption{\label{tableIII} Definitions of the interaction coefficients in Eqs.~\eqref{eq:He-e:Ab}--\eqref{eq:HF:Ab}.
$g_1 \equiv g^*$, $g_2 \equiv g$, and $\sigma^{(i)}_{\alpha \alpha'}$ is the Pauli matrix.}
\begin{ruledtabular}
\begin{tabular}{ll}
$U'(z_{1} z_{2} z_{3} z_{4})$ & $U'_{\k_{4} - \k_{1}} \delta_{\k_{1}+\k_{2}, \k_{3}+\k_{4}} \delta_{\alpha_{1},\alpha_{4}} \delta_{\alpha_{2},\alpha_{3}}$ \\
$g(z_{1} ; z_{2} z_{3})$ & $-g_{\alpha_{2}} \sigma^{(1)}_{\alpha_{1}\alpha_{2}} \sigma^{(1)}_{\alpha_{2}\alpha_{3}} \delta_{\k_{2}, \k_{3} - \k_{1}}$ \\
$\varsigma(z_{1} z_{2})$ & $\varsigma_{\alpha_{1}, \k_{1}} \delta_{\alpha_{1}, \alpha_{2}} $ \\
$\zeta(z_{1} z_{2})$ & $\zeta_{\k_{2}} \delta_{1,\alpha_{1}} \sigma^{(1)}_{2,\alpha_{2}} + \zeta_{-\k_{2}} \delta_{2,\alpha_{1}} \sigma^{(1)}_{1,\alpha_{2}} $ \\
$\eta_{z_{1}}(t)$ & $\eta_{\alpha_{1}, \k_{1}}(t)$\\
$U_{z_{1}z_{2}}(t)$ & $U_{\alpha_{1} \alpha_{2}, \k_{1} \k_{2}}(t)$\\
\end{tabular}
\end{ruledtabular}
%----------
% TABLE IV
%----------
\caption{\label{tableIV} Definitions of the contour-time interaction coefficients, where the indices $z_{j}$ and $\tau_{j}$ are compactly labeled by the number $j$.}
\begin{ruledtabular}
\begin{tabular}{ll}
$U'_{\C}(1234)$ & $U'(z_{1} z_{2} z_{3} z_{4}) \delta_{\C}(\tau_{1} \tau_{2}) \delta_{\C}(\tau_{2} \tau_{3}) \delta_{\C}(\tau_{3} \tau_{4})$ \\
$g_{\C}(1 ; 2 3)$ & $g(z_{1} ; z_{2} z_{3}) \delta_{\C}(\tau_{1} \tau_{2}) \delta_{\C}(\tau_{2} \tau_{3})$ \\
$\varsigma_{\C}(12)$ & $\varsigma(z_{1} z_{2}) \delta_{\C}(\tau_{1} \tau_{2})$ \\
$\zeta_{\C}(12)$ & $\zeta(z_{1} z_{2}) \delta_{\C}(\tau_{1} \tau_{2})$ \\
$\eta(1)$ & $\eta_{z_{1}}(\tau_1)$\\
$U(12)$ & $U_{z_{1} z_{2}}(\tau_1) \delta_{\C}(\tau_{1} \tau_{2})$ \\
\end{tabular}
\end{ruledtabular}
\end{table}

%----------
% Figure
%----------
%----------

%===========================
% Preliminary definitions and notations
%===========================
\subsection{Preliminary definitions and notations}\label{sec:preliminary}
In order to take the generating functional approach, we first introduce the following Hamiltonian,
%----------
% Equation
%----------
\begin{align}
\oH{total} &= \hat{H} + \oH{A}(t),
\label{eq:Htotal}
\end{align}
%----------
in the Schr\"odinger picture, where $\hat{H}$ is the Hamiltonian described in Subection~\ref{subsec:hamiltonians}, while the time-dependent $\oH{A}(t)$ is an auxiliary perturbing Hamiltonian to formally derive the NEGFs, the concept of which is based on the idea that the GFs are, in general, response functions to a certain kind of perturbations.
In this context, the auxiliary perturbing Hamiltonian $\oH{A}(t)$ is initially assumed, and then, set to zero after the formulation is completed.
Since we are interested in the electronic responses as well as the photonic ones, we define $\oH{A}(t)$ as
%----------
% Equation
%----------
\begin{align}
\oH{A}(t) = & \sum_{\alpha,\k} \eta_{\alpha,\k}(t) \oa_{\alpha,\k} \nonumber\\
       & + \sum_{\alpha,\alpha',\k,\k'} U_{\alpha \alpha', \k \k'}(t) \ocd_{\alpha,\k} \oc_{\alpha',\k'},
\label{eq:HF} 
\end{align}
%----------
where $\eta_{\alpha,\k}(t)$ and $U_{\alpha \alpha', \k \k'}(t) \equiv U_{\alpha \alpha', \k}(t) \delta_{\k \k'}$ are the auxiliary external source fields.
Note that $\oH{A}(t)$ does not have to be physical because it will be used purely for mathematical purpose and we will take the limit of $\oH{A}(t) \rightarrow 0$ ($\eta_{\alpha,\k}(t) \rightarrow 0$ and $U_{\alpha \alpha', \k \k'}(t) \rightarrow 0$) at the final stage of our formulation.
In Eq.~\eqref{eq:HF}, the following operators are also defined
%----------
% Equation
%----------
\begin{align}
\oa_{1,\k} \equiv \oa_{\k}, \ \oa_{2,\k} \equiv \oad_{-\k},
\label{eq:Nambu_operators} 
\end{align}
%----------
which allows us to derive the photon GF in the Nambu space in later discussion.

In order to keep the description of our formalism as simple as possible, let us introduce abridged notations preliminary to our treatments of the NEGFs.
By introducing $z_{i} \equiv \{ \alpha_{i}, \k_{i} \}$ with $\alpha_{i} \in \{ 1, 2 \}$, the interaction Hamiltonians of $\oH{e-e}$, $\oH{e-ph}$, $\oH{SR}$ and $\oH{A}(t)$ become
%----------
% Equation
%----------
\begin{align}
&\oH{e-e} = U'(z_{1} z_{2} z_{3} z_{4})\ocd_{z_{1}}\ocd_{z_{2}}\oc_{z_{3}}\oc_{z_{4}}/2,
\label{eq:He-e:Ab}\\
&\oH{e-ph} = g(z_{1} ; z_{2} z_{3})\oad_{z_{1}}\ocd_{z_{2}}\oc_{z_{3}}, 
\label{eq:He-ph:Ab}\\
&\oH{SR} = \varsigma(z_{1} z_{2}) (\ocd_{z_{1}}  \ob_{z_{2}} + \Hc) + \zeta(z_{1} z_{2}) \oPsi_{z_{1}} \oad_{z_{2}},
\label{eq:HSR:Ab}\\
&\oH{A}(t) = \eta_{z_{1}}(t) \oa_{z_{1}} + U_{z_{1} z_{2}}(t) \ocd_{z_{1}} \oc_{z_{2}},
\label{eq:HF:Ab} 
\end{align}
%----------
where $\oPsi_{1,\k} \equiv \oPsi_{\k}$ and $\oPsi_{2,\k} \equiv \oPsid_{-\k}$ in a similar manner to Eq.~\eqref{eq:Nambu_operators} and  a summation over repeated arguments $z_{i}$ are assumed.
One can easily confirm that Eqs.~\eqref{eq:He-e:Ab}--\eqref{eq:HF:Ab} are equivalent to Eqs.~\eqref{eq:He-e}, \eqref{eq:He-ph}, \eqref{eq:HSR} and \eqref{eq:HF}, respectively, with the interaction coefficients shown in Table~\ref{tableIII}.

In these notations, $\oa_{z}$ and $\oad_{z}$ are related with each other by
%----------
% Equation
%----------
\begin{align}
\oa_{z_1} = \sigma^{(1)}_{z_{1} z_{2}} \oad_{z_{2}}, 
\quad \oad_{z_1} = \sigma^{(1)}_{z_{1} z_{2}} \oa_{z_{2}},
\label{eq:relation}
\end{align}
%----------
when we define $\sigma^{(1)}_{z_{1} z_{2}}$ and $\sigma^{(3)}_{z_{1} z_{2}}$ through the Pauli matrices $\sigma^{(i)}_{\alpha_{1} \alpha_{2}}$ as
%----------
% Equation
%----------
\begin{align}
\sigma^{(1)}_{z_{1} z_{2}} \equiv \sigma^{(1)}_{\alpha_{1} \alpha_{2}} \delta_{\k_{1},-\k_{2}}, 
\quad \sigma^{(3)}_{z_{1} z_{2}} \equiv \sigma^{(3)}_{\alpha_{1} \alpha_{2}} \delta_{\k_{1},\k_{2}}.
\label{eq:Pauli}
\end{align}
%----------
The commutation relations are then given by
%----------
% Equation
%----------
\begin{align}
[\oa_{z_1}, \oad_{z_2}] = \sigma^{(3)}_{z_{1} z_{2}}, 
\quad [\oa_{z_1}, \oa_{z_2}] = \sigma^{(3)}_{z_{1} z'_{1}} \sigma^{(1)}_{z'_{1} z_{2}},
\label{eq:commutation}
\end{align}
%----------
due to Eq.~\eqref{eq:Nambu_operators}.
Note that $\oPsi_{z}$ and $\oPsid_{z}$ also satisfy similar equations to Eqs.~\eqref{eq:relation} and \eqref{eq:commutation}.

By the way, in the limit of $\oH{A}(t) \rightarrow 0$, an expectation value $O(t)$ of any operator $\hat{O}^{\text{S}}(t)$ is, in general, given by
%----------
% Equation
%----------
\begin{align}
& O(t) = \text{Tr} \left[\hat{u}(t_0t) \hat{O}^{\text{S}}(t) \hat{u}(tt_0) \hat{\rho}_0 \right] \nonumber\\
& \quad = \text{Tr} \left[
\bar{\T} \left \{ e^{ - \ii \int^{t_0}_{t} \dd t' \hat{H}(t') }  \right\}
\hat{O}^{\text{S}}(t) 
\T \left\{ e^{ - \ii \int^t_{t_0} \dd t' \hat{H}(t') }  \right\} \hat{\rho}_0 \right],
\label{eq:Expectation}
\end{align}
%----------
where $\hat{\rho}_0$ is an arbitrary initial state at an initial time $t_0$, $\T$ ($\bar{\T}$) is the chronological (anti-chronological) time ordering operator and
$\hat{u}(t_2 t_1)$ is the evolution operator defined as
%----------
% Equation
%----------
\begin{align*}
\hat{u}(t_2 t_1) \equiv \left \{
	\begin{array}{ll}
	\T \exp \left\{ -\ii \int^{t_{2}}_{t_{1}} \dd t' \hat{H}(t') \right\} & \quad t_{2} > t_{1} \\
	\bar{\T} \exp \left\{ +\ii \int^{t_{1}}_{t_{2}} \dd t' \hat{H}(t') \right\} & \quad t_{1} > t_{2} \\
	\end{array}
\right..
\end{align*}
%----------
The superscript `S' emphasizes that the operator is described in the Schr\"odinger picture.
In the second line of Eq.~\eqref{eq:Expectation}, the mathematical structure is notable because the products of operators are finally arranged, from right to left,  in temporal order of $t_0 \rightarrow t \rightarrow t_0$ according to the position of the time arguments.
In this context, it is convenient to consider the operator ordering on the closed-time contour $\C \equiv \C_1 + \C_2$ (Figure~\ref{fig11:Overview}; inset) by introducing the contour time $\tau$.
By defining $\hat{H}(\tau) \equiv \hat{H}(t)$ and $\hat{O}^{\text{S}}(\tau) \equiv \hat{O}^{\text{S}}(t)$, Eq.~\eqref{eq:Expectation} is, then, compactly rewritten as
%----------
% Equation
%----------
\begin{align}
& O(t) = \text{Tr} \left[ \Tc \left\{ e^{- \ii \int_{\C} \dd \tau' \hat{H}(\tau')} \hat{O}^{\text{S}}(\tau) \right\} \hat{\rho}_0 \right],
\end{align}
%----------
where $\int_{\C} \dd \tau$ is the integral along the closed-time contour $\C$ and $\Tc$ ($\Tcbar$) denotes the chronological (anti-chronological) contour-time ordering operator.
The GFs defined on such a closed-time path corresponds to the NEGFs; see Refs.~\onlinecite{Rammer07,Stefanucci13}, for examples.

In order to take the generating functional approach with $\oH{A}(t) \ne 0$, however, it is further required to define the contour interaction picture for an operator $\hat{O}^{\text{S}} (\tau)$ as 
%----------
% Equation
%----------
\begin{align}
\hat{O}^{\text{I}} (\tau) \equiv \hat{\U}(\tau_{0} \tau) \hat{O}^{\text{S}} (\tau) \hat{\U}(\tau \tau_{0}),
\label{eq:int picture}
\end{align}
%----------
with the contour evolution operator~\cite{Stefanucci13}
%----------
% Equation
%----------
\begin{align}
\hat{\U}(\tau_{2}\tau_{1}) \equiv \left \{
	\begin{array}{ll}
	\Tc \exp \left\{ -\ii \int^{\tau_{2}}_{\tau_{1}} \dd \tau \hat{H}(\tau) \right\} & \tau_{2}~\text{later than}~\tau_{1} \\
	\Tcbar \exp \left\{ +\ii \int^{\tau_{1}}_{\tau_{2}} \dd \tau \hat{H}(\tau) \right\} & \tau_{1}~\text{later than}~\tau_{2} \\
	\end{array}
\right.,
\label{eq:evolution}
\end{align}
%----------
because the equations of motion for the NEGFs will be discussed on the closed-time contour. 
The superscript `I' signifies the contour interaction picture, in a similar manner to the superscript `S' for the Schr\"odinger picture.
In Eq.~\eqref{eq:evolution}, $\hat{H} (\tau)$ corresponds to $\hat{H} = \oH{S} + \oH{R} + \oH{SR}$ in Eq.~\eqref{eq:Htotal}, which does not explicitly depend on the contour time.
For the reader's convenience, we briefly summarize the fundamental features of $\hat{\U}$ in Appendix~\ref{app:Contour evolution operator}.

For later use, by defining $j \equiv \{z_{j}, \tau_{j} \} = \{\alpha_{j}, \k_{j}, \tau_{j} \}$, we here introduce the contour-time interaction coefficients, as shown in Table~\ref{tableIV}, with the contour-time delta function~\cite{Rammer07}
%----------
% Equation
%----------
\begin{align}
\delta_{\C}(\tau_{1}\tau_{2}) & \equiv \left \{
	\begin{array}{ll}
	\delta(\tau_{1}\tau_{2}) & \text{for}~\tau_{1} \in \C_{1}, \tau_{2} \in \C_{1} \\
	-\delta(\tau_{1}\tau_{2}) & \text{for}~\tau_{1} \in \C_{2}, \tau_{2} \in \C_{2} \\
	0 & \text{for others}
	\end{array}
\right..
\end{align}
%----------
These definitions and notations will considerably reduce our effort to formally describe the NEGFs in the subsequent sections.

%===========================
% Generating functional
%===========================
\subsection{Generating functional}\label{subsec:Generating functional}
We can now introduce the generating functional $W$ as
%----------
% Equation
%----------
\begin{align}
W \equiv \ln \Ex{\hat{S}_{\C}},
\label{eq:W}
\end{align}
%----------
with the $S$-matrix operator
%----------
% Equation
%----------
\begin{align}
\hat{S}_{\C} & \equiv \Tc \exp \left\{ -\ii \int_{\C} \dd \tau \hat{H}^{\text{I}}_{\A} (\tau) \right\}.
\label{eq:Sc}
\end{align}
%----------
Here, $\Ex{\cdots} = \Tr{\cdots\hat{\rho}_{0}}$ denotes the expectation value.
With the definition of the contour interaction picture [Eqs.~\eqref{eq:int picture} and \eqref{eq:evolution}], 
substitution of Eq.~\eqref{eq:HF} into Eq.~\eqref{eq:Sc} yields
%----------
% Equation
%----------
\begin{align}
\hat{S}_{\C}  =  \Tc \exp \left\{  - \ii \eta(1) \oa(1) -\ii U(11') \ocd(1)\oc(1')  \right\},
\label{eq:Sc2}
\end{align}
%----------
where the definitions of $\eta(1)$ and $U(11')$ can be found in Table~\ref{tableIV} and 
the superscript `I' is dropped when the contour evolution of an operator is evidently in the contour interaction picture.
We also assume that repeated arguments $j$ are integrated by $\int_{\C} \dd j \equiv \sum_{z_{j}} \int_{\C} \dd \tau_{j}$ unless otherwise stated;
$\eta(1) \oa(1)$ means $\int_{\C} \dd 1 \eta(1) \oa(1)$, for example.
Thus, $W$ can obviously be regarded as a functional of the auxiliary source fields $W[\eta, U]$.

%===========================
% Definitions of the NEGFs
%===========================
\subsection{Definitions of the NEGFs}\label{subsec:NEGFs}
One of the advantages of the generating functional is that the NEGFs can be systematically defined through the functional derivative of $W[\eta, U]$.
For example, by using relations
%----------
% Equation
%----------
\begin{align}
\bar{\eta}(1) \equiv \sigma^{(1)}_{\C}(12) \eta(2), 
\quad \eta(1) = \sigma^{(1)}_{\C}(12) \bar{\eta}(2),
\label{eq:eta_def}
\end{align}
%----------
with a definition of a Pauli-type matrix function
%----------
% Equation
%----------
\begin{align}
\sigma_{\C}^{(i)}(12) \equiv \sigma^{(i)}_{z_{1} z_{2}}\delta_{\C}(\tau_{1}\tau_{2}),
\label{eq:Pauli-like}
\end{align}
%----------
one readily obtains
%----------
% Equation
%----------
\begin{align}
&a_{\C}(1) \equiv \ii \frac{\delta W}{\delta \eta(1)} = \Ex{\Tc [\oa(1)]}_{\A}, \nonumber \\
&a_{\C}^{*}(1) \equiv \ii \frac{\delta W}{\delta \bar{\eta}(1)} = \Ex{\Tc [\oad(1)]}_{\A},
\label{eq:ac}
\end{align}
%----------
through the standard functional derivative technique.
In the derivation, we have used 
%----------
% Equation
%----------
\begin{align}
\oa(1) = \sigma^{(1)}_{\C}(12) \oad(2), 
\quad \oad(1) = \sigma^{(1)}_{\C}(12) \oa(2),
\end{align}
%----------
derived from Eqs.~\eqref{eq:relation} and \eqref{eq:Pauli-like}, and
%----------
% Equation
%----------
\begin{align}
\frac{\delta \bar{\eta}(1)}{\delta \eta(1')} = \sigma_{\C}^{(1)}(11') = \frac{\delta \eta(1)}{\delta \bar{\eta}(1')},
\end{align}
%----------
derived from Eq.~\eqref{eq:eta_def} with the chain rule of the functional derivative.
The auxiliary expectation value is also introduced in Eq.~\eqref{eq:ac} as 
%----------
% Equation
%----------
\begin{align}
\Ex{\Tc [\cdots ]}_{\A} & \equiv \frac{1}{ \Ex{\hat{S}_{\C}} } \Ex{ \Tc[ \hat{S}_{\C} \cdots ] }.
\label{eq:auxiliary ex}
\end{align}
%----------
Note, however, that $\Ex{\Tc [\cdots ]}_{\A}$ reduces to the standard statistical expectation value $\Ex{ \Tc [\cdots] } = \Tr{ \Tc [\cdots] \hat{\rho}_{0} }$ in the limit of the vanishing auxiliary source fields.
In this context, $a_{\C}(1)$ and $a^{*}_{\C}(1)$ correspond to the response functions to the auxiliary source fields and can be regarded as the single contour-time NEGFs for the cavity photon amplitude.
$a_{\C}(1)$ and $a^{*}_{\C}(1)$ will take non-zero values only when the macroscopic coherence is developed through the phase symmetry breaking.

The photon GF is then introduced as 
%----------
% Equation
%----------
\begin{align}
D_{\C}(11') & \equiv \ii  \frac{\delta^2 W}{\delta \eta(1) \delta \bar{\eta}(1')} 
   = \frac{\delta a_{\C}(1)}{\delta \bar{\eta}(1')} 
   = \frac{\delta a^{*}_{\C}(1')}{\delta \eta(1)},
\label{eq:Dc}
\end{align}
%----------
which, from Eqs.~\eqref{eq:W} and \eqref{eq:Sc2}, can be described as
%----------
% Equation
%----------
\begin{align}
D_{\C}(11') & = -\ii \Ex{ \Tc [ \varDelta\oa(1)\varDelta\oad(1) ] }_{\A} \nonumber \\
& = -\ii \{ \Ex{ \Tc [ \oa(1)\oad(1') ] }_{\A}  - a_{\C}(1) a^{*}_{\C}(1') \},
\label{eq:Dc2}
\end{align}
%----------
where $\varDelta\oa(1)$ and $\varDelta\oad(1)$ are the fluctuation operators
%----------
% Equation
%----------
\begin{align}
&\varDelta\oa(1) \equiv \oa(1) -  a_{\C}(1), \nonumber \\
&\varDelta\oad(1) \equiv \oad(1) -  a^{*}_{\C}(1).
\label{eq:fluctuation}
\end{align}
%----------
It is important to note that, in Eqs.~\eqref{eq:Dc2} and \eqref{eq:fluctuation}, the condensed part of the photon operator is separated from the non-condensed part.
Essentially identical treatments are well-known in the weakly interacting Bose condensed systems.~\cite{Hohenberg65,Abrikosov75, Fetter71, Griffin09}

In a similar manner, the single-particle GF $G_{\C}(11')$ and the two-particle GF $K_{\C}(11'22')$ for the electronic system can be introduced as
%----------
% Equation
%----------
\begin{align}
G_{\C}(11') &\equiv -\frac{\delta W}{\delta U(1'1)} = -\ii \Ex{\Tc [ \oc(1) \ocd(1') ]}_{\A},
\label{eq:Gc}\\
K_{\C}(11'22') &\equiv \frac{\delta^2 W}{\delta U(1'1) \delta U(2'2)} \nonumber \\
& = -\frac{\delta G_{\C}(11')}{\delta U(2'2)}  
 = -\frac{\delta G_{\C}(22')}{\delta U(1'1)}.
\label{eq:Kc}
\end{align}
%----------
It follows from Eqs.~\eqref{eq:W} and \eqref{eq:Sc2} that~\cite{Baym61,Baym62}
%----------
% Equation
%----------
\begin{align}
K_{\C}(11'22') = G_{\C}(11'22')-G_{\C}(11')G_{\C}(22'),
\label{eq:Kc2}
\end{align}
%----------
where $G_{\C}(11'22')$ is defined as
%----------
% Equation
%----------
\begin{align}
G_{\C}(11'22') \equiv (-\ii)^2 \Ex{ \Tc [ \oc(1)\oc(2)\ocd(2')\ocd(1') ] }_{\A}.
\label{eq:two-particle Gc}
\end{align}
%----------
Since our system is the interacting Bose-Fermi mixture, however, it might be insufficient to prepare only the bosonic and fermionic NEGFs described above.
As an intermediate NEGF, we here introduce a photon-assisted electronic GF as
%----------
% Equation
%----------
\begin{multline}
P_{\C}(22';1) \equiv -\ii \frac{\delta^2 W}{\delta\bar{\eta}(1) \delta U(2'2)} = \ii \frac{\delta G_{\C}(22')}{\delta\bar{\eta}(1)}\\
= -\ii \Ex{\Tc [ \oad(1)\oc(2)\ocd(2') ] }_{\A} - a^{*}_{\C}(1) G_{\C}(22'),
\label{eq:Pc}
\end{multline}
%----------
the name of which is due to the formal similarity to the photon-assisted polarization in the cluster expansion method.~\cite{Kira98,Kira99,Gies07} 
As we shall see later, the correlations between photons and electrons (holes) are essentially included in this GF.

%----------
% Figure
%----------
\begin{figure*}[!tb] 
\centering
\includegraphics[width=.90\textwidth, clip]{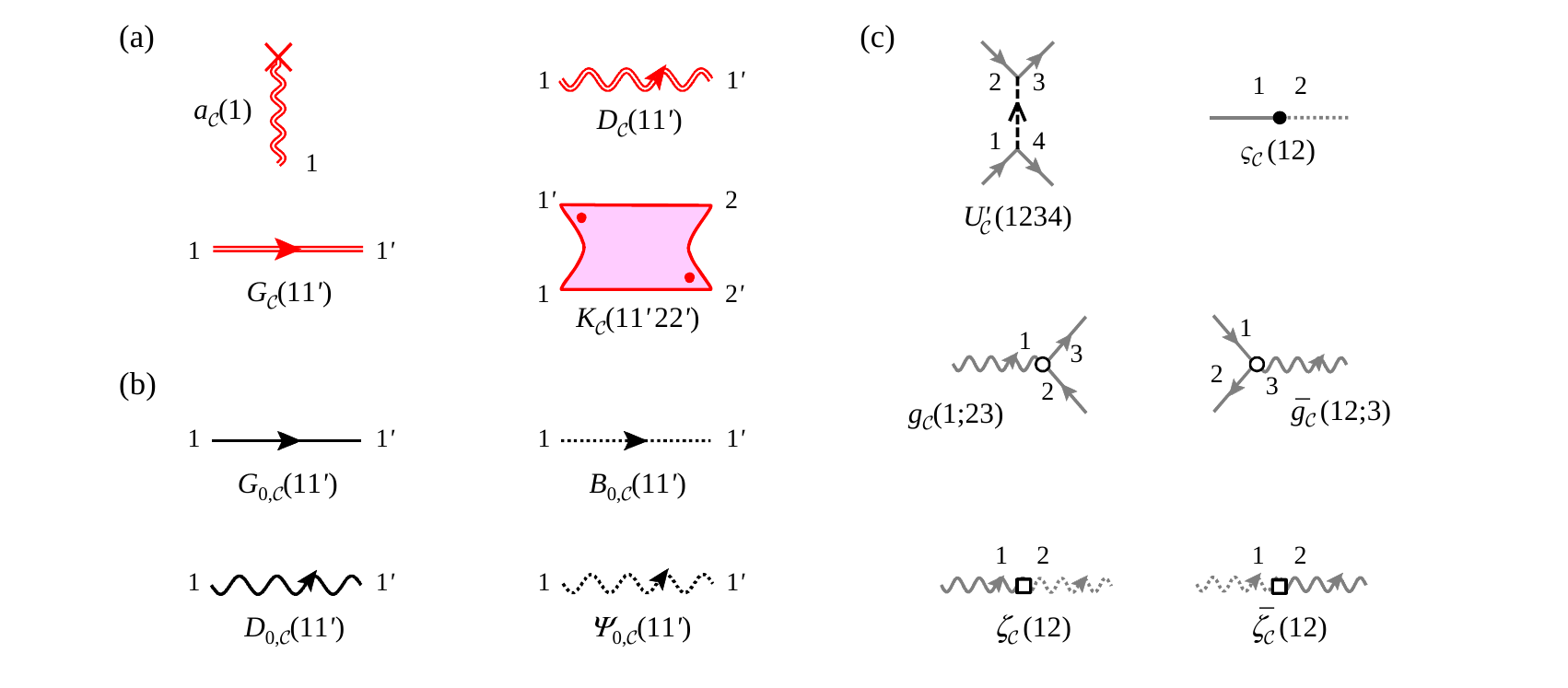} 
\caption{(Color online) Diagrammatic representations of (a) the dressed NEGFs, (b) the bare NEGFs and (c) the interaction coefficients (the bare vertices).
In panel (a), $a_{\C}(1)$ is the photon amplitude in the cavity and $D_{\C}(11')$ describes the photon GF.
$G_{\C}(11')$ and $K_{\C}(11'22')$ are the single-particle and two-particle GFs for electrons, respectively.
$\bullet$ in $K_{\C}(11'22')$ means that the edge is directed outward.
In panel (b), $G_{0,\C}(11')$ and $D_{0,\C}(11')$ describe the bare NEGFs for electrons and photons, while $B_{0,\C}(11')$ and $\varPsi_{0,\C}(11')$ are those for electrons in the pumping baths and photons in the free-space vacuum reservoir, respectively.
In panel (c), the interaction coefficients (the bare vertices) are defined in Table~\ref{tableIV} and \ref{tableVI}.}
\label{fig12:Diagram_NEGFs}
\end{figure*}
%----------
%----------
% TABLE V
%----------
\begin{table}[!tb]
\caption{\label{tableV} The inverse of the bare NEGFs $X^{-1}_{0,\C}$.}
\begin{ruledtabular}
\begin{tabular}{ll}
$D^{-1}_{0,\C}(11')$ & $ \ii \partial_{\tau_1} \sigma^{(3)}_{\C}(11') -\xi_{\ph, \k_{1}} \delta_{\C}(11') $ \\
$\varPsi^{-1}_{0,\C}(11')$ &  $ \ii \partial_{\tau_1} \sigma^{(3)}_{\C}(11') -\xi_{\ph, \k_{1}}^{\text{B}} \delta_{\C}(11') $  \\
$G^{-1}_{0,\C}(11')$ & $ \left( \ii \partial_{\tau_1} - \xi_{z_{1}} \right) \delta_{\C}(11') $ \\
$B^{-1}_{0,\C}(11')$ & $ \left( \ii \partial_{\tau_1} - \xi_{z_{1}}^{\text{B}} \right) \delta_{\C}(11') $  \\
\end{tabular}
\end{ruledtabular}
\end{table}
%----------

We have, thus, described the definitions of the NEGFs, the diagrams of which are shown in Figure~\ref{fig12:Diagram_NEGFs}(a).
However, these NEGFs are basically categorized as the fully dressed diagrams. 
For the study of their equations of motion, the bare (non-interacting) NEGFs are, if defined, favorable to describe the contour-time free evolution of the particles.~\cite{Rammer07}
For this purpose, in Table~\ref{tableV}, the inverse of the bare NEGFs $X^{-1}_{0,\C}$ are introduced by taking into account only the non-interacting part of our Hamiltonian in the Heisenberg equations of motion, the definitions of which will come into clearer view in the next subsection.
The bare NEGFs $X_{0,\C}$ are then introduced as a function that satisfies the following relation:
%----------
% Equation
%----------
\begin{align}
X^{-1}_{0,\C}(12)X_{0,\C}(21') = \delta_{\C}(11') = X_{0,\C}(12)X^{-1}_{0,\C}(21'),
\label{eq:Bare_NEGFs}
\end{align}
%----------
where, in a similar manner to Eq.~\eqref{eq:Pauli-like}, the delta function is given by
%----------
% Equation
%----------
\begin{align}
\delta_{\C}(11') &\equiv \delta_{z_{1} z'_{1}} \delta_{\C}(\tau_{1} \tau'_{1}) \nonumber \\
&= \delta_{\alpha_{1} \alpha_{1}}  \delta_{\k_{1} \k'_{1}}  \delta_{\C}(\tau_{1} \tau'_{1}).
\label{eq:Delta_c}
\end{align}
%----------
In this paper, we diagrammatically represent the bare NEGFs, as shown in Figure~\ref{fig12:Diagram_NEGFs}(b).

%=======================================
% Equations of motion for the NEGFs
%=======================================
\subsection{Equations of motion for the NEGFs}\label{subsec:Equations of motion for the NEGFs}
In the previous subsection, we have explained the definitions of NEGFs in our generating functional approach.
The equations of motion for these NEGFs are now ready to be studied, some of which are identical to the Dyson equations in differential forms.~\cite{Rammer07} 
In the followings, we describe the way to obtain the equations of motion, the self-energies, and their diagrammatic representation in our formalism.

%---------------------------------------
% Equations of motion for the photon amplitude
%---------------------------------------
\subsubsection{Photon amplitude}\label{subsubsec:photon amplitude}
We first study the equations of motion for the photon amplitude $a_{\C}(1)$.
By differentiating $a_{\C}(1)$ with respect to $\tau_{1}$, we obtain 
%----------
% Equation
%----------
\begin{align}
\ii &\frac{\partial}{\partial \tau_{1}} a_{\C}(1) = \ii \frac{\partial}{\partial \tau_{1}} \left\{ \frac{1}{ \Ex{\hat{S}_{\C}} } \Ex{ \hat{S}(\tau_{0} \tau_{1}) \oa_{z_{1}}(\tau_{1}) \hat{S}(\tau_{1} \tau_{0}) } \right\} \nonumber \\
&= \EX{\Tc \left[ \ii \frac{\partial \oa_{z_{1}}(\tau_{1})}{\partial \tau_{1}} \right] }_{\A} + \Ex{\Tc [ \oa_{z_{1}}(\tau_{1}) , \oH{A}^{\text{I}}(\tau_{1}) ] }_{\A} \nonumber \\
&= \Ex{\Tc [ \oa_{z_{1}}(\tau_{1}) , \oH{total}^{\text{I}}(\tau_{1}) ] }_{\A}.
\label{eq:ac_dynamics1}
\end{align}
%----------
In the first line, together with Eqs.~\eqref{eq:ac} and \eqref{eq:auxiliary ex}, we have used a description of 
%----------
% Equation
%----------
\begin{align}
\hat{S}_{\C} = \hat{S}(\tau_0, \infty)\hat{S}(\infty, \tau_0),
\end{align}
%----------
by introducing a contour evolution operator
%----------
% Equation
%----------
\begin{align}
\hat{S}(\tau_{2}\tau_{1}) \equiv \left \{
	\begin{array}{ll}
	\Tc \exp \left\{ -\ii \int^{\tau_{2}}_{\tau_{1}} \dd \tau \hat{H}^{\text{I}}_{\A}(\tau) \right\} & \tau_{2}~\text{later than}~\tau_{1} \\
	\Tcbar \exp \left\{ +\ii \int^{\tau_{1}}_{\tau_{2}} \dd \tau \hat{H}^{\text{I}}_{\A}(\tau) \right\} & \tau_{1}~\text{later than}~\tau_{2} \\
	\end{array}
\right..
\label{eq:Sevolution}
\end{align}
%----------
We note that $\hat{S}(\tau_{2}\tau_{1})$ has analogous properties to $\hat{\U}(\tau_{2}\tau_{1})$ summarized in Appendix~\ref{app:Contour evolution operator}; the second and third lines of Eq.~\eqref{eq:ac_dynamics1} are obtained by using Eqs.~\eqref{eq:A1} and \eqref{eq:A2} with replacing $\hat{\U} \rightarrow \hat{S}_{\C}$ and $\hat{H}(\tau) \rightarrow \hat{H}^{\text{I}}_{\A}(\tau)$. 
Here, Eq.~\eqref{eq:ac_dynamics1} means that the contour evolution of $a_{\C}(1)$ can simply be described by the contour Heisenberg equations.~\cite{Stefanucci13}
A straightforward application of the commutation relations [Eq.~\eqref{eq:commutation}] to Eq.~\eqref{eq:ac_dynamics1} then gives
%----------
% Equation
%----------
\begin{multline}
D^{-1}_{0,\C}(11') a_{\C}(1') \\ = -\ii g_{\C}(1;23)G_{\C}(32) + \zeta_{\C}(12)\varPsi_{\C}(2) + \bar{\eta}(1),
\label{eq:ac_dynamics2}
\end{multline}
%----------
where $\varPsi_{\C}(1) \equiv \Ex{\Tc [ \hat{\varPsi}(1) ]}_{\A} $.
It is now evident that $D^{-1}_{0,\C}$ defined in Table~\ref{tableV} arises originally from the Heisenberg equation of motion with the non-interacting part of our Hamiltonian, as described above.
We also note that formal notational simplicity is seen in Eq.~\eqref{eq:ac_dynamics2} by virtue of the preliminary arrangements in Subsection~\ref{sec:preliminary}.

Exactly in the same manner, for the contour evolution of $\varPsi_{\C}(1)$, we can obtain
%----------
% Equation
%----------
\begin{align}
\varPsi^{-1}_{0,\C}(11') \varPsi_{\C}(1') = \bar{\zeta}_{\C}(11') a_{\C}(1') \equiv J_{\text{B}}(1),
\label{eq:Psi_dynamics}
\end{align}
%----------
where $\bar{\zeta}_{\C}(12)$ is defined in Table~\ref{tableVI}.
Here, by taking the functional derivative with respect to $J_{\text{B}}$, we have
%----------
% Equation
%----------
\begin{align}
\varPsi^{-1}_{0,\C}(11')\varPsi_{0,\C}(1'2) = \delta_{\C}(12) = \varPsi_{0,\C}(11')\varPsi^{-1}_{0,\C}(1'2),
\label{eq:bare_Psi_def2}
\end{align}
%----------
with a definition of
%----------
% Equation
%----------
\begin{align}
\varPsi_{0,\C}(12) \equiv \frac{ \delta \varPsi_{\C}(1) }{ \delta J_{\text{B}}(2) }.
\label{eq:bare_Psi_def}
\end{align}
%----------
Since Eq.~\eqref{eq:bare_Psi_def2} is now identical to Eq.~\eqref{eq:Bare_NEGFs}, Eq.~\eqref{eq:bare_Psi_def} is adequate for the definition of $\varPsi_{0,\C}(12)$.

We have thus obtained the equations of motion for the photon amplitudes in the cavity $a_{\C}(1)$ [Eq.~\eqref{eq:ac_dynamics2}] and in the vacuum photon bath $\varPsi_{\C}(1)$ [Eq.~\eqref{eq:Psi_dynamics}].
Naturally, these equations are coupled with each other.
However, we are mainly interested in the dynamics of the variables in the system.
In order to eliminate the dynamics of $\varPsi_{\C}(1)$ from Eq.~\eqref{eq:ac_dynamics2}, a formal solution of Eq.~\eqref{eq:Psi_dynamics}
%----------
% Equation
%----------
\begin{align}
\varPsi_{\C}(1) = \varPsi_{0,\C}(12) \bar{\zeta}_{\C}(23) a_{\C}(3),
\label{eq:formal_solution}
\end{align}
%----------
is available, derived by multiplying Eq.~\eqref{eq:Psi_dynamics} by $\varPsi_{0,\C}(21)$ from the left and using Eq.~\eqref{eq:bare_Psi_def2}.
This equation physically suggests that the photon amplitude observed in the vacuum photon bath corresponds to the freely propagated field after escaping from the cavity through the coupling constant.
As a result, substitution of Eq.~\eqref{eq:formal_solution} into Eq.~\eqref{eq:ac_dynamics2} yields
%----------
% Equation
%----------
\begin{align}
\tilde{D}^{-1}_{0,\C}(11') a_{\C}(1')  = -\ii g_{\C}(1;23) G_{\C}(32) + \bar{\eta}(1),
\label{eq:Dyson eq. in differential form}
\end{align}
%----------
with a definition of the inverse of the {\em partially dressed} photon GF
%----------
% Equation
%----------
\begin{align}
\tilde{D}^{-1}_{0,\C}(11') \equiv D^{-1}_{0,\C}(11') - \zeta_{\C}(12) \varPsi_{0,\C}(23) \bar{\zeta}_{\C}(31'),
\label{eq:Inverse of partially dressed photon GF}
\end{align}
%----------
the meaning of which will soon become apparent.
These two equations play the most fundamental roles when we discuss the dynamics of the cavity photon amplitude, which reproduce the Heisenberg-Langevin type equation of motion [Eq.~\eqref{eq:MSBEs--a0}], as seen in Subsection~\ref{subsec:EOM1}.

\begin{figure}[!tb] 
\centering
\includegraphics[width=.45\textwidth, clip]{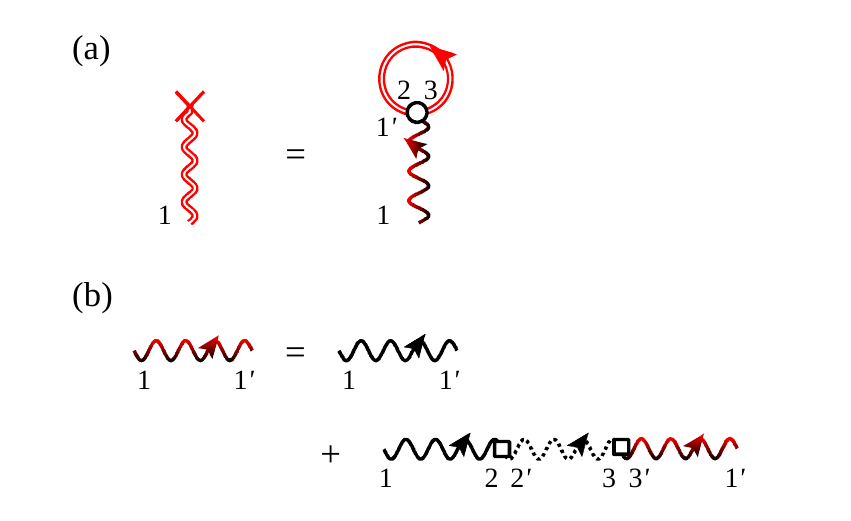} 
\caption{(Color)
Diagrammatic representations of (a) Eq.~\eqref{eq:Dyson eq.} and (b) Eq.~\eqref{eq:Dyson eq.2} in the limit of the vanishing auxiliary source fields. 
These are equivalent to Eqs.~\eqref{eq:Dyson eq. in differential form} and \eqref{eq:Inverse of partially dressed photon GF}, respectively.
The Heisenberg-Langevin type equation of motion for the cavity photon amplitude [Eq.~\eqref{eq:MSBEs--a0}] can be recovered from these diagrams, as seen in Subsection~\ref{subsec:EOM1}.
}
\label{fig13:Amplitude}
\end{figure}

In order to obtain the diagrammatic representations, we define the partially dressed photon GF as
%----------
% Equation
%----------
\begin{align}
\tilde{D}_{0,\C}(12) \equiv \frac{\delta a_{\C}(1)}{\delta J_{\text{S}}(2)},
\label{eq:partially dressed photon GF}
\end{align}
%----------
where, in a similar manner to Eq.~\eqref{eq:Psi_dynamics}, $J_{\text{S}}(1)$ is introduced as the right-hand side of Eq.~\eqref{eq:Dyson eq. in differential form};
%----------
% Equation
%----------
\begin{align}
J_{\text{S}}(1) \equiv -\ii g_{\C}(1;23) G_{\C}(32) + \bar{\eta}(1).
\label{eq:Js}
\end{align}
%----------
It is then obvious that $\tilde{D}_{0,\C}(12)$ satisfies 
%----------
% Equation
%----------
\begin{align}
\tilde{D}^{-1}_{0,\C}(11')\tilde{D}_{0,\C}(1'2) = \delta_{\C}(12) = \tilde{D}_{0,\C}(11')\tilde{D}^{-1}_{0,\C}(1'2),
\label{eq:partially dressed photon GF2}
\end{align}
%----------
by taking the functional derivative of Eq.~\eqref{eq:Dyson eq. in differential form} with respect to $J_{\text{S}}$.
This means that Eq.~\eqref{eq:partially dressed photon GF} is indeed appropriate for the definition of $\tilde{D}_{0,\C}(12)$ because Eq.~\eqref{eq:partially dressed photon GF2} takes the same form as Eq.~\eqref{eq:Bare_NEGFs} when the bare NEGF is replaced by the partially dressed one.
As a result, Eqs.~\eqref{eq:Dyson eq. in differential form} and \eqref{eq:Inverse of partially dressed photon GF} can be rewritten respectively as 
%----------
% Equation
%----------
\begin{align}
&a_{\C}(1)  = -\ii \tilde{D}_{0,\C}(11') g_{\C}(1';23) G_{\C}(32) \nonumber\\
& \hspace{4cm} + \tilde{D}_{0,\C}(11')\bar{\eta}(1'),
\label{eq:Dyson eq.}\\
&\tilde{D}_{0,\C}(11') = D_{0,\C}(11') \nonumber\\
& \quad + D_{0,\C}(12) \zeta_{\C}(22') \varPsi_{0,\C}(2'3) \bar{\zeta}_{\C}(33') \tilde{D}_{0,\C}(3'1').
\label{eq:Dyson eq.2}
\end{align}
%----------
Note that Eq.~\eqref{eq:Dyson eq.2} takes the form of the Dyson equation,~\cite{Rammer07,Stefanucci13} with the self-energy $\varSigma_{\C}^{\kappa}$ made of the bare photon bath GF
%----------
% Equation
%----------
\begin{align}
\varSigma_{\C}^{\kappa}(23') \equiv \zeta_{\C}(22') \varPsi_{0,\C}(2'3) \bar{\zeta}_{\C}(33').
\label{eq:Sigma_Decay}
\end{align}
%----------
The superscript $\kappa$ indicates that this self-energy describe the effect of the cavity photon loss in later discussion.
By introducing the graphical representations of the interaction coefficients (the bare vertices) as shown in Figure~\ref{fig12:Diagram_NEGFs}(c),  
Eqs.~\eqref{eq:Dyson eq.} and \eqref{eq:Dyson eq.2} are drawn diagrammatically in Figure~\ref{fig13:Amplitude}, which are equivalent to Eqs.~\eqref{eq:Dyson eq. in differential form} and \eqref{eq:Inverse of partially dressed photon GF}.
Here, we emphasize that the tail of the tadpole diagram in Figure~\ref{fig13:Amplitude}(a) is not the fully dressed photon GF $D_{\C}$ but the partially dressed one $\tilde{D}_{0,\C}$.
Such a diagrammatic structure is not evident if the standard diagrammatic technique is employed because there are several choices to replace the skeleton diagram by the fully dressed one or partially dressed ones. 
We also remark that the bare photon bath GF $\varPsi_{0,\C}$ in Eq.~\eqref{eq:Dyson eq.2} {\em cannot} be replaced by the fully dressed one $\varPsi_{\C}$ in order to avoid the double counting of the photon GF.
However, even without considering these problems, the generating functional approach allows us to naturally derive the equations of motion, as presented above.

%----------
% TABLE VI
%----------
\begin{table}[!tb]
\caption{\label{tableVI} Definitions of additional interaction coefficients helpful to understand our formalism.}
\begin{ruledtabular}
\begin{tabular}{ll}
$\bar{g}_{\C}(1 2 ; 3)$ & $g_{\C}(2';12) \sigma^{(1)}_{\C}(2'3)$ \\
$\bar{\zeta}_{\C}(12)$ & $\sigma^{(1)}_{\C}(11') \zeta_{\C}(1'2') \sigma^{(1)}_{\C}(2'2)$ \\
\end{tabular}
\end{ruledtabular}
\end{table}
%----------

%----------
% Figure
%----------
%----------

%---------------------------------------
% single-particle GF
%---------------------------------------
\subsubsection{Single-particle GF}
\begin{figure}[!tb] 
\centering
\includegraphics[width=.45\textwidth]{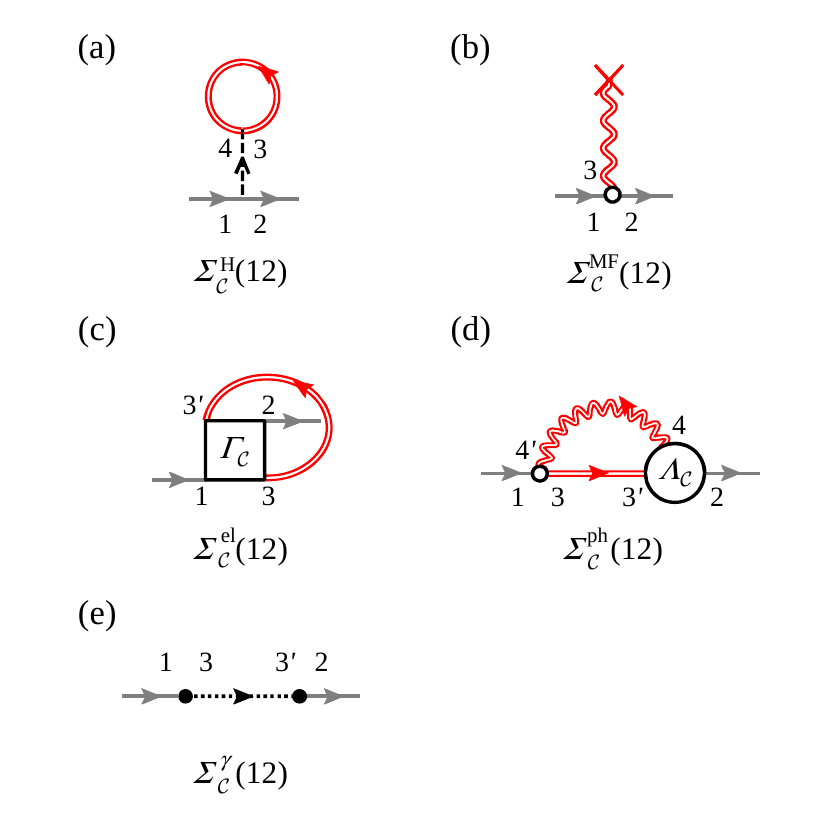} 
\caption{(Color online) Self-energy diagrams for the single-particle GF $G_{\C}$.
The Hartree term [Eq.~\eqref{eq:Sigma_H}] and the mean-field potential formed by the cavity photon field [Eq.~\eqref{eq:Sigma_MF}] are shown in panel (a) and (b), respectively.
Correlations beyond these effects, Eqs.~\eqref{eq:Sigma_el} and \eqref{eq:Sigma_ph}, are shown in panel (c) and (d).
The interaction between the \mbox{e-h} system and the pumping baths can be found in panel (e) [Eq.~\eqref{eq:Sigma_bath}].
}
\label{fig14:Selfenergies}
\end{figure}

The dynamics of the single-particle GF $G_{\C}(11')$, however, should be given to study the behavior of the cavity photon amplitude $a_{\C}(1)$, as seen in Eq.~\eqref{eq:Dyson eq. in differential form}.
To obtain the contour evolution of $G_{\C}(11')$, we differentiate $G_{\C}(11')$ with respect to $\tau_{1}$, in a similar manner to Eq.~\eqref{eq:ac_dynamics1}.
As a result, we can find 
%----------
% Equation
%----------
\begin{multline}
\ii \frac{\partial}{\partial \tau_{1}} G_{\C}(11') = \delta_{\C}(11') 
\\ - \ii \EX{ \Tc \left[ [ \oc_{z_{1}}(\tau_{1}), \oH{total}^{\text{I}}(\tau_{1}) ]\ocd_{z'_{1}}(\tau'_{1})  \right] }_{\A}.
\label{eq:Dynamics of Gc}
\end{multline}
%----------
In the derivation, from Eqs.~\eqref{eq:auxiliary ex} and \eqref{eq:Gc}, we have used that $G_{\C}(11')$ can be rewritten as 
%----------
% Equation
%----------
\begin{align}
&G_{\C}(11')  \nonumber \\
& = \frac{-\ii}{\Ex{\hat{S}_{\C}}} 
\theta_{\C}(\tau_{1} \tau'_{1}) \Ex{ \hat{S}(\tau_{0} \tau_{1}) \oc_{z_{1}}(\tau_{1}) \hat{S}(\tau_{1} \tau'_{1}) \ocd_{z'_{1}}(\tau'_{1}) \hat{S}(\tau'_{1} \tau_{0}) } \nonumber \\
& + \frac{\ii}{\Ex{\hat{S}_{\C}}}
\theta_{\C}(\tau'_{1} \tau_{1}) \Ex{ \hat{S}(\tau_{0} \tau'_{1}) \ocd_{z'_{1}}(\tau'_{1}) \hat{S}(\tau'_{1} \tau_{1}) \oc_{z_{1}}(\tau_{1}) \hat{S}(\tau_{1} \tau_{0}) },
\label{eq:Gc_rewritten}
\end{align}
%----------
where $\theta_{\C}(\tau_{1} \tau'_{1}) = 1$ if $\tau_{1}$ is later than $\tau'_{1}$ on the contour and zero otherwise; the Heaviside function on the contour.~\cite{Stefanucci13}
The delta function in Eq.~\eqref{eq:Dynamics of Gc} arises from the Heaviside functions in Eq.~\eqref{eq:Gc_rewritten}.
A straightforward calculation of the commutation relation in Eq.~\eqref{eq:Dynamics of Gc}, then, yields 
%----------
% Equation
%----------
\begin{align}
G^{-1}_{0,\C}(12) &G_{\C}(21')  = 
\delta_{\C}(11') +U(12)G_{\C}(21') \nonumber \\
& + [\varSigma^{\H}_{\C}(12) + \varSigma_{\C}^{\MF}(12) ]G_{\C}(21') \nonumber \\
& - \ii U'_{\C}(1234) K_{\C}(41'32) \nonumber\\
& + g_{\C}(2;13) P_{\C}(31';2) \nonumber\\
& - \ii \varsigma_{\C}(12) \Ex{ \Tc [ \ob(2)\ocd(1') ] }_{\A},
\label{eq:DynamicsGc}
\end{align}
%----------
where $\varSigma^{\H}_{\C}(12)$ denotes the self-energy called the Hartree term,
%----------
% Equation
%----------
\begin{align}
\varSigma^{\H}_{\C}(12) \equiv -\ii U'_{\C}(1432)G_{\C}(34),
\label{eq:Sigma_H}
\end{align}
%----------
and $\varSigma_{\C}^{\MF}(12)$ is the self-energy of the mean-field potential formed by the cavity photon field,
%----------
% Equation
%----------
\begin{align}
\varSigma_{\C}^{\MF}(12) \equiv \bar{g}_{\C}(12;3) a_{\C}(3).
\label{eq:Sigma_MF}
\end{align}
%----------
These self-energies are diagrammatically shown in Figures~\ref{fig14:Selfenergies}(a) and \ref{fig14:Selfenergies}(b), respectively.
It is then obvious that the electronic correlations beyond the Hartree term [Eq.~\eqref{eq:Sigma_H}] are described by $K_{\C}$, while those with photons beyond the mean-field potential [Eq.~\eqref{eq:Sigma_MF}] are expressed by $P_{\C}$.
A source of spontaneous emission from the \mbox{e-h} system to the cavity, for example, is provided from $P_{\C}$ as is well-known in the cluster expansion approach; see Ref.~\onlinecite{Kira99} for details.
However, at this stage, these correlation terms prevent the diagrammatic description with the self-energies even though Eq.~\eqref{eq:DynamicsGc} is analogous to the Dyson equation in the differential form.

%----------
% Figure
%----------
%----------

%----------
% Figure
%----------
\begin{figure}[!tb] 
\centering
\includegraphics[width=.45\textwidth]{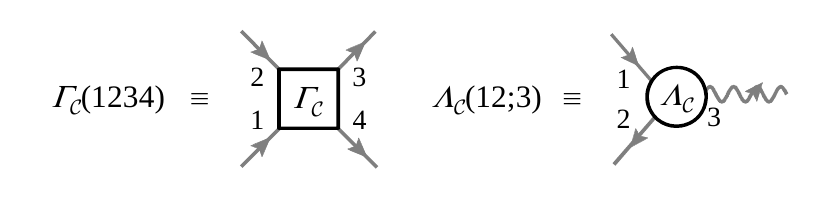} 
\caption{Definitions of the vertex functions.
}
\label{fig15:DressedVertices}
\end{figure}
%----------

As a next step, we therefore explain the way to formally obtain the self-energies in our formalism by using the chain rule of the functional derivative.~\cite{Martin59, Baym61,Baym62,Dominicis64,Hohenberg65,Kragler80,Henneberger88,Fujikawa05}
For this purpose, it is convenient to introduce the inverse of the single-particle GF $G^{-1}_{\C}$ that satisfies
%----------
% Equation
%----------
\begin{align}
G^{-1}_{\C}(12)G_{\C}(21') = \delta_{\C}(11') = G_{\C}(12)G^{-1}_{\C}(21'),
\label{eq:InverseGc}
\end{align}
%----------
in the same manner as Eq.~\eqref{eq:Bare_NEGFs}.
By taking the variation of both sides in Eq.~\eqref{eq:InverseGc}, one finds
%----------
% Equation
%----------
\begin{align}
\delta G_{\C}(11') = -G_{\C}(12) \delta G^{-1}_{\C}(22') G_{\C}(2'1'),
\label{eq:UsefulRelation}
\end{align}
%----------
which allows us to transform the correlations of $K_{\C}$ and $P_{\C}$ into the following forms:
%----------
% Equation
%----------
\begin{align}
&K_{\C}(41'32)= G_{\C}(45) \frac{\delta G_{\C}^{-1}(56)}{\delta U(23)} G(61'),
\label{eq:Kc_temp1}\\
&P_{\C}(31';2) = -\ii G_{\C}(34) \frac{ \delta G_{\C}^{-1}(44') }{\delta \bar{\eta}(2)} G_{\C}(4'1') \nonumber \\
& \quad =-\ii G_{\C}(34) D_{\C}(52) \frac{ \delta G_{\C}^{-1}(44') }{\delta a_{\C}(5)} G_{\C}(4'1'), 
\end{align}
%----------
where the chain rule of the functional derivative as well as Eqs.~\eqref{eq:Dc}, \eqref{eq:Kc} and \eqref{eq:Pc} have been used.
By substituting them into Eq.~\eqref{eq:DynamicsGc}, we can find
%----------
% Equation
%----------
\begin{align}
- \ii U'(1234)K_{\C}(41'32) &= \varSigma^{\text{el}}_{\C}(12) G_{\C}(21'),
\label{eq:Sigma_el_conversion} \\
g_{\C}(2;13)P_{\C}(31';2) &= \varSigma^{\text{ph}}_{\C}(12) G_{\C}(21'),
\label{eq:Sigma_ph_conversion} 
\end{align}
%----------
with the self-energies
\begin{align}
\varSigma^{\text{el}}_{\C}(12) &\equiv \ii G_{\C}(33') \varGamma_{\C}(13'23),
\label{eq:Sigma_el}\\
\varSigma^{\text{ph}}_{\C}(12) &\equiv \ii g_{\C}(4';13) D_{\C}(44') G_{\C}(33') \varLambda_{\C}(3'2;4),
\label{eq:Sigma_ph}
\end{align}
%----------
and the vertex functions
%----------
% Equation
%----------
\begin{align}
\varGamma_{\C}(1234) &\equiv -U'(12'3'4) \frac{\delta G^{-1}_{\C}(23)}{\delta U(2'3')},
\label{eq:GammaC}\\
\varLambda_{\C}(12;3) &\equiv - \frac{\delta G^{-1}_{\C}(12)}{\delta a_{\C}(3)}.
\label{eq:LambdaC}
\end{align}
%----------
The terms arising from $K_{\C}$ and $P_{\C}$ in Eq.~\eqref{eq:DynamicsGc} can, thus, be described by the self-energies with the vertex functions.
The corresponding diagrammatic representations are shown in Figures~\ref{fig14:Selfenergies}(c),~\ref{fig14:Selfenergies}(d) and \ref{fig15:DressedVertices}.

Finally, it is straightforward to obtain
%----------
% Equation
%----------
\begin{align}
B^{-1}_{0,\C}(12) \Ex{ \Tc [ \ob(2)\ocd(1') ] }_{\A} = \ii \varsigma_{\C}(21) G_{\C}(21'),
\end{align}
%----------
by differentiating $\Ex{ \Tc [ \ob(2)\ocd(1') ] }_{\A}$ with respect to $\tau_{2}$ in a similar manner to Eqs.~\eqref{eq:ac_dynamics1} and \eqref{eq:Dynamics of Gc}.
As a result, we can rewrite the last term in Eq.~\eqref{eq:DynamicsGc} as
%----------
% Equation
%----------
\begin{align}
- \ii \varsigma_{\C}(12) \Ex{ \Tc [ \ob(2)\ocd(1') ] }_{\A} = \varSigma^{\gamma}_{\C}(12) G_{\C}(21'),
\label{eq:Sigma_bath_conversion}
\end{align}
%----------
with a definition of 
%----------
% Equation
%----------
\begin{align}
\varSigma^{\gamma}_{\C}(12) \equiv \varsigma_{\C}(13) B_{0,\C}(33') \varsigma_{\C}(23'),
\label{eq:Sigma_bath}
\end{align}
%----------
the diagram of which is in Figure~\ref{fig14:Selfenergies}(e).
The superscript $\gamma$ signifies the thermalization effect in a similar manner to Eq.~\eqref{eq:Sigma_Decay}.
The Dyson equation is then obtained in the differential form by inserting Eqs.~\eqref{eq:Sigma_el_conversion}, \eqref{eq:Sigma_ph_conversion} and \eqref{eq:Sigma_bath_conversion} into Eq.~\eqref{eq:DynamicsGc};
%----------
% Equation
%----------
\begin{multline}
G^{-1}_{0,\C}(12) G_{\C}(21') = \delta_{\C}(11') + \varSigma_{\C}(12)G_{\C}(21') \\
+ U(12)G_{\C}(21'),
\label{eq:Dyson_differeintial}
\end{multline}
%----------
where $\varSigma_{\C} \equiv \varSigma^{\H}_{\C} +\varSigma^{\MF}_{\C} + \varSigma^{\text{el}}_{\C} + \varSigma^{\text{ph}}_{\C} + \varSigma^{\gamma}_{\C}$.
This equation plays a key role when we derive the equation of motion for the polarization function $p_{\k}$ [Eq.~\eqref{eq:MSBEs--pk}] as well as the distribution functions of electrons $n_{\e,\k}$ and holes $n_{\h,\k}$ [Eq.~\eqref{eq:MSBEs--nehk}]; see also Subsection~\ref{subsec:EOM2}. 
Eq~\eqref{eq:Dyson_differeintial} is, of course, equivalent to
%----------
% Equation
%----------
\begin{multline}
G_{\C}(11') = G_{0,\C}(11') + G_{0,\C}(12)\varSigma_{\C}(22')G_{\C}(2'1'),
\label{eq:Dyson}
\end{multline}
%----------
in the limit of the vanishing auxiliary source fields and can be drawn diagrammatically in Figure~\ref{fig16:DysonEquations}(a).

The equation of motion for the single-particle GF $G_{\C}$ can thus be obtained in the form of the Dyson equation successfully.
However, there arise further needs to study 
\begin{enumerate}
\renewcommand{\labelenumi}{\roman{enumi})}
\item the photon GF $D_{\C}(11')$ [Eq.~\eqref{eq:Sigma_ph}],
\item the vertex functions of $\varGamma_{\C}(1234)$ [Eq.~\eqref{eq:Sigma_el}] and $\varLambda_{\C}(12;3)$ [Eq.~\eqref{eq:Sigma_ph}].
\end{enumerate}
Although not yet encountered, 
\begin{enumerate}
\renewcommand{\labelenumi}{\roman{enumi})}
\setcounter{enumi}{2}
\item the two-particle GF $K_{\C}(11'22')$,
\end{enumerate}
should also be discussed.
In the following subsubsections, we discuss the equations of motion for these NEGFs.

%---------------------------------------
% Equations of motion for the photon GF
%---------------------------------------
\subsubsection{Photon GF}\label{subsubsec:Photon GF}

%----------
% Figure
%----------
\begin{figure}[!tb] 
\centering
\includegraphics[width=.45\textwidth]{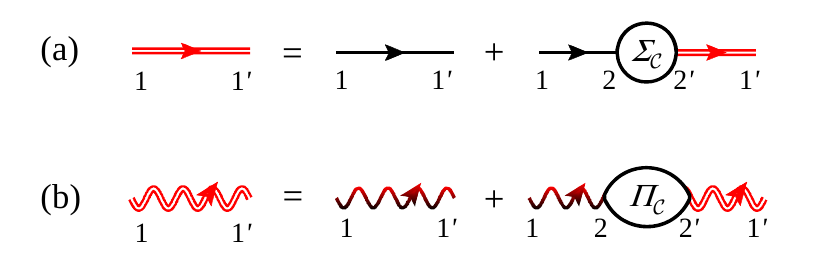} 
\caption{(Color) Diagrammatic descriptions of the Dyson equations for (a) $G_{\C}(11')$ [Eq.~\eqref{eq:Dyson_differeintial} or Eq.~\eqref{eq:Dyson}] and (b) $D_{\C}(11')$ [Eq.~\eqref{eq:Dyson_Dc}].
$\varSigma_{\C}$ is the summation of the self-energies shown in Figure~\ref{fig14:Selfenergies}, while $\varPi_{\C}$ is in Figure~\ref{fig17:PIc}.
}
\label{fig16:DysonEquations}
\end{figure}
%----------
% Figure
%----------
\begin{figure}[!tb] 
\centering
\includegraphics[width=.45\textwidth]{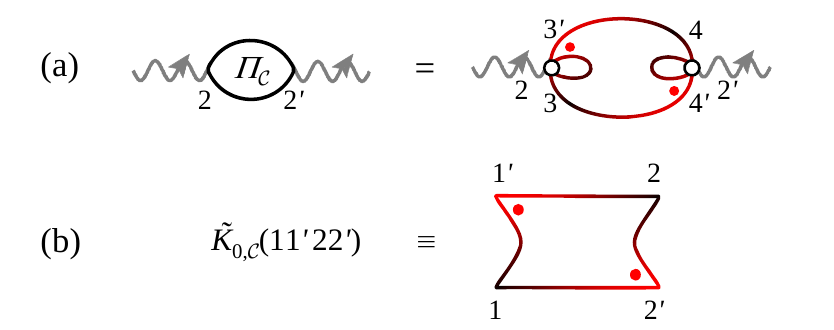} 
\caption{(Color online)
(a) the self-energy $\varPi_{\C}$ for the photon GF [Eq.~\eqref{eq:Pi_c}]; (b) the definition of the partially dressed two-particle GF $\tilde{K}_{0,\C}$.
}
\label{fig17:PIc}
\end{figure}
%----------
%----------
% Figure
%----------
\begin{figure*}[!tb] 
\centering
\includegraphics[width=.90\textwidth, clip]{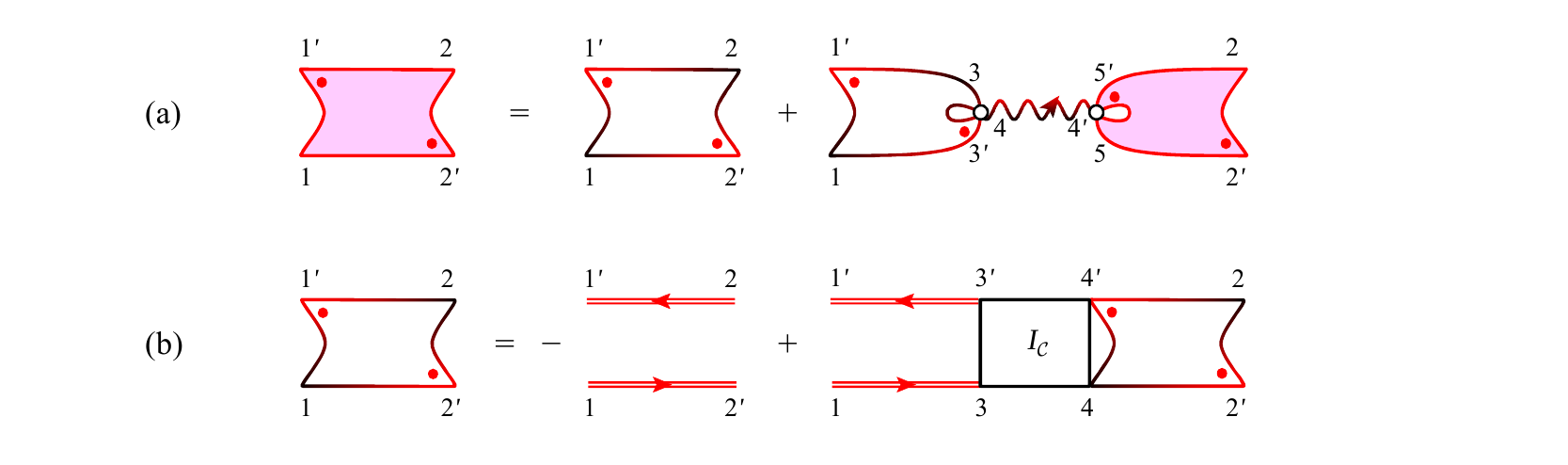} 
\caption{(Color)
Diagrammatic descriptions of the Bethe-Salpeter equations for (a) the fully dressed two-particle GF $K_{\C}(11'22')$ [Eq.~\eqref{eq:BSE1}] and (b) the partially dressed one $\tilde{K}_{0,\C}(11'22')$ [Eq.~\eqref{eq:BSE2}].
}
\label{fig18:BSE}
\end{figure*}
%----------

By using Eqs.~\eqref{eq:Dc} and \eqref{eq:partially dressed photon GF}, the chain rule of the functional derivative allows us to rewrite the photon GF as 
%----------
% Equation
%----------
\begin{align}
D_{\C}(11') = \frac{\delta a_{\C}(1)}{\delta J_{\text{S}}(2)} \frac{\delta J_{\text{S}}(2)}{\delta \bar{\eta}(1')}
= \tilde{D}_{0,\C}(12) \frac{\delta J_{\text{S}}(2)}{\delta \bar{\eta}(1')},
\label{eq:temp0}
\end{align}
%----------
where, from Eq.~\eqref{eq:Js}, 
\begin{align}
\frac{\delta J_{\text{S}}(2)}{\delta \bar{\eta}(1')} = \delta_{\C}(21') - \ii g_{\C}(2;33') \frac{\delta G_{\C}(33')}{\delta \bar{\eta}(1')}.
\label{eq:temp1}
\end{align}
By introducing an effective potential $\tilde{U}(12)$ and a {\em partially dressed} two-particle GF $\tilde{K}_{0,\C}(11'22')$ as
%----------
% Equation
%----------
\begin{align}
&\tilde{U}(12) \equiv U(12) + \bar{g}_{\C}(12;3) a_{\C}(3),
\label{eq:EffectiveU} \\
&\tilde{K}_{0,\C}(11'22') \equiv - \frac{\delta G_{\C}(11')}{\delta \tilde{U}(2'2)},
\label{eq:Kc_partial} 
\end{align}
%----------
the last term in Eq.~\eqref{eq:temp1} can be calculated as 
%----------
% Equation
%----------
\begin{align}
\frac{\delta G_{\C}(33')}{\delta \bar{\eta}(1')} &= \frac{\delta G_{\C}(33')}{\delta \tilde{U}(4'4)}  \frac{\delta \tilde{U}(4'4)}{\delta \bar{\eta}(1')} \nonumber \\
&= -\tilde{K}_{0,\C}(33'44') \bar{g}_{\C}(4'4;5) D_{\C}(51').
\label{eq:temp2}
\end{align}
%----------
The roles of Eqs.~\eqref{eq:EffectiveU} and \eqref{eq:Kc_partial} will be illustrated in the next subsubsection.
Here, by putting Eq.~\eqref{eq:temp1} into Eq.~\eqref{eq:temp0} with Eq.~\eqref{eq:temp2}, one finds the Dyson equation 
%----------
% Equation
%----------
\begin{align}
D_{\C}(11') = \tilde{D}_{0,\C}(11') + \tilde{D}_{0,\C}(12) \varPi_{\C}(22') D_{\C}(2'1'),
\label{eq:Dyson_Dc}
\end{align}
%----------
with the self-energy defined as 
%----------
% Equation
%----------
\begin{align}
\varPi_{\C}(22') = \ii g_{\C}(2;3'3) \tilde{K}_{0,\C}(33'44') \bar{g}_{\C} (4'4;2').
\label{eq:Pi_c}
\end{align}
%----------
We can now successfully represent the diagrams of these equations in Figures~\ref{fig16:DysonEquations}(b) and \ref{fig17:PIc}(a) with a definition of $\tilde{K}_{0,\C}(11'22')$ in Figure~\ref{fig17:PIc}(b).
We note that, in Figure~\ref{fig17:PIc}, the self-energy $\varPi_{\C}$ is not described by the fully dressed two-particle GF $K_{\C}$ but by the partially dressed two-particle GF $\tilde{K}_{0,\C}$, which has an effect to avoid the double counting problem in the calculation of the photon GF.
This feature will become more evident by studying the equations of motion for $K_{\C}$ and $\tilde{K}_{0,\C}$ in the next subsubsection.
Eqs.~\eqref{eq:Dyson_Dc} and \eqref{eq:Pi_c} are important for the study the emission spectrum [Eq.~\eqref{eq:Sinc2}] as discussed in Section~\ref{sec:Spectral} and Appendix~\ref{app:Formalism for Spectra}.

%---------------------------------------
% Two-particle GF 
%---------------------------------------
\subsubsection{Two-particle GF}
The chain rule of the functional derivative can also make a connection between $K_{\C}$ and $\tilde{K}_{0,\C}$ in a similar manner to Eq.~\eqref{eq:temp0}.
From Eqs.~\eqref{eq:Kc} and \eqref{eq:Kc_partial}, $K_{\C}$ can be written as
%----------
% Equation
%----------
\begin{align}
K_{\C}(11'22') = \tilde{K}_{0,\C}(11'33') \frac{\delta \tilde{U}(3'3)}{\delta U(2'2)},
\end{align}
%----------
where, from Eqs.~\eqref{eq:EffectiveU} and \eqref{eq:Js}, one can find
%----------
% Equation
%----------
\begin{align}
&\frac{\delta \tilde{U}(3'3)}{\delta U(2'2)}  \nonumber \\
& \quad = \delta_{\C}(3'2')\delta_{\C}(32) + \bar{g}_{\C}(3'3;4) \tilde{D}_{0,\C}(44')  \frac{\delta J_{\text{S}}(4')}{\delta U(2'2)}  \nonumber \\
& \quad = \delta_{\C}(3'2')\delta_{\C}(32) \nonumber \\
& \qquad + \ii \bar{g}_{\C}(3'3;4) \tilde{D}_{0,\C}(44') g_{\C}(4';5'5) K_{\C}(55'22').
\label{eq:tildeU/U}
\end{align}
%----------
As a result, we obtain 
%----------
% Equation
%----------
\begin{align}
&K_{\C}(11'22') = \tilde{K}_{0,\C}(11'22') \nonumber \\
&+ \ii \tilde{K}_{0,\C}(11'33') \bar{g}_{\C}(3'3;4) \tilde{D}_{0,\C}(44') g_{\C}(4';5'5) K_{\C}(55'22'),
\label{eq:BSE1}
\end{align}
%----------
the diagram of which is shown in Figure~\ref{fig18:BSE}(a).
Note that Eq.~\eqref{eq:BSE1} is a kind of the Bethe-Salpeter equation~\cite{Haug84} (BSE) for the two-particle GF $K_{\C}$.
In contrast, the equation of motion for the partially dressed GF $\tilde{K}_{0,\C}$ can be obtained by applying Eq.~\eqref{eq:UsefulRelation} to Eq.~\eqref{eq:Kc_partial} as
%----------
% Equation
%----------
\begin{align}
&\tilde{K}_{0,\C}(11'22') = G_{\C}(13)G_{\C}(3'1') \frac{\delta G^{-1}_{\C}(33')}{\delta \tilde{U}(2'2)} \nonumber \\
& \quad = -G_{\C}(12')G_{\C}(21') - G_{\C}(13)G_{\C}(3'1')\frac{\delta \tilde{\varSigma}_{\C}(33')}{\delta \tilde{U}(2'2)}  \nonumber \\
& \quad = -G_{\C}(12')G_{\C}(21') \nonumber\\
& \qquad\quad + G_{\C}(13)G_{\C}(3'1') I_{\C}(33'44') \tilde{K}_{0,\C}(44'22'),
\label{eq:BSE2}
\end{align}
%----------
where
%----------
% Equation
%----------
\begin{align}
I_{\C}(33'44') \equiv \frac{\delta \tilde{\varSigma}_{\C}(33')}{\delta G_{\C}(44')}.
\label{eq:IntegralKernel}
\end{align}
%----------
In the second line of Eq.~\eqref{eq:BSE2}, we have used the following equation derived from Eqs.~\eqref{eq:Sigma_MF}, \eqref{eq:Dyson_differeintial} and \eqref{eq:EffectiveU}:
%----------
% Equation
%----------
\begin{align}
G^{-1}_{\C} =  G^{-1}_{0, \C} - U - \varSigma_{\C} = G^{-1}_{0, \C} - \tilde{U} - \tilde{\varSigma}_{\C},
\label{eq:G_inverse}
\end{align}
%----------
with 
%----------
% Equation
%----------
\begin{align*}
\tilde{\varSigma}_{\C} \equiv \varSigma_{\C} - \varSigma^{\MF}_{\C}.
\end{align*}
%----------
Again, Eq.~\eqref{eq:BSE2} is a kind of the BSE and diagrammatically shown in Figure~\ref{fig18:BSE}(b).
Here, the integration kernel $I_{\C}$ can be determined if the self-energy diagrams (or the vertex functions) are truncated at a certain level, which allows us to consider the rational BSEs required for the calculations of the emission spectrum as well as the gain-absorption spectrum; see Appendix~\ref{app:Formalism for Spectra}.

It is also instructive to mention that, from Figure~\ref{fig18:BSE}, the fully dressed $K_{\C}$ can be decomposed into two categories of diagrams; the chain diagram connected by $\tilde{D}_{0,\C}$ is included or not.
The former evidently causes the double counting problem if used for the self-energy $\varPi_{\C}$ because the chain diagrams are also generated from the Dyson equation in Figure~\ref{fig16:DysonEquations}(b), while the latter does not cause such a problem and, in fact, corresponds to $\tilde{K}_{0,\C}$.
This is the reason why the partially dressed $\tilde{K}_{0,\C}$ appears in the description of $\varPi_{\C}$ [Eq.~\eqref{eq:Pi_c}].
In this context, $K_{\C}$ should not be confused with $\tilde{K}_{0,\C}$.
Now, it is evident that, although $\tilde{K}_{0,\C}$ is enough to study the photon GF as shown in Figure~\ref{fig17:PIc}(a), we have to take the fully dressed $K_{\C}$ in the calculations of the gain-absorption spectrum [Eq.~\eqref{eq:Susceptibility}], as described in Section~\ref{sec:Spectral}.

%---------------------------------------
% Vertex functions
%---------------------------------------
\subsubsection{Vertex functions}\label{subsubsec:Vertex functions}
\begin{figure}[!tb] 
\centering
\includegraphics[width=.45\textwidth, clip]{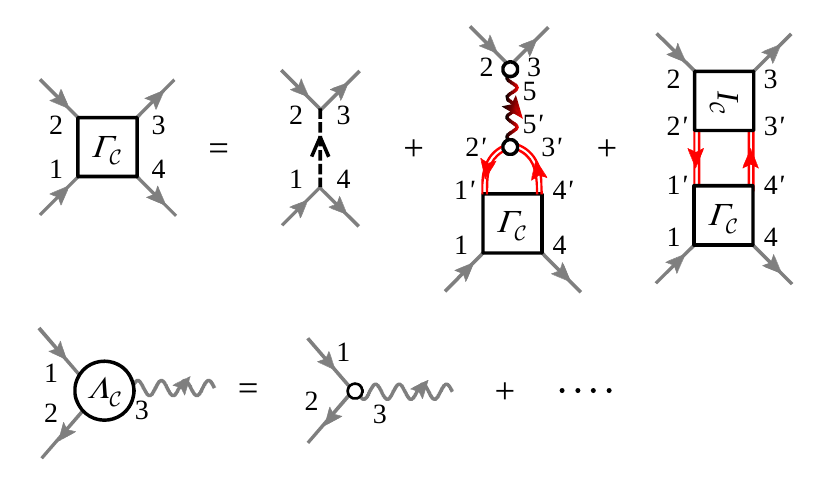} 
\caption{(Color)
Diagrammatic representations of the vertex functions $\varGamma_{\C}$ [Eq.~\eqref{eq:GammaC_Eq}] and $\varLambda_{\C}$ [Eq.~\eqref{eq:LambdaC_Eq}].
}
\label{fig19:VertexEquation}
\end{figure}

Finally, we study the vertex functions $\varGamma_{\C}$ and $\varLambda_{\C}$.
By substituting Eq.~\eqref{eq:G_inverse} into Eqs.~\eqref{eq:GammaC} and \eqref{eq:LambdaC},
%----------
% Equation
%----------
\begin{align}
&\varGamma_{\C}(1234) = U'_{\C}(12'3'4) \left\{ \frac{\delta \tilde{U}(23)}{\delta U(2'3')} + \frac{\delta \tilde{\varSigma}_{\C}(23)}{\delta U(2'3')}  \right\}, \\
&\varLambda_{\C}(12;3) = \frac{\delta \tilde{U}(12)}{\delta a_{\C}(3)} + \frac{\delta \tilde{\varSigma}_{\C}(12)}{\delta a_{\C}(3)}.
\end{align}
%----------
We then obtain for the vertex function $\varGamma_{\C}$
%----------
% Equation
%----------
\begin{align}
&\varGamma_{\C}(1234) = U'_{\C}(1234) \nonumber\\
& \quad - \ii  \bar{g}_{\C}(23;5) \tilde{D}_{0,\C}(55') g_{\C}(5';3'2') \nonumber\\
& \qquad\qquad\qquad \times G_{\C}(2'1')G_{\C}(4'3') \varGamma_{\C}(11'4'4) \nonumber \\
& \quad + I_{\C}(232'3') G_{\C}(2'1')G_{\C}(4'3') \varGamma_{\C}(11'4'4),
\label{eq:GammaC_Eq}
\end{align}
%----------
where we have used Eq.~\eqref{eq:tildeU/U} with Eq.~\eqref{eq:Kc_temp1}. 
In the same manner, by using Eq.~\eqref{eq:EffectiveU}, 
%----------
% Equation
%----------
\begin{align}
&\varLambda_{\C}(12;3) = \bar{g}_{\C}(12;3) + \frac{\delta \varSigma_{\C}^{\text{el}}(12)}{\delta a_{\C}(3)} + \frac{\delta \varSigma_{\C}^{\text{ph}}(12)}{\delta a_{\C}(3)}.
\label{eq:LambdaC_Eq}
\end{align}
%----------
We note that, for our purpose,  there is no need to expand the second and third terms in Eq.~\eqref{eq:LambdaC_Eq} that yield higher order terms in the interaction coefficients.
As a result, the diagrammatic representations for Eqs.~\eqref{eq:GammaC_Eq} and \eqref{eq:LambdaC_Eq} are shown in Figure~\ref{fig19:VertexEquation}.

%----------
% Figure
%----------
%----------

In summary, in this section, we have constructed a general formalism to treat the semiconductor \mbox{e-h-p} systems, based on the generating functional approach.
The equations of motion for the relevant NEGFs are derived with the diagrammatic representations, as shown in Figures~\ref{fig13:Amplitude}, \ref{fig16:DysonEquations} and \ref{fig18:BSE}, some of which are identical to the Dyson equations and the Bethe-Salpeter equations.
The self-energies for the single-particle GF $G_{\C}$ and the photon GF $D_{\C}$ are summarized in Figures~\ref{fig14:Selfenergies} and \ref{fig17:PIc}, while the vertex functions are in Figure~\ref{fig19:VertexEquation}.
The partially dressed NEGFs $\tilde{D}_{0,\C}$ and $\tilde{K}_{0,\C}$ have been naturally introduced as a result of the generating functional formalism, which play a key role to correctly describe the equations of motion with avoiding the double counting problem.
Figure~\ref{fig11:Overview} thus shows the summary of this section.

In the next section, we transform the NEGFs into the real-time matrix representations and derive a time-dependent framework that generalizes the \mbox{MSBEs} under the RTA.
This framework will give a starting point to study the cooperative phenomena in a unified view.

%----------
% Figure
%----------
\begin{figure*}[!tb] 
\centering
\includegraphics[width=.90\textwidth, clip]{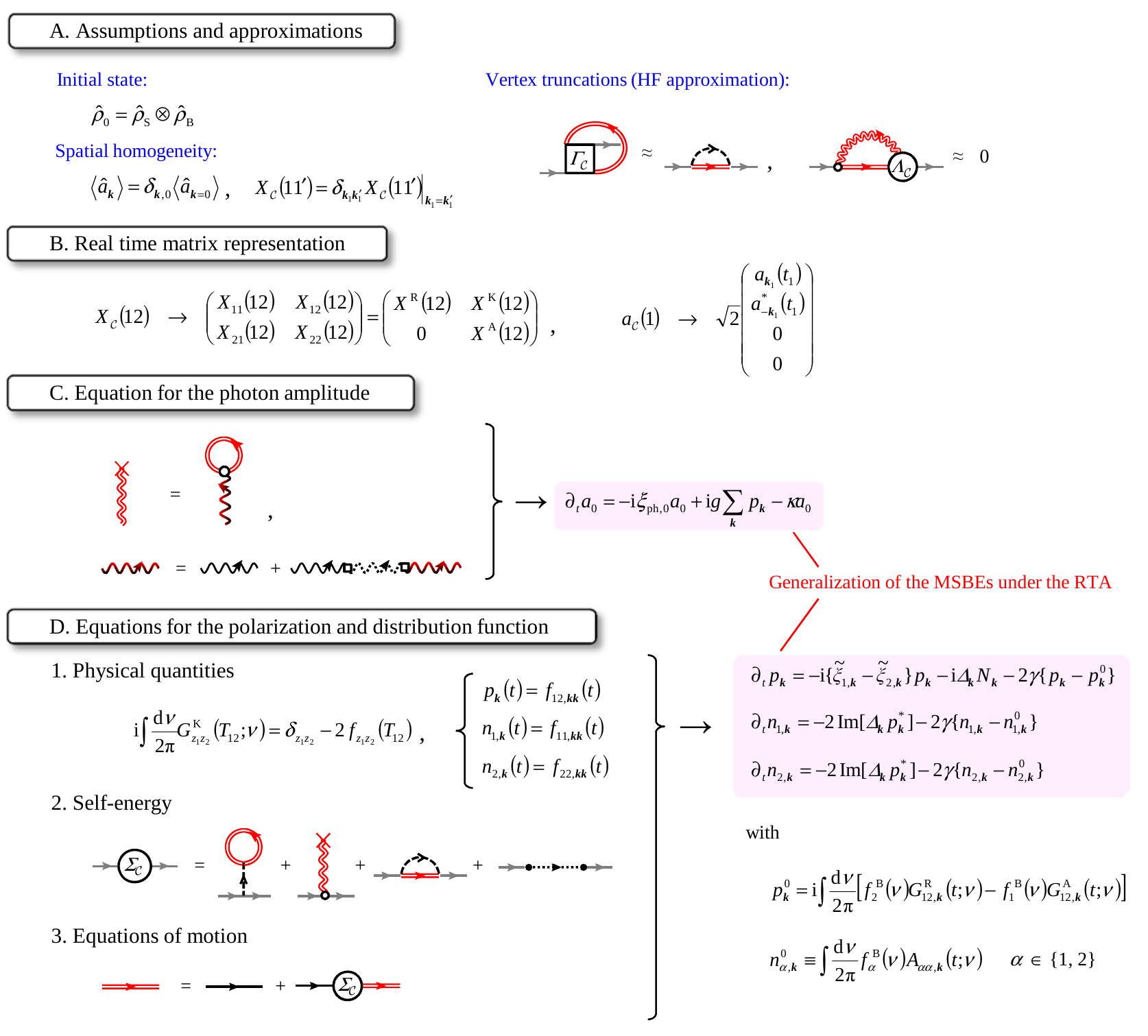} 
\caption{(Color) Our scheme for the derivation of equations of motion, depicted along the structure of Section~\ref{sec:real time}.
$A_{\alpha \alpha, \k}(t;\nu)$ and $G^{\R/\A}_{12,\k}(t;\nu)$ describe the time-dependent renormalization of the electronic band structures.
}
\label{fig20:Scheme}
\end{figure*}
%----------

%======================================================
% real-time formalism
%======================================================
\section{real-time formalism}\label{sec:real time}
The main purpose of this section is to derive the time-dependent simultaneous equations of motion for the physical quantities of the cavity photon amplitude $a_{0}(t)$, the polarization function $p_{\k}(t)$, and the distribution functions of electrons in the conduction band $n_{1,\k}(t)$ and in the valence band $n_{2,\k}(t)$ based on the formalism presented in Section~\ref{sec:Generating Functional}.
The scheme of our derivation is shown in Figure~\ref{fig20:Scheme}, again depicted along the structure of this section.
We first illustrate the required assumptions and approximations in Subsection~\ref{Assumptions}.
We then derive the equation of motion for the photon amplitude in Subsection~\ref{subsec:EOM1} and the polarization and distribution functions in Subsection~\ref{subsec:EOM2} by using the results shown in Section~\ref{sec:Generating Functional}.
As a result, it is found that the time-dependent renormalization of the electronic band structures should be traced together with the evolution of the relevant physical quantities, at least in principle.
By transforming the results, we finally obtain the generalization of the MSBEs, namely, the key results [Eqs.~\eqref{alleq:MSBEs} with Eqs.~\eqref{alleq:p0k2-n0k2}--\eqref{eq:A}] shown in Subsection~\ref{subsec:Key results}.

%=======================================
% Assumptions and approximations
%=======================================
\subsection{Assumptions and approximations}\label{Assumptions}
In order to treat the open-dissipative nature, we first assume that the system and the reservoirs are initially uncorrelated and the initial state $\hat{\rho}_{0}$ can be described by
%----------
% Equation
%----------
\begin{align}
\hat{\rho}_{0} = \hat{\rho}_{\text{S}} \otimes \hat{\rho}_{\text{B}},
\label{eq:Born}
\end{align}
%----------
where $\hat{\rho}_{\text{S}}$ is an arbitrary initial state of the system, while $\hat{\rho}_{\text{B}}$ is the direct product of the pumping baths in thermal equilibrium and the vacuum photon bath in Figure~\ref{fig01:Model}(b).
The separable assumption of Eq.~\eqref{eq:Born} is sometimes called the Born approximation.
We note that, in our formalism, the system and the baths do not have to be separable in the process of evolution in contrast to the QME approach~\cite{Nakatani10,Nakatani10E,Breuer02,Shirai14,Yuge14} and will be entangled.

In addition, for simplicity, we also assume the spatial homogeneity of the system.
We therefore consider the macroscopic coherence of the photon amplitude developed only for the $\k = 0$ state
%----------
% Equation
%----------
\begin{align}
a_{\k} \equiv \Ex{\oa_{\k}} = \delta_{\k, 0}\Ex{\oa_{\k=0}} = \delta_{\k, 0} a_0,
\label{eq:Assumption}
\end{align}
%----------
and the NEGFs satisfying
%----------
% Equation
%----------
\begin{align}
&X_{\C}(11')=\delta_{\k_{1} \k'_{1}} X_{\C}(\tau_{1} \tau'_{1}; \alpha_{1} \alpha'_{1}; \k_{1}\k_{1}),
\label{eq:Conservation}
\end{align}
%----------
due to the momentum conservation law.

Under these assumptions, only the self-energies of the first order in $U'_{\C}$ and $g_{\C}$ are taken into account. 
In this context, we truncate the vertex functions as
%----------
% Equation
%----------
\begin{align}
\varGamma_{\C}(1234) \simeq U'_{\C}(1234), \quad \varLambda_{\C}(12;3) \simeq 0.
\label{eq:Vertex_Approx}
\end{align}
%----------
As a result, $\varSigma^{\text{el}}_{\C}$ and $\varSigma^{\text{ph}}_{\C}$ are approximated as shown in the vertex truncations of Figure~\ref{fig20:Scheme}.
This means that we neglect the correlations beyond the HF approximation for the Coulomb interaction and those beyond the MF potential formed by the photon amplitude.
In this limit, it can be found that the equations of motion for $a_{\C}$ (Figure~\ref{fig13:Amplitude}) and $G_{\C}$ (Figures~\ref{fig14:Selfenergies} and \ref{fig16:DysonEquations}(a)) are closed even without $D_{\C}$ and $K_{\C}$.
In the followings, therefore, we focus on the dynamics of $a_{\C}$ and $G_{\C}$ derived in the previous section.
Note, however, that $D_{\C}$ and $K_{\C}$ can be studied by using the solution of $a_{\C}$ and $G_{\C}$, which is required to study the emission spectrum [Eq.~\eqref{eq:Sinc2}] and the gain-absorption spectrum [Eq.~\eqref{eq:Susceptibility}]; see Appendix~\ref{app:Formalism for Spectra} for the evaluation of $D_{\C}$ and $K_{\C}$.

%=======================================
% Equation for the photon amplitude
%=======================================
\subsection{Equation for the photon amplitude}\label{subsec:EOM1}
We now consider the equation of motion for the photon amplitude $a_{0}(t)$ by using Eqs.~\eqref{eq:Dyson eq. in differential form} and \eqref{eq:Inverse of partially dressed photon GF}, or equivalently Eqs.~\eqref{eq:Dyson eq.}--\eqref{eq:Sigma_Decay} (Figure~\ref{fig13:Amplitude}).
In the limit of $\oH{\A}(t) \rightarrow 0$, Eqs.~\eqref{eq:Dyson eq. in differential form} and \eqref{eq:Inverse of partially dressed photon GF} can be rewritten in the RAK basis (see Appendix~\ref{app:Real-time matrix}) as
%----------
% Equation
%----------
\begin{multline}
\left\{ D^{-1}_{0,kl}(11') -\varSigma_{kl}^{\kappa}(11') \right\} a_{l}(1') \\
 = - \ii g(1;23) L_{kl} \sigma^{(3)}_{ll'} \bar{G}_{ll'}(32),
\label{eq:Dyson eq. for the photon amplitude1}
\end{multline}
%----------
where, in the case of the real-time RAK representation, the times in $j = \{\alpha_j, \bm{k}, t_j \}$ are the standard real time,
%----------
% Equation
%----------
\begin{align*}
g(1;23) \equiv g(z_1;z_2 z_3)\delta(t_1 t_2)\delta(t_2 t_3),
\end{align*}
%----------
and it is assumed that the repeated arguments $j$ are integrated by $\int \dd j \equiv \sum_{z_j} \int^{\infty}_{-\infty} \dd t_j$.
By using the definitions of $\zeta_{\C}$ and $\bar{\zeta}_{\C}$ with Eqs.~\eqref{eq:kappa}, \eqref{eq:DOS_ph}, \eqref{eq:Sigma_Decay} and \eqref{eq:Conservation}, the self-energy $\varSigma^{\kappa}_{kl}(11')$ can be described as
%----------
% Equation
%----------
\begin{align}
\varSigma^{\R / \A}_{\kappa}(11') & \equiv \varSigma_{11 / 22}^{\kappa}(11') \nonumber\\
&= \delta_{\k_1 \k'_1} \delta(t_1 t'_1)
\begin{pmatrix}
\mp \ii \kappa &  \\
 & \pm \ii \kappa
\end{pmatrix},
\label{eq:SigmaRA_decay} \\
\varSigma^{\K}_{\kappa}(11') & \equiv \varSigma_{12}^{\kappa}(11') \nonumber\\
&= \delta_{\k_1 \k'_1} \delta(t_1 t'_1)
\begin{pmatrix}
- 2\ii \kappa &  \\
 & -2 \ii \kappa
\end{pmatrix},
\label{eq:SigmaK_decay}
\end{align}
%----------
in the RAK basis.
Here, the $2\times2$ matrix arises from the arguments of $\alpha_1$ and $\alpha_2$, i.e. the Nambu space for photons.
As a result, Eq.~\eqref{eq:Dyson eq. for the photon amplitude1} becomes 
%----------
% Equation
%----------
\begin{multline}
\begin{pmatrix}
\ii \partial_{t_1} - \xi_{\ph,\k_1} + \ii \kappa& \\
& -\ii \partial_{t_1} - \xi_{\ph,\k_1} - \ii \kappa \\
\end{pmatrix}
\begin{pmatrix}
a_{\k_1}(t_1) \\
a^{*}_{-\k_1}(t_1)\\
\end{pmatrix}
\\ 
= \delta_{\k_1,0} \sum_{\k_2} 
\begin{pmatrix}
- g p_{\k_2}(t_1) \\
- g^{*} p^{*}_{\k_2}(t_1)\\
\end{pmatrix},
\end{multline}
%----------
where the third and fourth elements in the four-component vector [Eq.~\eqref{eq:Vector}] are omitted because these are always zero in the RAK basis. 
In the derivation, we have used $D^{-1}_{0}$ defined in Table~\ref{tableVIII} (Appendix~\ref{app:Real-time matrix}) and $p_{\k}(t) = -\ii G^{<}_{12,\k}(tt)$ [see also Eq.~\eqref{eq:PhysicalQuantities1}].
We then obtain the equation of motion for $a_0(t)$ as 
%----------
% Equation
%----------
\begin{align}
\partial_{t} a_0 = -\ii \xi_{\ph,0}a_0 + \ii g \sum_{\k} p_{\k} - \kappa a_0,
\label{eq:Heisenberg-Langevin}
\end{align}
%----------
which is a member of our self-consistent equations of motion [Eq.~\eqref{eq:MSBEs--a0}].
Although Eq.~\eqref{eq:Heisenberg-Langevin} is the same as obtained by the Heisenberg-Langevin approach,~\cite{Scully97} we now know the diagrammatic representation to derive Eq.~\eqref{eq:Heisenberg-Langevin}, as shown in Figure~\ref{fig13:Amplitude}, which in turn clarifies our standpoint for the study of the single-particle GF.

%=======================================
% Equation for the polarization and distribution functions
%=======================================
\subsection{Equations for the polarization and distribution functions}\label{subsec:EOM2}
We now discuss the equations of motion for $p_{\k}(t)$, $n_{1,\k}(t)$ and $n_{2,\k}(t)$ by using the Dyson equation [Figure~\ref{fig16:DysonEquations}(a)] with the self-energies in Figure~\ref{fig14:Selfenergies} and the vertex truncations in Figure~\ref{fig20:Scheme}.
For this purpose, in the subsequent subsubsections, we will relate the physical quantities $p_{\k}(t)$, $n_{1,\k}(t)$ and $n_{2,\k}(t)$ to the GFs in the Wigner representations.~\cite{Wigner32, Rammer07}
We then describe the self-energies in the Wigner coordinates and give finally our time-dependent formalism to study the cooperative phenomena.

%---------------------------------------
% Physical quantities
%---------------------------------------
\subsubsection{Physical quantities}\label{subsubsec:Physical quantities}
The physical quantities of $p_{\k}(t)$, $n_{1,\k}(t)$ and $n_{2,\k}(t)$ defined in Section~\ref{subsec:the BCS theory and the MSBEs} can be related to the lesser GF (see also Table~\ref{tableVII} in Appendix~\ref{app:Real-time matrix}) as
%----------
% Equation
%----------
\begin{align}
\begin{pmatrix}
n_{1,\k}(t) & p_{\k}(t)\\
p^{*}_{\k}(t) & n_{2,\k}(t)
\end{pmatrix} 
= -\ii
\begin{pmatrix}
G^{<}_{11,\k}(t t) & G^{<}_{12,\k}(t t)\\
G^{<}_{21,\k}(t t) & G^{<}_{22,\k}(t t)
\end{pmatrix}.
\label{eq:PhysicalQuantities1}
\end{align}
%----------
In the RAK basis, we obtain
%----------
% Equation
%----------
\begin{align}
\begin{pmatrix}
1-2n_{1,\k}(t) & -2p_{\k}(t)\\
-2p^{*}_{\k}(t) & 1-2n_{2,\k}(t)
\end{pmatrix} 
= \ii
\begin{pmatrix}
G^{\K}_{11,\k}(t t) & G^{\K}_{12,\k}(t t)\\
G^{\K}_{21,\k}(t t) & G^{\K}_{22,\k}(t t)
\end{pmatrix},
\label{eq:PhysicalQuantities2}
\end{align}
%----------
because 
%----------
% Equation
%----------
\begin{align}
\ii G^{\K}(12)|_{t_1=t_2}  &= \ii \left( G^{>}(12) + G^{<}(12) \right)|_{t_1=t_2} \nonumber \\
&= \delta_{z_1 z_2} + 2\ii G^{<}(12)|_{t_1=t_2},
\label{eq:temp3}
\end{align}
%----------
is satisfied from Eq.~\eqref{eq:RAK2} with the equal-time anti-commutation relations of the Fermi operators.

In contrast, the time-dependent renormalization of the electronic band structures can be discussed through the single-particle spectral function $A(12)$ defined as~\cite{Rammer07}
%----------
% Equation
%----------
\begin{align}
A(12) & \equiv \ii \left( G^{>}(12) - G^{<}(12) \right) \nonumber\\
&= \ii \left( G^{\R}(12) - G^{\A}(12) \right),
\label{eq:SpectralFunc}
\end{align}
%----------
where
%----------
% Equation
%----------
\begin{align}
A(12)|_{t_1=t_2} = \delta_{z_1 z_2},
\label{eq:temp4}
\end{align}
%----------
is satisfied again from the equal-time anti-commutation relations of the Fermi operators.
The retarded and advanced GFs are therefore essential to describe the renormalization of the electronic band structures, the explicit treatment of which is one of the important advantages of the RAK basis.

In order to proceed further, we here introduce the relative time $t_{12} \equiv t_{1} - t_{2}$ and the average time $T_{12} \equiv (t_{1} + t_{2})/2$ and perform the Fourier transformation with respect to the relative time; the Wigner representation~\cite{Wigner32, Rammer07}  (see also Appendix~\ref{app:Wigner}).
By using Eq.~\eqref{eq:EqualTimeWigner}, we can rewrite Eq.~\eqref{eq:temp3} as
%----------
% Equation
%----------
\begin{align}
\ii  \int^\infty_{-\infty} \frac{\dd \nu}{2 \pi}  G^{\K}_{z_1 z_2}(T_{12}; \nu) &= \delta_{z_1 z_2} - 2 f_{z_1 z_2}(T_{12}),
\label{eq:f-GK_relation}
\end{align}
%----------
with a definition of
%----------
% Equation
%----------
\begin{align}
f_{z_1 z_2}(T_{12}) & \equiv -\ii G^{<}(12)|_{t_1=t_2} \nonumber \\ 
& = - \ii \int^{\infty}_{-\infty} \frac{\dd \nu}{2 \pi} G^{<}_{z_1 z_2}(T_{12}; \nu).
\label{eq:f-Glesser_relation}
\end{align}
%----------
This means that the $\nu$-integrated Keldysh GF [Eq.~\eqref{eq:f-GK_relation}] is directly related to $p_{\k}(t)$, $n_{1,\k}(t)$ and $n_{2,\k}(t)$ in the Wigner representation.
For example, $p_{\k}(t) = f_{12, \k \k}(t)$ when $z_1 = \{\alpha_1, \k_1 \} = \{1, \k \}$ and $z_2 = \{\alpha_2, \k_2 \} = \{2, \k \}$ because of Eq.~\eqref{eq:PhysicalQuantities1} and the first line of Eq.~\eqref{eq:f-Glesser_relation}.
In the same manner, we obtain $n_{1,\k}(t) = f_{11, \k \k}(t)$ and $n_{2,\k}(t) = f_{22, \k \k}(t)$.

In contrast, Eq.~\eqref{eq:temp4} becomes
%----------
% Equation
%----------
\begin{align}
\int^{\infty}_{-\infty} \frac{\dd \nu}{2 \pi} A_{z_1 z_2}(T_{12};\nu) = \delta_{z_1 z_2},
\label{eq:SumRule}
\end{align}
%---------
which corresponds to the sum rule of the single-particle spectral function.
The renormalized band structure as well as the physical quantities $p_{\k}(t)$, $n_{1,\k}(t)$ and $n_{2,\k}(t)$ are, thus, closely linked with the GFs in the RAK basis.

%---------------------------------------
% Self-energies
%---------------------------------------
\subsubsection{Self-energy}\label{subsubsec:Self-energies in the Wigner}
In order to derive the equations of motion for $p_{\k}(t)$, $n_{1,\k}(t)$ and $n_{2,\k}(t)$, however, we also have to transform the contour self-energy $\varSigma_{\C}(12)$ into the real-time matrix representation in the RAK basis, in the same manner as presented in Appendix~\ref{app:Real-time matrix}.
As a result of straightforward calculations (see Appendix~\ref{app:Self-energies}), the self-energies in Figure~\ref{fig14:Selfenergies} under the HF approximation (Figure~\ref{fig20:Scheme}) give the total self-energy in the Wigner representation as
%----------
% Equation
%----------
\begin{widetext}
\begin{align}
\varSigma(T_{12} ; \nu ; \k_1 \k_2) 
&= 
\left(
\begin{array}{c|c}
\varSigma^{\R}(T_{12}) & \varSigma^{\K}(\nu) \\ \hline
 & \varSigma^{\A}(T_{12})
\end{array}
\right) \nonumber \\
&= \delta_{\k_1 \k_2} 
\left(
\begin{array}{cc|cc}
\varSigma^{\BGR}_{1, \k_1}(T_{12}) - \ii \gamma & - \varDelta_{\k_1}(T_{12}) & -2 \ii \gamma F^{\B}_{1}(\nu) & \\
- \varDelta^{*}_{\k_1}(T_{12}) & \varSigma^{\BGR}_{2, \k_1}(T_{12}) - \ii \gamma & & -2 \ii \gamma F^{\B}_{2} (\nu) \\ \hline
&&\varSigma^{\BGR}_{1, \k_1}(T_{12}) + \ii \gamma & - \varDelta_{\k_1}(T_{12}) \\
&&- \varDelta^{*}_{\k_1}(T_{12}) & \varSigma^{\BGR}_{2, \k_1}(T_{12}) + \ii \gamma
\end{array}
\right), 
\label{eq:Self-energy}
\end{align}
\end{widetext}
%----------
where $\varDelta_{\k}(T_{12})$ denotes the renormalized Rabi frequency
%----------
% Equation
%----------
\begin{align}
\varDelta_{\k}(T_{12}) \equiv g^{*} a_0(T_{12}) + \sum_{\k'} U'_{\k' - \k} p_{\k'} (T_{12}),
\label{eq:Rabi}
\end{align}
%----------
describing the effect of forming the \mbox{e-h} pairs~\cite{Schmitt-Rink88, Yamaguchi13, Yamaguchi14} and $\varSigma^{\BGR}_{\alpha, \k}(T_{12})$ is the Coulomb-induced renormalization of the single-particle energy
%----------
% Equation
%----------
\begin{align}
\varSigma^{\BGR}_{\alpha, \k}(T_{12}) \equiv - \sum_{\k'} U'_{\k' - \k} n_{\alpha, \k'} (T_{12}),
\end{align}
%----------
including the band-gap renormalization (BGR) in semiconductor physics.
These are nothing but the variables defined in Eq.~\eqref{eq:HMF}.
$F^{\B}_{\alpha}(\nu)$ denotes the distribution of the pumping bath
%----------
% Equation
%----------
\begin{align}
F^{\B}_{\alpha}(\nu) \equiv 1 - 2 f^{\B}_{\alpha}(\nu),
\label{eq:FB}
\end{align}
%----------
through the Fermi distribution function
%----------
% Equation
%----------
\begin{align}
f^{\B}_{1/2}(\nu) \equiv \frac{1}{ 1 + \exp ( \beta ( \nu - \mu^{\B}_{1/2} \pm \mu/2 ) )},
\label{eq:Fermi_B}
\end{align}
%----------
where $\beta \equiv 1/T$ and $\mu^{\B}_{\alpha}$ are the inverse temperature and the chemical potential of the pumping bath, respectively.
We note that, in Eq.~\eqref{eq:Self-energy}, the retarded and advanced parts are independent of the frequency $\nu$, while the Keldysh part does not include the average time $T_{12}$.
In other words, in our treatments, the memory effect (the frequency dependence) is not taken into account in the retarded and advanced part but it remains in the Keldysh part due to the correlations with the pumping baths.~\cite{Keeling10,Yamaguchi12}
These behaviors arise solely from our truncations of the vertex functions shown in Figure~\ref{fig20:Scheme}; see also Appendix~\ref{app:Self-energies}.
These correlations as well as the renormalization of the band structures are essential for our theory to especially recover the equilibrium phases because the carriers have to be redistributed in the renormalized band according to their energies in equilibrium, the information of which, however, cannot be obtained instantaneously (with no memory time) due to the uncertainty principle.
As a consequence, the memory effect has to be taken into account, at least in the Keldysh part of the self-energy.

%---------------------------------------
% Equations of motion
%---------------------------------------
\subsubsection{Equations of motion}\label{subsubsec:EOM}
By using the real-time matrix form in the RAK basis, the Dyson equation [Eq.~\eqref{eq:Dyson_differeintial} or \eqref{eq:Dyson}] becomes
%----------
% Equation
%----------
\begin{align}
G^{-1}_{0}(12)G_{ii'}(21') = \delta_{ii'} \delta(11') + \varSigma_{ij}(12)G_{ji'}(21'),
\label{eq:left-Dyson}
\end{align}
%----------
in the limit of $\hat{H}_{\A}(\tau) \rightarrow 0$, where $\delta(11') \equiv \delta_{z_1 z'_1} \delta(t_1 t'_1)$.
Equivalently, we can find 
%----------
% Equation
%----------
\begin{align}
G_{ii'}(12)G^{-1}_{0}(21') = \delta_{ii'} \delta(11') + G_{ij}(12)\varSigma_{ji'}(21').
\label{eq:right-Dyson}
\end{align}
%----------
In a similar manner to study the Boltzmann equation by the NEGF technique,~\cite{Rammer07} we perform the subtraction of the left and right Dyson equations, Eqs.~\eqref{eq:left-Dyson} and \eqref{eq:right-Dyson}, and then, after taking the Keldysh part ($i = 1$, $i' = 2$), we obtain 
%----------
% Equation
%----------
\begin{align}
[G^{-1}_{0} \stackrel{\otimes}{,} G^{\K}]_{-}
= & \varSigma^{\R} \otimes G^{\K} + \varSigma^{\K} \otimes G^{\A} \nonumber \\
& - G^{\K} \otimes \varSigma^{\A} - G^{\R} \otimes \varSigma^{\K},
\label{eq:left-right Dyson}
\end{align}
%----------
where $\otimes$ denotes the summation over the possible internal degrees of freedom and $[X \stackrel{\otimes}{,} Y]_{-} \equiv X \otimes Y - Y \otimes X$.
It follows from Eq.~\eqref{eq:GF-Wigner} that
\begin{widetext}
%----------
% Equation
%----------
\begin{multline}
( \ii \partial_{T_{12}} - \xi_{z_1} + \xi_{z_2} ) G^{\K}_{z_1 z_2} (T_{12}; \nu) 
= \varSigma^{\R}_{z_1 z_3} (T_{12}) \star G^{\K}_{z_3 z_2} (T_{12}; \nu) 
+ \varSigma^{\K}_{z_1 z_3} (\nu) \star G^{\A}_{z_3 z_2} (T_{12}; \nu) \\
- G^{\K}_{z_1 z_3} (T_{12}; \nu) \star \varSigma^{\A}_{z_3 z_2} (T_{12})
- G^{\R}_{z_1 z_3} (T_{12}; \nu) \star \varSigma^{\K}_{z_3 z_2} (\nu) ,
\label{eq:left-right Dyson in Wigner}
\end{multline}
%----------
where $\varSigma^{\R/\A/\K}_{z_1 z_2}$ corresponds to the element of the matrix in Eq.~\eqref{eq:Self-energy} and $\star$ denotes the Moyal product.
By integrating both sides of Eq.~\eqref{eq:left-right Dyson in Wigner} in terms of $\nu$ after applying the gradient approximation for the Moyal products (see Appendix~\ref{app:Wigner}), we get 
%----------
% Equation
%----------
\begin{multline}
\delta_{\k_1 \k_2} (\ii \partial_{t} - \xi_{z_1} + \xi_{z_2}) \{\delta_{\alpha_1 \alpha_2} - 2 f_{\alpha_1 \alpha_2, \k_1}(t) \} \\
= \delta_{\k_1 \k_2} \sum_{\alpha_3} \Biggl\{ 
\varSigma^{\R}_{\alpha_1 \alpha_3, \k_1} (t) \{ \delta_{\alpha_3 \alpha_2} - 2f_{\alpha_3 \alpha_2, \k_1}(t) \} 
+ \ii \int^{\infty}_{-\infty} \frac{\dd \nu}{2 \pi} \varSigma^{\K}_{\alpha_1 \alpha_3, \k_1}(\nu) G^{\A}_{\alpha_3 \alpha_2, \k_1}(t;\nu) \\
- \{ \delta_{\alpha_1 \alpha_3} - 2 f_{\alpha_1 \alpha_3, \k_1}(t) \} \varSigma^{\A}_{\alpha_3 \alpha_2, \k_1}(t) 
- \ii \int^{\infty}_{-\infty} \frac{\dd \nu}{2 \pi} G^{\R}_{\alpha_1 \alpha_3, \k_1}(t;\nu) \varSigma^{\K}_{\alpha_3 \alpha_2, \k_1}(\nu)
\Biggr\} ,
\label{eq:temp5}
\end{multline}
%----------
\end{widetext}
where $X^{\mathcal{Z}}_{z_1 z_2} = X^{\mathcal{Z}}_{\alpha_1 \alpha_2 , \k_1 \k_2} = \delta_{\k_1 \k_2} X_{\alpha_1 \alpha_2, \k_1} $ and $f_{z_1 z_2} = f_{\alpha_1 \alpha_2, \k_1 \k_2} = \delta_{\k_1 \k_2}  f_{\alpha_1 \alpha_2, \k_1}$ have been used with Eq.~\eqref{eq:f-GK_relation} and the average time $T_{12}$ has been replaced simply by $t$.
By taking $(\alpha_1, \alpha_2) = (1,2)$, Eq.~\eqref{eq:temp5} results in 
%----------
% Equation
%----------
\begin{multline}
\partial_{t} p_{\k}(t) = - \ii \{ \tilde{\xi}_{1,\k}(t) - \tilde{\xi}_{2,\k}(t) \} p_{\k}(t)  \\
 - \ii \varDelta_{\k}(t) N_{\k}(t) - 2 \gamma \{ p_{\k} (t) - p^0_{\k}(t) \},
\label{eq:EOM_pk1}
\end{multline}
%----------
where $\tilde{\xi}_{\alpha, \k}(t) \equiv \xi_{\alpha, \k} + \varSigma^{\BGR}_{\alpha, \k} (t)$ is the Coulomb-renormalized single-particle energy, $N_{\k}(t) \equiv n_{1,\k} - n_{2, \k}$ denotes the population inversion, and $p^{0}_{\k}(t)$ is defined as
%----------
% Equation
%----------
\begin{align}
p^{0}_{\k}(t) \equiv \ii \int^{\infty}_{-\infty} \frac{\dd \nu}{2 \pi } [f^{\B}_{2}(\nu) G^{\R}_{12,\k}(t;\nu) - f^{\B}_{1}(\nu) G^{\A}_{12,\k}(t;\nu)].
\label{eq:p0k1}
\end{align}
%----------
In the derivation, we have used $p_{\k}(t) = f_{12,\k}(t)$ from Eqs.~\eqref{eq:PhysicalQuantities1} and \eqref{eq:f-Glesser_relation}.
In the same manner, since  $n_{\alpha, \k}(t) = f_{\alpha \alpha,\k}(t)$, Eq.~\eqref{eq:temp5} for $(\alpha_1, \alpha_2) = (1,1)$ and $(2, 2)$ leads to 
%----------
% Equation
%----------
\begin{align}
\partial_{t} n_{\alpha, \k}(t) = -2 \Im [\varDelta_{\k}(t) p^{*}_{\k}(t)] - 2\gamma \{ n_{\alpha,\k}(t) - n^{0}_{\alpha,\k}(t) \},
\label{eq:EOM_nk1}
\end{align}
for $\alpha \in \{1, 2 \}$ with 
\begin{align}
n^{0}_{\alpha,\k}(t) \equiv \int^{\infty}_{-\infty} \frac{\dd \nu}{2 \pi} f_{\alpha}^{\B}(\nu) A_{\alpha \alpha, \k}(t;\nu).
\label{eq:n0k1}
\end{align}
%----------
Note that Eqs.~\eqref{eq:EOM_pk1} and \eqref{eq:EOM_nk1} have the well-known forms of the \mbox{MSBEs} under the RTA \cite{Chow02, Henneberger92, Yamaguchi12} if $n^{0}_{\alpha, \k}$ is replaced by the Fermi distribution with $p^{0}_{\k} = 0$.
In general, the \mbox{MSBEs} under the RTA can describe the physics of semiconductor lasers but cannot describe those of the BEC and BCS states because the equations of motion cannot recover the (quasi-)equilibrium physics in the steady state.~\cite{Yamaguchi13, Yamaguchi14}
However, the key point here is that $p^{0}_{\k}$ and $n^{0}_{\alpha, \k}$ are defined by Eqs.~\eqref{eq:p0k1} and \eqref{eq:n0k1}, respectively, and the effects of the time-dependent renormalization of the electronic band structures are taken into account through $G^{\R/\A}$ and $A$.

In a similar manner to Eq.~\eqref{eq:left-right Dyson}, the subtraction of the left and right Dyson equations for the retarded part ($i = 1, i' =1$) gives $[G^{-1}_{0} \stackrel{\otimes}{,} G^{\R}] = \varSigma^{\R} \otimes G^{\R} - G^{\R} \otimes \varSigma^{\R}$ and the Wigner representation becomes
%----------
% Equation
%----------
\begin{multline}
(\ii \partial_{t} - \xi_{\alpha_1, \k} + \xi_{\alpha_2, \k})G^{\R}_{\alpha_1 \alpha_2,\k} (t;\nu) \\
= \sum_{\alpha_3} \left\{ \varSigma^{\R}_{\alpha_1 \alpha_3, \k}(t) \star G^{\R}_{\alpha_3 \alpha_2, \k}(t;\nu) \right. \\
\left. - G^{\R}_{\alpha_1 \alpha_3, \k}(t;\nu) \star \varSigma^{\R}_{\alpha_3 \alpha_2, \k}(t) \right\}.
\end{multline}
%----------
Within the gradient approximation, we therefore obtain
%----------
% Equation
%----------
\begin{align}
& G^{-1}_{0, \k} G^{\R}_{\k}(t;\nu) - G^{\R}_{\k}(t;\nu) [ G^{-1}_{0, \k} ]^{\dagger} \nonumber \\
& = \varSigma^{\R}_{\k}(t) G^{\R}_{\k}(t;\nu) - G^{\R}_{\k}(t;\nu)\varSigma^{\R}_{\k}(t) \nonumber \\
& \quad - \frac{\ii}{2} \left\{ \partial_{t}\varSigma^{\R}_{\k}(t) \partial_{\nu}G^{\R}_{\k}(t;\nu) + \partial_{\nu}G^{\R}_{\k}(t;\nu) \partial_{t}\varSigma^{\R}_{\k}(t)  \right\},
\label{eq:EOM_GR2}
\end{align}
%----------
with
%----------
% Equation
%----------
\begin{align}
G^{-1}_{0, \k} \equiv
\begin{pmatrix}
\frac{\ii}{2} \partial_{t} + \nu -\xi_{1,\k} & 0 \\
0 & \frac{\ii}{2} \partial_{t} + \nu -\xi_{2,\k} 
\end{pmatrix},
\label{eq:EOM_GR_Inverse1}
\end{align}
%----------
in the $2 \times 2$ matrix representation.
$G^{\A}$ and $A$ can then be obtained from the relation $G^{\A }_{z_1 z_2}(t;\nu) = G^{\R*}_{z_2 z_1}(t;\nu)$ and Eq.~\eqref{eq:SpectralFunc}, respectively.
As a result, Eqs.~\eqref{eq:Heisenberg-Langevin}, \eqref{eq:EOM_pk1} and \eqref{eq:EOM_nk1} are closed simultaneously with Eqs.~\eqref{eq:p0k1}, \eqref{eq:n0k1}, and \eqref{eq:EOM_GR2}.
The time-dependent renormalization of the electronic band structures is thus taken into account.
These are the main results of Section~\ref{sec:real time} summarized in Figure~\ref{fig20:Scheme}, which generalize the standard \mbox{MSBEs} under the RTA.
By transforming them into the \mbox{e-h} picture ({Table~\ref{tableI}), our key results [Eqs.~\eqref{alleq:MSBEs} with Eqs.~\eqref{alleq:p0k2-n0k2}--\eqref{eq:A}] can successfully obtained.

%======================================================
%   Conclusions
%======================================================
\section{Conclusions}\label{sec:Conclusions}
In this paper, we have presented a unified theory to study the relationship of the cooperative phenomena spontaneously developed in semiconductor \mbox{e-h-p} systems.
Starting from the microscopic Hamiltonian, as a key result of our theory, we presented a time-dependent formalism for the photon amplitude $a_{0}$, the polarization function $p_{\k}$ and the distributions of electrons in the conduction band $n_{\e,\k}$ and holes in the valence band $n_{\h,\k}$, based on the generating functional approach [Eqs.~\eqref{alleq:MSBEs} with Eqs.~\eqref{alleq:p0k2-n0k2}--\eqref{eq:A}].
The simultaneous equations of motion keep a similar form to the \mbox{MSBEs} under the RTA [Eq.~\eqref{alleq:MSBEs}] but the key differences are the following two points; (i) the \mbox{e-h} pairing effect is taken into account in the band renormalization and (ii) the thermalization by the pumping baths is treated in the non-Markovian way.
The first one is evidently important because the \mbox{e-h} BCS gap, for example, must be included in the theory.
The second one, on the other hand, plays a crucial role to describe the redistributions of carriers in the renormalized bands.
In our view, the non-Markov treatment is required because the particle energies cannot be measured instantaneously (or in the Markovian way) due to the uncertainty principle.
These are one of our key results (Subsec.~\ref{subsec:Key results}) and enable us to discuss the cooperative phenomena in a unified view.

As an important application, we have studied the BEC-BCS-LASER crossover in the exciton-polariton systems.~\cite{Yamaguchi12,Yamaguchi13}
The steady-state phase diagrams then revealed that the system has rich and distinct ordered phases depending on the cavity photon loss, the detuning, and the pumping strength.
At the same time, we also stressed that, whenever the phase symmetry is broken, the pairing gap is opened at least in principle by a similar mechanism to the Mollow triplet in resonance fluorescence.
This claim is important because it means that there exist bound \mbox{e-h} pairs even in the standard lasing regime, in contrast to earlier expectations.~\cite{Yamaguchi13}
Furthermore, the bound \mbox{e-h} pairs are expected also in the Fermi-edge SF.
In this context, our theory revealed that the \mbox{e-h} BCS phase can indeed be developed after the Fermi-edge SF.
These results strongly encourage the experimental discovery of the \mbox{e-h} BCS phase in the context the Fermi-edge SF because the presence of the \mbox{e-h} BCS phase is still very much an open question.

Aside from this, under the steady-state condition, we have also presented the formalism to analyze the emission spectrum and the gain-absorption spectrum, again based on the generating functional approach; the fully and partially dressed two-particle GFs have essential roles.
For the emission spectra, we then discussed the origin of the spectral structures by introducing the energy- and momentum-resolved distributions.
The physical picture is again similar to the Mollow triplet and the side peaks have information about the carrier distributions in the renormalized bands.
In the gain-absorption spectra, on the other hand, we pointed out that the results are affected not only by the distributions in the renormalized bands but also by the phase difference between the developed order in the system and the coherent probe field.
This result is physically not surprising because there are two relevant phases in the gain-absorption spectra; one is the phase of the spontaneous coherence in the system and the other is the phase of the external probe field.
However, there has been no such claim in the past, to our knowledge.
In addition, we have also noted that the gain-absorption spectrum is one of the important ways to verify the lasing gap.

We finally described a general framework based on the generating functional approach that systematically gives the coupled equations of motion for the NEGFs.
As a result, the partially dressed NEGFs are naturally introduced to avoid the double counting problems.
This is one of the most important advantages to take such an approach.

We have thus developed a prototypical theory to study the relationship of the cooperative phenomena in a unified view.
However, there remain non-trivial issues on this formalism.
For example, the effect of the spontaneous emission~\cite{Kira99,Scully97,Gross82} is still unclear even though this directly determines the statistical behavior of the SF.
Pure dephasing has a possibility to significantly modify the intensity ratio of the side peaks, as pointed out in the cavity quantum electrodynamics using a single quantum dot.~\cite{Yamaguchi12-2, *Yamaguchi08, Valle11}
The \mbox{e-h} center-of-mass fluctuation~\cite{Ohashi02, *Ohashi03} and the mass imbalance~\cite{Hanai13} are also important, as discussed in the ultra-cold atomic systems.

In this context, it will be fruitful to discuss these issues in the future.
It is also interesting to study the equilibrium-to-nonequilibrium change of the vortex formation~\cite{Coullet89,Lagoudakis08,Carusotto13} and the Andreev reflection~\cite{Andreev64,Gennes66} without the homogeneous assumptions [Eqs.~\eqref{eq:Assumption} and \eqref{eq:Conservation}].~\cite{Hess96}
In such a case, the recent theoretical advancements on the ultracold atomic systems~\cite{Francesco14, *Francesco14-2} would also be helpful where, based on the NEGF approach, the quantum kinetics is studied. 
We believe that our approach stimulates a different class of studies and paves the way to providing a bridge between the equilibrium and the nonequilibrium physics.

%======================================================
%   Acknowledgments
%======================================================
\begin{acknowledgments}
We thank H. Akiyama for pointing out the importance of the phase difference on the gain-absorption spectrum. 
We are also grateful to T. Kato, J. Kono, T. Horikiri, Y. Shikano, Y. Matsuo, M. Bamba, T. Yuge, and K. Asano for fruitful discussions.
This work was supported by KAKENHI (Grant No. 26287087) and funded by ImPACT Program of Council for Science, Technology and Innovation (Cabinet Office, Government of Japan).
K.K. is supported by the Project for Developing Innovation Systems of MEXT, Japan.
\end{acknowledgments}

\appendix

%======================================================
% Effective distribution function
%======================================================
\begin{widetext}
\section{Effective distribution function}\label{app:Effective distribution}
To keep the paper as self-contained as possible, we here derive Eq.~\eqref{alleq:BCSform} from Eq.~\eqref{alleq:MSBEs} and \eqref{alleq:p0k2-n0k2} with Eq.~\eqref{eq:GR_SS} under the steady-state condition even though the equations are equivalent to our previous work.~\cite{Yamaguchi12, Yamaguchi13}
For this purpose, we write Eq.~\eqref{eq:GR_SS} as 
%----------
% Equation
%----------
\begin{align}
&G^{\R}_{\k}(\nu) = \frac{1}{|\mathcal{D}_{\k}(\nu)|^2} 
\begin{pmatrix}
\mathcal{D}^{*}_{\k}(\nu) (\nu+\tilde{\xi}_{\h,\k} + \ii \gamma ) & -\mathcal{D}^{*}_{\k}(\nu) \varDelta_{\k}  \\
-\mathcal{D}^{*}_{\k}(\nu) \varDelta^{*}_{\k}  & \mathcal{D}^{*}_{\k}(\nu) (\nu - \tilde{\xi}_{\e, \k} + \ii \gamma)
\end{pmatrix},
\label{eq:GR-temp}
\end{align}
%----------
where 
%----------
% Equation
%----------
\begin{align}
\mathcal{D}_{\k}(\nu) \equiv (\nu - \tilde{\xi}^{-}_{\eh,\k} + E_{\k} + \ii \gamma) (\nu - \tilde{\xi}^{-}_{\eh,\k} - E_{\k} + \ii \gamma ).
\end{align}
%----------
From Eq.~\eqref{eq:A}, we then obtain 
%----------
% Equation
%----------
\begin{align}
A_{\k}(\nu) &= \frac{-2}{|\mathcal{D}_{\k}(\nu)|^2} 
\begin{pmatrix}
\Im [ \mathcal{D}^{*}_{\k}(\nu) (\nu+\tilde{\xi}_{\h,\k} + \ii \gamma ) ] & \Im [\mathcal{D}_{\k}(\nu)]  \varDelta_{\k}  \\
\Im [\mathcal{D}_{\k}(\nu)] \varDelta^{*}_{\k}  & \Im [\mathcal{D}^{*}_{\k}(\nu) (\nu - \tilde{\xi}_{\e, \k} + \ii \gamma)]  
\end{pmatrix}, 
\nonumber \\
&= \frac{-2}{|\mathcal{D}_{\k}(\nu)|^2} 
\begin{pmatrix}
-\gamma \{ (\nu + \tilde{\xi}_{\h,\k})^2 + \gamma^2 + |\varDelta_{\k}|^2  \} & 2 \gamma \varDelta_{\k} (\nu - \tilde{\xi}^{-}_{\eh,\k} )  \\
2 \gamma \varDelta^{*}_{\k} (\nu - \tilde{\xi}^{-}_{\eh,\k} ) & - \gamma \{ (\nu - \tilde{\xi}_{\e,\k} )^2 + \gamma^2 + |\varDelta_{\k}|^2 \}
\end{pmatrix}.
\label{eq:A-temp}
\end{align}
%----------
We note that the diagonal element $A_{11/22, \k}(\nu)$ is equivalent to Eq.~\eqref{eq:A_SS}.
In the followings, we first derive the $\nu$-integral forms of the population inversion $N_{\k}$ and the polarization $p_{\k}$ because Eq.~\eqref{alleq:BCSform} is described by the integration in terms of $\nu$.
From Eqs.~\eqref{eq:MSBEs--nehk} and \eqref{eq:n0k2}, the population inversion $N_{\k} = n_{\e,\k} + n_{\h,\k} - 1$ is described as
%----------
% Equation
%----------
\begin{align}
N_{\k} = - \frac{2}{\gamma} \Im[\varDelta_{\k} p^{*}_{\k}] 
+ \int \frac{\dd \nu}{2 \pi} [f^{\B}_{\e}(\nu) A_{11,\k}(\nu) - \{1-f^{\B}_{\h} (-\nu) \} A_{22,\k}(\nu)],
\label{eq:Nk-temp}
\end{align}
%----------
where $\int \frac{\dd \nu}{2 \pi} A_{\alpha_1 \alpha_2, \k}(\nu) = \delta_{\alpha_1 \alpha_2}$ has been used as a result of Eq.~\eqref{eq:SumRule}.
In a similar manner, Eqs.~\eqref{eq:MSBEs--pk} and \eqref{eq:p0k2} yield 
%----------
% Equation
%----------
\begin{align}
p_{\k} = - \frac{\varDelta_{\k}}{2(\tilde{\xi}^{+}_{\eh,\k} - \ii \gamma)} N_{\k}
+ \frac{\gamma}{\tilde{\xi}^{+}_{\eh,\k} - \ii \gamma} 
 \int \frac{\dd \nu}{2 \pi} 
\left[ \{ 1- f^{\B}_{\h}(-\nu) \} G^{\R}_{12,\k}(\nu) - f^{\B}_{\e}(\nu) [G^{\R}_{21,\k}(\nu)]^{*}  \right],
\label{eq:pk-temp}
\end{align}
%----------
By inserting Eq.~\eqref{eq:pk-temp} into Eq.~\eqref{eq:Nk-temp} with Eqs.~\eqref{eq:GR-temp} and \eqref{eq:A-temp}, we find
%----------
% Equation
%----------
\begin{align}
N_{\k} = \int \frac{\dd \nu}{2 \pi}  \frac{2 \gamma}{\mathcal{|D}_{\k}(\nu)|^2} 
\left[ f^{\B}_{\e}(\nu) \{ (\nu+\tilde{\xi}_{\h,\k})^2 + \gamma^2 - |\varDelta_{\k}|^2 \}  + \{ f^{\B}_{\h}(-\nu)-1 \} \{ (\nu - \tilde{\xi}_{\e,\k} )^2 + \gamma^2 -|\varDelta_{\k}|^2  \}\right].
\label{eq:Nk-temp2}
\end{align}
%----------
As a result, by substituting Eq.~\eqref{eq:Nk-temp2} into Eq.~\eqref{eq:pk-temp}, we obtain
%----------
% Equation
%----------
\begin{align}
p_{\k} = \varDelta_{\k} \int \frac{\dd \nu}{2 \pi}  \frac{2 \gamma}{\mathcal{|D}_{\k}(\nu)|^2} 
\left[ \{ f^{\B}_{\h}(-\nu)-1 \} ( \nu - \tilde{\xi}_{\e,\k} - \ii \gamma ) - f^{\B}_{\e}(\nu)(\nu + \tilde{\xi}_{\h,\k} + \ii \gamma) \right].
\label{eq:pk-temp2}
\end{align}
%----------
In the derivation of Eqs.~\eqref{eq:Nk-temp2} and \eqref{eq:pk-temp2}, the following equations have been used:
%----------
% Equation
%----------
\begin{align*}
&\pm |\varDelta_{\k}|^2 \Im[\mathcal{D}_{\k}(\nu) (\tilde{\xi}^{+}_{\eh,\k} \mp \ii \gamma)] + \{ (\tilde{\xi}^{+}_{\eh,\k})^2 + \gamma^2 \} \Im[ \mathcal{D}^{*}_{\k}(\nu) ( \nu \mp \tilde{\xi}_{\e/\h, \k} + \ii \gamma ) ] 
= - \gamma (E^{2}_{\k} + \gamma^2) \{ (\nu \mp \tilde{\xi}_{\e/\h,\k} )^2 + \gamma^2 - |\varDelta_{\k}|^2 \}, \\
&(\nu + \tilde{\xi}_{\h,\k})^2 + \gamma^2 - |\varDelta_{\k}|^2 - \mathcal{D}_{\k}(\nu) = 2 (\tilde{\xi}^{+}_{\eh,\k} - \ii \gamma) (\nu + \tilde{\xi}_{\h,\k} + \ii \gamma), \\
&(\nu - \tilde{\xi}_{\e,\k})^2 + \gamma^2 -|\varDelta_{\k}|^2 - \mathcal{D}^{*}_{\k}(\nu) = - 2(\tilde{\xi}_{\eh,\k} - \ii \gamma) (\nu - \tilde{\xi}_{\e,\k} - \ii \gamma).
\end{align*}
%----------
\end{widetext}
From Eq.~\eqref{eq:A-temp}, we can write Eq.~\eqref{eq:pk-temp2} as
%----------
% Equation
%----------
\begin{align}
p_{\k} = \int \frac{\dd \nu}{2 \pi} f^{\SS}_{\eh,\k}(\nu)A_{12,\k}(\nu),
\label{eq:pk}
\end{align}
%----------
with the definition of $f^{\SS}_{\eh,\k}(\nu)$ [Eq.~\eqref{alleq:f_SS}].
From Eq.~\eqref{eq:MSBEs--a0} and $\varDelta_{\k} = g^{*}a_{0} + \sum_{\k'} U'_{\k' - \k} p_{\k'}$, we obtain
%----------
% Equation
%----------
\begin{align}
\varDelta_{\k} = \sum_{\k} U^{\eff, \kappa}_{\k', \k} p_{\k'},
\end{align}
%----------
and therefore, Eq.~\eqref{eq:BCSform1} can be found by inserting Eq.~\eqref{eq:pk}.
By multiplying $\varDelta_{\k}$ by the complex conjugate of Eq.~\eqref{eq:pk-temp2}, we further find
%----------
% Equation
%----------
\begin{align*}
&\frac{1}{\gamma} \Im[\varDelta_{\k} p^{*}_{\k}] = \\
& \qquad  |\varDelta_{\k}|^2 \int \frac{\dd \nu}{2 \pi} \frac{2\gamma}{|\mathcal{D}_{\k}(\nu)|^2} \{ f^{\B}_{\e}(\nu) + f^{\B}_{\h}(-\nu) -1 \}.
\end{align*}
%----------
As a result, by substituting this equation into Eq.~\eqref{eq:MSBEs--nehk} with Eq.~\eqref{eq:A-temp}, Eq.~\eqref{eq:BCSform2} can be derived with the definitions of Eqs.~\eqref{alleq:f_SS} and \eqref{eq:eta}.

We have thus derived Eq.~\eqref{alleq:BCSform}.
We note that the procedure shown here is basically the same as the Appendix~II in Ref.~\onlinecite{Yamaguchi14}.

%======================================================
% the photon fraction
%======================================================
\section{Photonic fraction}\label{app:Photonic fraction}
In order to measure the photonic effect, we have defined the photonic fraction $F_{\ph}$ as 
%----------
% Equation
%----------
\begin{align*}
F_{\ph} \equiv \frac{n_{\ph}}{n_{\ph}+n^{\eff}_{\text{car}}},
\end{align*}
%----------
where $n_{\ph} \equiv |a_0|^2$ is the coherent photon number and $n^{\eff}_{\text{car}} \equiv n_{\text{car}} - n^{\text{inc}}_{\text{car}}$ is the effective carrier number with
%----------
% Equation
%----------
\begin{align*}
&n_{\text{car}} \equiv \frac{1}{2 }\sum_{\k} (n_{\e, \k}+n_{\h, \k}), \\
&n^{\text{inc}}_{\text{car}} \equiv \frac{1}{2 }\sum_{\k}(n^{\text{inc}}_{\e,\k} + n^{\text{inc}}_{\h,\k}).
\end{align*}
%----------
Here, the incoherent carrier number is introduced because the carriers can be excited even though the Fermi level does not reach the lowest energy level of the system due to the broadening by $\gamma$.
In other words, the carriers are excited incoherently even before the occurrence of the condensation.
Such carriers, therefore, should be eliminated for the evaluation of the photonic fraction, especially in the low density regime.

%----------
% Figure
%----------
\begin{figure}[!tb] 
\centering
\includegraphics[width=.45\textwidth, clip]{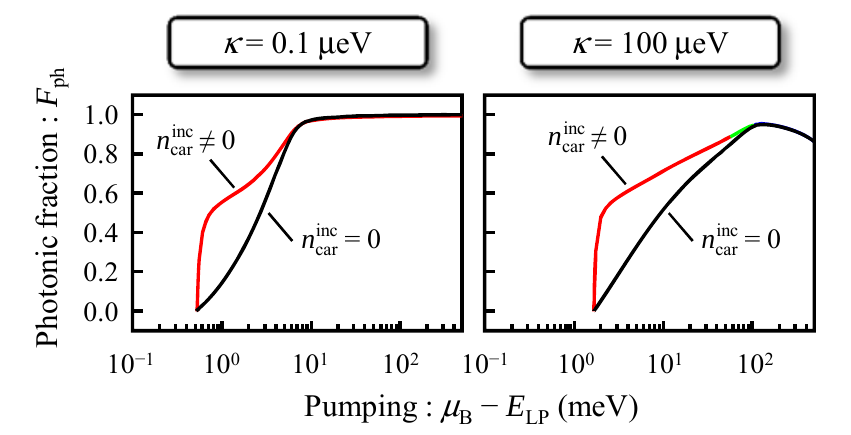} 
\caption{(Color online) Comparisons between the results with $n^{\text{inc}}_{\text{car}} \ne 0$ and $n^{\text{inc}}_{\text{car}} = 0$.
The parameters are the same as Figures~\ref{fig05:Results}(a) and \ref{fig05:Results}(b) with the detuning of 0~meV.
}
\label{fig21:Comparison}
\end{figure}
%----------

We have therefore determined $n^{\text{inc}}_{\text{car}}$ by the value right before the condensation, namely the solution for $n_{\e/\h,\k} = n^{0}_{\e/\h,\k} $ [Eq.~\eqref{eq:MSBEs--nehk} with $\partial_t = 0$ in the limit of $a_{0} \to 0$ and $p_{\k} \to 0$]; 
%----------
% Equation
%----------
\begin{align}
n^{\text{inc}}_{\e/\h, \k} = \int \frac{\dd \nu}{\pi} \frac{\gamma}{ (\nu - \tilde{\epsilon}_{\e/\h,\k})^2 + \gamma^2 } \frac{1}{1+ \exp \{ \beta(\nu - \mu^{\B}_{\e/\h}) \} },
\label{eq:n_inc}
\end{align}
%----------
with using the critical value of $\mu^{\B}_{\e/\h}$.
Notice that this value is equivalent to the carrier number excited before the Fermi level reaches the lowest energy level of the system.
Figure~\ref{fig21:Comparison} shows comparisons between the results with $n^{\text{inc}}_{\text{car}} \ne 0$ and $n^{\text{inc}}_{\text{car}} = 0$.
If $n^{\text{inc}}_{\text{car}}$ is not taken into account, the photonic fraction does not go to $\simeq 0.5$ even in the low density regime for the polariton BEC.

%======================================================
% Contour evolution operator
%======================================================
\section{Contour evolution operator}\label{app:Contour evolution operator}
The contour evolution operator defined in Eq.~\eqref{eq:evolution} has the properties of
%----------
% Equation
%----------
\begin{align}
&\hat{\U}(\tau \tau) =1, \nonumber \\
&\hat{\U}(\tau_3 \tau_2)\hat{\U}(\tau_2 \tau_1) = \hat{\U}(\tau_3 \tau_1), \nonumber \\
&\hat{\U}^{-1}(\tau_2 \tau_1) = \hat{\U}^{\dagger}(\tau_2 \tau_1) = \hat{\U}(\tau_1 \tau_2),
\label{eq:A1}
\end{align}
%----------
and, if $\tau_2$ is later than $\tau_1$ on the closed-time contour, the temporal differentiation of Eq.~\eqref{eq:evolution} yields
%----------
% Equation
%----------
\begin{align}
\ii \partial_{\tau_2} \hat{\U}(\tau_2 \tau_1) = \hat{H}(\tau_2) \hat{\U}(\tau_2 \tau_1), \nonumber \\
- \ii \partial_{\tau_2} \hat{\U}(\tau_1 \tau_2) = \hat{\U}(\tau_1 \tau_2) \hat{H}(\tau_2). 
\label{eq:A2}
\end{align}
%----------
These features are well-known in a similar manner to the real-time evolution operator.~\cite{Stefanucci13}
We also note that $\hat{S}_{\C}(\tau_2 \tau_1)$ defined in Eq.~\eqref{eq:Sevolution} has the same properties when we replace $\hat{\U} \rightarrow \hat{S}_{\C}$ and $\hat{H}(\tau) \rightarrow \hat{H}^{\text{I}}_{\A}(\tau)$ in Eqs.~\eqref{eq:A1} and \eqref{eq:A2}.

%=======================================
% Real-time matrix representations
%=======================================
\section{Real-time matrix representations}\label{app:Real-time matrix}
To study the real-time dynamics of the physical quantities, it is convenient to write the two-time NEGFs $X_{\C}(12)$ as $\bar{X}_{kl}(12)$ when $\tau_{1}$ and $\tau_{2}$ are on the contour of $\C_{k}$ and $\C_{l}$, respectively.~\cite{Keldysh64-2, Rammer07}
The times in $\bar{X}_{kl}(12)$ are now the standard real time, $j =\{\alpha_{j}, \k_{j}, t_{j} \}$, and 
each component of $\bar{X}_{kl}(12)$ corresponds to 
%----------
% Equation
%----------
\begin{align}
\begin{pmatrix}
\bar{X}_{11}(12) & \bar{X}_{12}(12)\\
\bar{X}_{21}(12) & \bar{X}_{22}(12)
\end{pmatrix} 
\equiv
\begin{pmatrix}
X^{\text{c}}(12) & X^{<}(12)\\
X^{>}(12) & X^{\bar{\text{c}}}(12)
\end{pmatrix}.
\label{eq:Causal}
\end{align}
%----------
In the case of $G_{\C}(12)$, for example, the definitions are summarized in Table~\ref{tableVII}.
The GFs in the so-called RAK basis is then obtained by the transformation~\cite{Rammer07} 
%----------
% Equation
%----------
\begin{align}
X_{kl}(12) = L_{kk'} \sigma^{(3)}_{k'm'} \bar{X}_{m'l'}(12) L^{-1}_{l'l},
\label{eq:RAK2}
\end{align}
%----------
where the summation over the repeated contour indices is assumed with the definitions of
%----------
% Equation
%----------
\begin{align}
L \equiv \frac{1}{\sqrt{2}}
\begin{pmatrix}
1 & -1\\
1 & 1
\end{pmatrix}, 
\quad 
L^{-1} = \frac{1}{\sqrt{2}}
\begin{pmatrix}
1 & 1\\
-1 & 1
\end{pmatrix},
\end{align}
%----------
and 
%----------
% Equation
%----------
\begin{align}
\begin{pmatrix}
X_{11}(12) & X_{12}(12)\\
X_{21}(12) & X_{22}(12)
\end{pmatrix} 
\equiv
\begin{pmatrix}
X^{\R}(12)  & X^{\K}(12)\\
0 & X^{\A}(12)
\end{pmatrix}.
\label{eq:RAK1}
\end{align}
%----------
Here, $X^{\R}$, $X^{\A}$ and $X^{\K}$ are called retarded, advanced and Keldysh GFs, respectively.
The two-time NEGFs can thus be transformed into the $2\times2$ matrix due to the contour indices.

%----------
% TABLE VII
%----------
\begin{table}[!tb]
\caption{\label{tableVII} The single-particle GFs in the real-time matrix representation.
$\T$ and $\bar{\T}$ are the chronological and anti-chronological time ordering operators, respectively.}
\begin{ruledtabular}
\begin{tabular}{lll}
causal GF & $G^{\text{c}}(12)$ & $ -\ii \Ex{\T [ \oc_{\alpha_1,\k_1}(t_1)\ocd_{\alpha_2,\k_2}(t_2) ] }$ \\
anti-causal GF & $G^{\bar{\text{c}}}(12)$ & $ -\ii \Ex{\bar{\T} [ \oc_{\alpha_1,\k_1}(t_1)\ocd_{\alpha_2,\k_2}(t_2) ] } $\\
lesser GF & $G^{<}(12)$ & $ +\ii \Ex{\ocd_{\alpha_2,\k_2}(t_2)\oc_{\alpha_1,\k_1}(t_1)} $\\
greater GF & $G^{>}(12)$ & $ -\ii \Ex{\oc_{\alpha_1,\k_1}(t_1)\ocd_{\alpha_2,\k_2}(t_2)}  $\\
\end{tabular}
\end{ruledtabular}
%----------
% TABLE VIII
%----------
\caption{\label{tableVIII} The inverse of the bare GFs in the RAK basis.}
\begin{ruledtabular}
\begin{tabular}{ll}
$G^{-1}_{0}(12)$ & $ \left( \ii \partial_{t_1} - \xi_{z_{1}} \right) \delta_{z_1z_2} \delta(t_{1} t_{2}) $ \\
$D^{-1}_{0}(12)$ & $ \left( \ii \partial_{t_1} \sigma^{(3)}_{z_1z_2} -\xi_{\ph, \k_{1}} \delta_{z_1z_2} \right) \delta(t_{1} t_{2})  $ \\
\end{tabular}
\end{ruledtabular}
\end{table}
%----------

In the right-hand side of Eqs.~\eqref{eq:Causal} and \eqref{eq:RAK1}, however, each component of $X^{\mathcal{Z}}(12)$ can further be regarded as a $2\times2$ matrix due to the indices of $\alpha_1$ and $\alpha_2$, called the Nambu space.
We describe this element of the matrix as 
%----------
% Equation
%----------
\begin{align}
X^{\mathcal{Z}}(12) = \delta_{\k_1 \k_2}X^{\mathcal{Z}}_{\alpha_1 \alpha_2, \k_1}(t_1t_2),
\label{eq:MatrixElement}
\end{align}
%----------
where we have used Eq.~\eqref{eq:Conservation}.
This means that $X_{\C}(12)$ is practically equivalent to a $4\times4$ matrix in the real-time representation, sometimes called the Nambu-Keldysh matrix.

In addition to the real-time representation of the two-time NEGFs, we describe the single-time NEGF $X_{\C}(1)$ as  $\bar{X}_{k}(1)$ when $\tau_1$ is on the contour $\C_{k}$.
We then define $X_{k}(1)$ in the RAK basis as 
%----------
% Equation
%----------
\begin{align}
X_{k}(1) \equiv L_{kk'} \sigma^{(3)}_{k'l'}\bar{X}_{l'}(1),
\end{align}
%----------
which allows us to describe $a_{\C}(1)$ by a four-component vector
%----------
% Equation
%----------
\begin{align}
a_{\C}(1) \longrightarrow \sqrt{2}  \left( 
\begin{array}{c}
a_{\k_1}(t_1)\\
a^{*}_{-\k_1}(t_1)\\
0\\
0
\end{array}
 \right),
\label{eq:Vector}
\end{align}
%----------
through Eq.~\eqref{eq:Nambu_operators}.
However, it is obvious that only the first component is independent in Eq.~\eqref{eq:Vector}, which means that $a_0(t)$ is directly related to the dynamics of $a_{\C}(1)$.

Finally, we remark that the inverse of the bare NEGFs summarized in Table~\ref{tableV} can also be described in the RAK basis.
Straightforward transformations using Eqs.~\eqref{eq:Causal} and \eqref{eq:RAK2} yield
%----------
% Equation
%----------
\begin{align}
&G^{-1}_{0,kl}(12) = \delta_{kl}G^{-1}_{0}(12), \nonumber\\
&D^{-1}_{0,kl}(12) = \delta_{kl}D^{-1}_{0}(12),
\label{eq:RealTimeInverse}
\end{align}
%----------
where $G^{-1}_{0}(12)$ and $D^{-1}_{0}(12)$ are defined in Table~\ref{tableVIII}.
The temporal differentiations included in $G^{-1}_{0}(12)$ and $D^{-1}_{0}(12)$ enable us to describe the dynamics of the relevant physical quantities, as seen in Section~\ref{sec:real time}.

%======================================================
% Wigner coordinates
%======================================================
\section{Wigner representations}\label{app:Wigner}
In this Appendix, we briefly explain the Wigner representation~\cite{Rammer07, Wigner32} for an arbitrary two-time function $f(t_i t_j)$.
By introducing the relative time $t_{ij} \equiv t_{i} - t_{j}$ and the average time $T_{ij} \equiv (t_{i} + t_{j})/2$,
the Wigner representation can be defined by the Fourier transformation with respect to the relative time
%----------
% Equation
%----------
\begin{align}
f(T_{ij}; \nu) \equiv \int^\infty_{-\infty} \dd t_{ij} \exp (\ii \nu t_{ij}) f(t_i t_j),
\label{eq:Wigner}
\end{align}
%----------
which is simply described as $f(t_i t_j) \stackrel{\W}{\longrightarrow} f(T_{ij}; \nu)$ in this paper.
The inverse transformation evidently becomes
%----------
% Equation
%----------
\begin{align}
f(t_i t_j) = \int^\infty_{-\infty} \frac{\dd \nu}{2 \pi} \exp (- \ii \nu t_{ij}) f(T_{ij}, \nu).
\label{eq:InverseWigner}
\end{align}
%----------
It is then obvious that any functions under the equal-time condition can be rewritten by the $\nu$-integration as
%----------
% Equation
%----------
\begin{align}
f(t_i t_j)|_{t_i=t_j} = \int^\infty_{-\infty} \frac{\dd \nu}{2 \pi} f(T_{ij}, \nu),
\label{eq:EqualTimeWigner}
\end{align}
%----------
which is useful to describe the physical quantities through the equal-time GFs [Eqs.~\eqref{eq:PhysicalQuantities1} and \eqref{eq:PhysicalQuantities2}].
Moreover, in the Wigner representation, it is well-known that the convolution of two functions are transformed into the Moyal product denoted by $\star$,
%----------
% Equation
%----------
\begin{widetext}
\begin{align}
\int^{\infty}_{-\infty} \dd t_k f(t_i t_k) h(t_k t_j) \stackrel{\W}{\longrightarrow} & f(T_{ij};\nu) \star  h(T_{ij};\nu) \nonumber \\
& \equiv  f(T_{ij};\nu) \exp \left\{  \frac{\ii}{2} (\leftoperation{\partial}_{\nu} \rightoperation{\partial}_{T_{ij}}  - \leftoperation{\partial}_{T_{ij}} \rightoperation{\partial}_{\nu} ) \right\}
 h(T_{ij};\nu) \nonumber \\
& = f(T_{ij};\nu) h(T_{ij};\nu) 
+ \frac{\ii}{2} f(T_{ij};\nu) (\leftoperation{\partial}_{\nu} \rightoperation{\partial}_{T_{ij}}  - \leftoperation{\partial}_{T_{ij}} \rightoperation{\partial}_{\nu} )
 h(T_{ij};\nu) 
+ \cdots ,
\label{eq:Moyal}
\end{align}
%----------
whereas the temporal differentiations turn into
%----------
% Equation
%----------
\begin{align}
\partial_{t_i} f(t_i t_j) \stackrel{\W}{\longrightarrow} \left\{ \frac{1}{2} \partial_{T_{ij}} -\ii \nu \right\} f(T_{ij}; \nu), 
\qquad \partial_{t_j} f(t_i t_j) \stackrel{\W}{\longrightarrow} \left\{ \frac{1}{2} \partial_{T_{ij}} +\ii \nu \right\} f(T_{ij}; \nu),
\label{eq:Wigner-differentiation}
\end{align}
\end{widetext}
%---------
because $\partial_{t_i} = \frac{1}{2} \partial_{T_{ij}} + \partial_{t_{ij}}$ and $\partial_{t_j} = \frac{1}{2} \partial_{T_{ij}} - \partial_{t_{ij}}$.
In particular, the approximation of taking the first two terms in the last line of Eq.~\eqref{eq:Moyal} is called the gradient approximation.

These features of the Wigner representation are used to derive Eq.~\eqref{eq:left-right Dyson in Wigner} in Section~\ref{subsubsec:EOM}, for example, giving
%----------
% Equation
%----------
\begin{align}
&G^{-1}_{0} \otimes X \stackrel{\W}{\longrightarrow}  \left\{ \frac{\ii}{2} \frac{\partial} {\partial T_{12}} + \nu -\xi_{z_1} \right\} X_{z_1 z_2}(T_{12}; \nu), \nonumber\\
&X \otimes G^{-1}_{0} \stackrel{\W}{\longrightarrow}  \left\{ -\frac{\ii}{2} \frac{\partial} {\partial T_{12}} + \nu -\xi_{z_2} \right\} X_{z_1 z_2}(T_{12}; \nu), \nonumber\\
& X \otimes Y \stackrel{\W}{\longrightarrow} X_{z_1 z_3}(T_{12}; \nu) \star Y_{z_3 z_2}(T_{12}; \nu).
\label{eq:GF-Wigner}
\end{align}
%---------

%======================================================
% Self-energies in the RAK basis
%======================================================
\section{Self-energies in the RAK basis}\label{app:Self-energies}
We here derive the total self-energy [Eq.~\eqref{eq:Self-energy}] as $\varSigma_{\C} = \varSigma^{\H}_{\C} + \varSigma^{\text{el}}_{\C} + \varSigma^{\MF}_{\C} + \varSigma^{\ph}_{\C} + \varSigma^{\gamma}_{\C}$.
In the followings, we therefore describe the respective contributions of $\varSigma^{\H}_{\C}$, $\varSigma^{\text{el}}_{\C}$, $\varSigma^{\MF}_{\C}$, $\varSigma^{\ph}_{\C}$ and $\varSigma^{\gamma}_{\C}$ shown in Figure~\ref{fig14:Selfenergies} under the HF approximation in Figure~\ref{fig20:Scheme}.

%=======================================
% Contribution from the Hartree term
%=======================================
\subsection{Contributions from $\varSigma^{\H}_{\C}$ and $\varSigma^{\ph}_{\C}$}\label{subapp:Hartree}
We first focus on the contributions from $\varSigma^{\H}_{\C}(12)$ and $\varSigma^{\ph}_{\C}$.
In the case for  $\varSigma^{\ph}_{\C}$, by inserting Eq.~\eqref{eq:Vertex_Approx} into Eq.~\eqref{eq:Sigma_ph}, one readily finds that the contribution from $\varSigma^{\ph}_{\C}(12)$ becomes zero.
In the case for $\varSigma^{\H}_{\C}(12)$, on the other hand, by substituting the definition of $U'_{C}$ (Table~\ref{tableIV}) into Eq.~\eqref{eq:Sigma_H}, we obtain
%----------
% Equation
%----------
\begin{align*}
\varSigma^{\H}_{\C}(12) = -\ii \delta_{\C}(\tau_1 \tau_2) \sum_{z_3, z_4} U'(z_1 z_4 z_3 z_2) G_{\C}(\tau_1 \tau_1; z_3 z_4).
\end{align*}
%---------
Under the assumption of Eq.~\eqref{eq:Conservation}, we then obtain 
%----------
% Equation
%----------
\begin{multline*}
\varSigma^{\H}_{\C}(12) = -\ii \delta_{\C}(\tau_1 \tau_2) U'_{\k_2 - \k_1} \delta_{\k_2, \k_1}  \\
\times  \sum_{\alpha_3, k_3} G_{\C}(\tau_1 \tau_1; \alpha_3 \alpha_3; \k_3 \k_3),
\end{multline*}
%---------
where we have used the definition of $U'$ in Table~\ref{tableIII}.
As a result, the contribution from $\varSigma^{\H}_{\C}(12)$ also becomes zero due to Eq.~\eqref{eq:Coulomb}.

%=======================================
% Contribution from the Fock term
%=======================================
\subsection{Contribution from $\varSigma^{\text{el}}_{\C}$}\label{subapp:Fock}
We next discuss the contribution from $\varSigma^{\text{el}}_{\C}(12)$.
By inserting Eq.~\eqref{eq:Vertex_Approx} into Eq.~\eqref{eq:Sigma_el}, we find 
%----------
% Equation
%----------
\begin{align*}
\varSigma^{\text{el}}_{\C}(12) = \ii \delta_{\C}(\tau_1 \tau_2) \sum_{z_3, z_4} U'(z_1 z_4 z_2 z_3)G_{\C}(\tau_1 \tau_1; z_3 z_4),
\end{align*}
%---------
with the definition of $U'_{\C}$ in Table~\ref{tableIV}.
We then obtain 
%----------
% Equation
%----------
\begin{align*}
\varSigma^{\text{el}}_{\C}(12) = \ii \delta_{\k_1 \k_2} \delta_{\C}(\tau_1 \tau_2) \sum_{\k'} U'_{\k' - \k_1} G_{\C}(\tau_1 \tau_1; \alpha_1 \alpha_2; \k'\k'),
\end{align*}
%---------
where  Eq.~\eqref{eq:Conservation} has been used with the definition of the definition of $U'$ in Table~\ref{tableIII}.
Based on Appendix~\ref{app:Real-time matrix}, this equation can be rewritten in the RAK basis [Eq.~\eqref{eq:RAK1}] as
%----------
% Equation
%----------
\begin{align*}
\varSigma^{\text{el}}_{ij}(12) = \ii \delta_{\k_1 \k_2} \delta(t_1 t_2) \delta_{ij} \sum_{\k'} U'_{\k' - \k_1} G^{<}(t_1 t_1; \alpha_1 \alpha_2; \k'\k').
\end{align*}
%---------
In the $2 \times 2$ matrix form, by applying the Wigner transformation, we therefore obtain $\varSigma^{\K}_{\text{el}} (T_{12}; \k_1 \k_2) = 0$ and 
%----------
% Equation
%----------
\begin{align}
&\varSigma^{\R}_{\text{el}} (T_{12}; \k_1 \k_2) = \varSigma^{\A}_{\text{el}} (T_{12}; \k_1 \k_2) \nonumber \\
& \quad = - \delta_{\k_1 \k_2} \sum_{\k'} U'_{\k' - \k_1} 
\begin{pmatrix}
n_{1,\k'}(T_{12}) & p_{\k'}(T_{12}) \\
p^{*}_{\k'}(T_{12}) & n_{2,\k'}(T_{12})
\end{pmatrix}, 
\end{align}
%---------
where Eq.~\eqref{eq:PhysicalQuantities1} has been used.

%=======================================
% Contribution from the MF term
%=======================================
\subsection{Contribution from $\varSigma^{\MF}_{\C}$}\label{subapp:MF}
In the case for $\varSigma^{\MF}_{\C}$, by inserting the definitions of $\bar{g}_{C}$ and $g_{C}$ (Tables~\ref{tableVI} and \ref{tableIV}) into Eq.~\eqref{eq:Sigma_MF}, we obtain 
%----------
% Equation
%----------
\begin{align*}
\varSigma^{\MF}_{\C}(12) = -  \sigma^{(1)}_{\alpha_1 \alpha_2} \delta_{\C}(\tau_1 \tau_2) g_{\alpha_1} a_{\alpha_1, \k_1-\k_2, \C}(\tau_1),
\end{align*}
%----------
where Eqs.~\eqref{eq:Pauli} and \eqref{eq:Pauli-like} have been used.
The real-time representation in the RAK basis then becomes
%----------
% Equation
%----------
\begin{align*}
\varSigma^{\MF}_{ij}(12) = - \sigma^{(1)}_{\alpha_1 \alpha_2} \delta_{ij} \delta( t_1 t_2) g_{\alpha_1} a_{\alpha_1, \k_1-\k_2}(t_1).
\end{align*}
%----------
The Wigner transformation therefore yields $\varSigma^{\K}_{\MF}(T_{12}; \k_1 \k_2) = 0$ and 
%----------
% Equation
%----------
\begin{align}
& \varSigma^{\R}_{\MF} (T_{12}; \k_1 \k_2) = \varSigma^{\A}_{\MF} (T_{12}; \k_1 \k_2) \nonumber \\
& \qquad\qquad = - \delta_{\k_1 \k_2}
\begin{pmatrix}
0 & g^{*} a_{0}(T_{12}) \\
g a^{*}_{0} (T_{12}) & 0
\end{pmatrix}, 
\end{align}
%----------
where we have used Eqs.~\eqref{eq:Nambu_operators} and \eqref{eq:Assumption}.

%=======================================
% Contribution from the pumping baths
%=======================================
\subsection{Contribution from $\varSigma^{\gamma}_{\C}$}\label{subapp:Bath}
Finally, we describe the contribution from $\varSigma^{\gamma}_{\C}(12)$.
By using the definition of $\varsigma_{\C}$ (Tables~\ref{tableIII} and \ref{tableIV}), we can write Eq.~\eqref{eq:Sigma_bath} as 
%----------
% Equation
%----------
\begin{align*}
\varSigma^{\gamma}_{\C}(12) 
= \sum_{\k_3, \k'_3} \varsigma_{\alpha_1 \k_1} \varsigma_{\alpha_2 \k_2} B_{0, \C}(\tau_1 \tau_2; \alpha_1 \alpha_2; \k_3 \k'_3),
\end{align*}
%----------
where $B_{0,\C}(12)$ is the bare NEGF of the pumping bath determined by Eq.~\eqref{eq:Bare_NEGFs} with $B^{-1}_{0,\C}$ in Table~\ref{tableV}.
In the RAK basis, it follows that
%----------
% Equation
%----------
\begin{align}
\varSigma^{\mathcal{Z}}_{\gamma}(12) = \sum_{\k_3, \k'_3} \varsigma_{\alpha_1 \k_1} \varsigma_{\alpha_2 \k_2} B^{\mathcal{Z}}_{0}(t_1 t_2; \alpha_1 \alpha_2; \k_3 \k'_3),
\label{eq:Sigma_gamma_RAK1}
\end{align}
%----------
for $\mathcal{Z} \in \{\R, \A, \K \}$.
We therefore require the bare NEGF $B^{\mathcal{Z}}_{0}(12)$ to obtain the self-energy.
For this purpose, we introduce the contour interaction picture with respect to $\oH{R}$ [Eq.~\eqref{eq:HR}],
%----------
% Equation
%----------
\begin{align*}
\hat{O}^{\R} (\tau) \equiv \hat{\U}_{\R}(\tau_{0} \tau) \hat{O}^{\text{S}} (\tau) \hat{\U}_{\R}(\tau \tau_{0}),
\end{align*}
%----------
in a similar manner to Eq.~\eqref{eq:int picture}.
Here, $\hat{\U}_{\R}(\tau_2 \tau_1)$ is defined by replacing $\hat{H} \to \hat{H}_{\R}$ in Eq.~\eqref{eq:evolution}.
$B_{0,\C}(12)$ can then be described as 
%----------
% Equation
%----------
\begin{align}
B_{0,\C}(12) = -\ii \Ex{\mathcal{T}_{\C} [\ob^{\R}(1) \ob^{\R \dagger}(2) ] }.
\label{eq:B0C}
\end{align}
%----------
One can easily check this is true because $\ii \partial_{\tau_1} \ob^{\R}(1) = \xi^{\B}_{z_1} \ob^{\R}(1)$.
In the RAK basis, Eq.~\eqref{eq:B0C} reads
%----------
% Equation
%----------
\begin{subequations}
\label{alleq:B0RAK}
\begin{align}
&B^{\R}_{0}(12) = - \ii \delta_{z_1 z_2} \theta(t_{12}) \exp(- \ii \xi^{\B}_{z_1} t_{12}), \\
&B^{\A}_{0}(12) = \ii \delta_{z_1 z_2} \theta(-t_{12}) \exp(- \ii \xi^{\B}_{z_1} t_{12}), \\
&B^{\K}_{0}(12) = - \ii \delta_{z_1 z_2} \left[1-2\Ex{ \obd_{z_1} \ob_{z_1} }  \right] \exp(- \ii \xi^{\B}_{z_1} t_{12}),
\end{align}
\end{subequations}
%----------
where $t_{12} = t_1 - t_2$ is the relative time.
Under the assumption of Eq.~\eqref{eq:Born}, the expectation value $\Ex{ \obd_{z_1} \ob_{z_1}}$ becomes
%----------
% Equation
%----------
\begin{align*}
\Ex{ \obd_{z_1} \ob_{z_1}} = \frac{1}{1+ \exp \{ \beta( \epsilon^{\B}_{\alpha_1, \k_1} - \mu^{\B}_{\alpha_1} ) \}}.
\end{align*}
%----------
Here, $\beta = 1/T$ and $\mu^{\B}_{\alpha_1}$ is the inverse temperature and the chemical potential of the pumping bath, respectively.
By inserting Eq.~\eqref{alleq:B0RAK} into Eq.~\eqref{eq:Sigma_gamma_RAK1} and applying Eqs.~\eqref{eq:gamma} and \eqref{eq:DOS_el}, we find 
%----------
% Equation
%----------
\begin{subequations}
\label{alleq:Sigma_gamma_RAK2}
\begin{align}
&\varSigma^{\R/\A}_{\gamma}(12) = \mp \ii \gamma \delta_{z_1 z_2}\delta(t_{12}), \\
&\varSigma^{\K}_{\gamma}(12) = - \ii 2\gamma \delta_{z_1 z_2} F^{\B}_{\alpha_1}(t_{12}), 
\end{align}
\end{subequations}
%----------
where we have introduced $F^{\B}_{\alpha_1}(t_{12}) \equiv \int\frac{\dd \nu}{2\pi} F^{\B}_{\alpha_1}(\nu) e^{- \ii \nu t_{12}} $ with the definition of Eqs.~\eqref{eq:FB} and \eqref{eq:Fermi_B}.
We note that, in the derivation of Eq.~\eqref{alleq:Sigma_gamma_RAK2}, only the contribution of $\k_1 = \k_2$ is taken into account to satisfy Eq.~\eqref{eq:Conservation}.
The Wigner transformation of Eq.~\eqref{alleq:Sigma_gamma_RAK2} then yields
%----------
% Equation
%----------
\begin{subequations}
\begin{align}
&\varSigma^{\R/\A}_{\gamma}(\k_1 \k_2) = \delta_{\k_1 \k_2}
\begin{pmatrix}
\mp \ii \gamma & 0 \\
0 & \mp \ii \gamma
\end{pmatrix}, \\
&\varSigma^{\K}_{\gamma} (\nu; \k_1 \k_2) = - \ii 2 \gamma \delta_{\k_1 \k_2}
\begin{pmatrix}
F^{\B}_{1}(\nu) & 0 \\
0 & F^{\B}_{2}(\nu)
\end{pmatrix}.
\end{align}
\end{subequations}
%----------

We have thus described the contributions from $\varSigma^{\H}_{\C}$, $\varSigma^{\text{el}}_{\C}$, $\varSigma^{\MF}_{\C}$, $\varSigma^{\ph}_{\C}$ and $\varSigma^{\gamma}_{\C}$ in the RAK basis.
The self-energy of Eq.~\eqref{eq:Self-energy} can easily be derived as a summation of these contributions.

%======================================================
%   Formalism for spectra
%======================================================
\section{Formalism for the spectral properties}\label{app:Formalism for Spectra}
In the main text, we have shown that the photon GF and the fully dressed two-particle GF are required to calculate the emission spectrum [Eq.~\eqref{eq:Sinc2}] and the gain-absorption spectrum [Eq.~\eqref{eq:Susceptibility}], respectively.
In this appendix, within the assumptions and approximations described in Subsection~\ref{Assumptions}, we show the way to estimate the emission spectrum and the gain-absorption spectrum, based on the generating functional approach (Section~\ref{sec:Generating Functional}).
We note that the steady state is assumed in this Appendix~\ref{app:Formalism for Spectra} for simplicity.

%=======================================
% Evaluation of the photon GFs
%=======================================
\subsection{Evaluation of the photon GF} \label{subapp:Evaluation of photon GF}
We first describe how to estimate the photon GF.
From Eqs.~\eqref{eq:Inverse of partially dressed photon GF} and \eqref{eq:Sigma_Decay}, the Dyson equation for the fully dressed photon GF [Eq.~\eqref{eq:Dyson_Dc}] can be described as 
%----------
% Equation
%----------
\begin{align}
D_{\C}^{-1}(11') &= \tilde{D}_{0,\C}^{-1}(11') - \varPi_{\C}(11') \nonumber \\ 
&=D^{-1}_{0,\C}(11') - \varSigma_{\C}^{\kappa}(11')  - \varPi_{\C}(11'),
\label{eq:DC_Inv}
\end{align}
%----------
when $D_{\C}^{-1}$ is introduced in the same manner as Eq.~\eqref{eq:InverseGc}.
Under the steady-state assumption, the Wigner transformation of the real-time matrix  (Appendices~\ref{app:Wigner} and \ref{app:Real-time matrix}) for Eq.~\eqref{eq:DC_Inv} yields the $4 \times 4$ matrix equation in the RAK basis,
%----------
% Equation
%----------
\begin{widetext}
\begin{align}
\begin{pmatrix}
D^{\R}_{\q}(\nu) & D^{\K}_{\q}(\nu)\\
0 & D^{\A}_{\q}(\nu)
\end{pmatrix}^{-1}
&=
\begin{pmatrix}
[D^{\R}_{\q}(\nu)]^{-1} & -[D^{\R}_{\q}(\nu)]^{-1} D^{\K}_{\q}(\nu) [D^{\A}_{\q}(\nu)]^{-1}\\
0 & [D^{\A}_{\q}(\nu)]^{-1}
\end{pmatrix}
\nonumber\\
&=
\begin{pmatrix}
D_{0,\q}^{-1}(\nu) -\varSigma^{\R}_{\kappa, \q}(\nu) -\varPi^{\R}_{\q}(\nu) & -\varSigma^{\K}_{\kappa, \q}(\nu) -\varPi^{\K}_{\q}(\nu) \\
0 & D_{0,\q}^{-1}(\nu) -\varSigma^{\A}_{\kappa, \q}(\nu) -\varPi^{\A}_{\q}(\nu)
\end{pmatrix},
\label{eq:XX}
\end{align}
%----------
where the $2 \times 2$ matrix form of Eq.~\eqref{eq:MatrixElement} has been applied to $\varSigma^{\mathcal{Z}}_{\kappa}(12)$ and $\varPi^{\mathcal{Z}}(12)$, $\varSigma^{\mathcal{Z}}_{\kappa, \q}(\nu)$ corresponds to the Wigner representation of Eqs.~\eqref{eq:SigmaRA_decay} and 
\eqref{eq:SigmaK_decay} with $\k_1 = \k'_1 = \q$, and $D_{0,\q}^{-1}(\nu)$ is obtained from Eqs.~\eqref{eq:RealTimeInverse} and \eqref{eq:Wigner-differentiation} as
%----------
% Equation
%----------
\begin{align}
D_{0,\q}^{-1}(\nu) = 
\begin{pmatrix}
\nu - \xi_{\ph, \q} & 0 \\
0 & - \nu - \xi_{\ph, \q}
\end{pmatrix}.
\end{align}
%----------
Note that the Wigner representation under the steady-state assumption is just the Fourier transformation with respect to the relative time.
The $2 \times 2$ matrices of $D^{\mathcal{Z}}_{\q}(\nu)$ are then given by
%----------
% Equation
%----------
\begin{subequations}
\label{alleq:photonGFs}
\begin{align}
&D_{\q}^{\R/\A}(\nu) =
\begin{pmatrix}
\nu - \xi_{\ph, \q} \pm \ii \kappa -\varPi^{\R/\A}_{11,\q}(\nu)& -\varPi^{\R/\A}_{12,\q}(\nu) \\
-\varPi^{\R/\A}_{21,\q}(\nu) & -\nu - \xi_{\ph,\q} \mp \ii \kappa -\varPi^{\R/\A}_{22,\q}(\nu)
\end{pmatrix}^{-1},
\label{eq:DRA} \\
&D_{\q}^{\K}(\nu) = D_{\q}^{\R}(\nu)
\begin{pmatrix}
- \ii 2 \kappa + \varPi^{\K}_{11,\q}(\nu) &  \varPi^{\K}_{12,\q}(\nu) \\
 \varPi^{\K}_{21,\q}(\nu) & - \ii 2 \kappa + \varPi^{\K}_{22,\q}(\nu)
\end{pmatrix}
D_{\q}^{\A}(\nu).
\end{align}
\end{subequations}
\end{widetext}
%----------
The self-energy function $\varPi_{\alpha_1 \alpha_2, \q}^{\mathcal{Z}}(\nu)$ is thus required to obtain $D^{\mathcal{Z}}_{\q}(\nu)$, which is related to the partially dressed two-particle GF as seen in Eq.~\eqref{eq:Pi_c}.
To evaluate $\varPi_{\alpha_1 \alpha_2, \q}^{\mathcal{Z}}(\nu)$, we insert the definitions of $g_{\C}$ and $\bar{g}_{\C}$ (Tables~\ref{tableIV} and \ref{tableVI}) into Eq.~\eqref{eq:Pi_c} as
%----------
% Equation
%----------
\begin{multline}
\varPi_{\C}(12) = \ii g(z_1;z'_3 z_3) \tilde{K}_{0,\C}(\tau_1 \tau_2 ; z_3 z'_3 z_4 z'_4) \\
\times \sigma^{(1)}_{z_5 z_2} g_{\C}(z_5;z'_4 z_4),
\end{multline}
%----------
where $\tilde{K}_{0,\C}(\tau_1 \tau_2 ; z_3 z'_3 z_4 z'_4) \equiv \tilde{K}_{0,\C}(\tau_1 \tau_1 \tau_2 \tau_2; z_3 z'_3 z_4 z'_4)$ for notational simplicity.
The real-time matrix in the RAK basis then becomes
%----------
% Equation
%----------
\begin{multline}
\varPi_{ij}(12) = \ii g(z_1;z'_3 z_3) \tilde{K}_{0,ij}(t_1 t_2 ; z_3 z'_3 z_4 z'_4) \\
\times \sigma^{(1)}_{z_5 z_2} g_{\C}(z_5;z'_4 z_4),
\label{eq:Pi_RAK}
\end{multline}
%----------
where $i$ and $j$ indicate the $ij$-component of the matrix as defined in Eq.~\eqref{eq:RAK1}.
As a result, with the definition of $g(z_1;z_2 z_3)$ in Table~\ref{tableIII}, the Wigner transformation of Eq.~\eqref{eq:Pi_RAK} gives $\varPi_{\alpha_1 \alpha_2, \q}^{\mathcal{Z}}(\nu)$ for $\mathcal{Z} \in \{\R, \A, \K\}$  as 
%----------
% Equation
%----------
\begin{align}
&\varPi_{\q}^{\mathcal{Z}}(\nu) = \nonumber \\
&\ii \sum_{\k_1, \k_2}
\begin{pmatrix}
|g|^2 \tilde{K}^{\mathcal{Z}}_{0,11,\q}(\nu; \k_1 \k_2) & g^2 \tilde{K}^{\mathcal{Z}}_{0,12,\q}(\nu; \k_1 \k_2)  \\
[g^{*}]^2 \tilde{K}^{\mathcal{Z}}_{0,21,\q}(\nu; \k_1 \k_2)  & |g|^2 \tilde{K}^{\mathcal{Z}}_{0,22,\q}(\nu; \k_1 \k_2)
\end{pmatrix},
\label{eq:Pi_RAK2}
\end{align}
%----------
where we have introduced a notation $\tilde{K}^{\mathcal{Z}}_{0,\q}(12) = \tilde{K}^{\mathcal{Z}}_{0,\alpha_1 \alpha_2, \q}(t_1t_2; \k_1 \k_2) \equiv \tilde{K}^{\mathcal{Z}}_{0}(t_1t_2; \alpha_1 \bar{\alpha}_1  \bar{\alpha}_2 \alpha_2;\k_1 + \q/2,\k_1 - \q/2, \k_2 -\q/2, \k_2 + \q/2)$ and $\bar{\alpha}_{1/2}$ denotes the inverse of $\alpha_{1/2}$; $\bar{\alpha}_{1} = 1$ for $\alpha_{1} = 2$, for example.
Eq.~\eqref{eq:Pi_RAK2} can be inserted into Eq.~\eqref{alleq:photonGFs} but the {\em partially dressed} two-particle GF $\tilde{K}^{\mathcal{Z}}_{0,\alpha_1 \alpha_2,\q}(\nu; \k_1 \k_2)$ is further required to discuss the emission spectrum.

%=======================================
% Evaluation of the two-particle GFs
%=======================================
\subsection{Evaluation of the two-particle GFs} \label{subapp:Evaluation of two-particle GF}
It is now obvious that the two kinds of the two-particle GFs,  namely, the partially dressed one $\tilde{K}^{\mathcal{Z}}_{0,\alpha_1 \alpha_2,\q}(\nu; \k_1 \k_2)$ and the fully dressed one $K^{\mathcal{Z}}_{\alpha_1 \alpha_2, \q}(\nu; \k_1 \k_2)$, are essential to study the emission spectrum [Eq.~\eqref{eq:Pi_RAK2}] and the gain-absorption spectrum [Eq.~\eqref{eq:Susceptibility}].
For this purpose, the BSEs [Figure~\ref{fig18:BSE}; Eqs.~\eqref{eq:BSE1} and \eqref{eq:BSE2}] is now available together with Eq.~\eqref{eq:IntegralKernel}.
With the self-energies shown in Figure~\ref{fig14:Selfenergies} under the HF approximation (Figure~\ref{fig20:Scheme}), the integration kernel $I_{\C}$ reduces to 
%----------
% Equation
%----------
\begin{align}
I_{\C}(11'22') = \ii U'_{\C}(12'1'2).
\label{eq:Ic}
\end{align}
%----------
Here, the contribution from the Hartree term [Figure~\ref{fig14:Selfenergies}(a)] is neglected for simplicity because the self-energy becomes zero due to Eq.~\eqref{eq:Coulomb} in the relevant Dyson equation; see also Appendix~\ref{subapp:Hartree}.
As a result, substitution of $U'_{\C}$ [Figure~\ref{fig12:Diagram_NEGFs}(c)] into Figure~\ref{fig18:BSE}(b) becomes equivalent to the ladder approximation and we find
%----------
% Equation
%----------
\begin{multline}
\tilde{K}_{0,\C}(11'22') = -G_{\C}(12')G_{\C}(21') \\
+ G_{\C}(13)G_{\C}(3'1') \ii U'_{\C}(34'3'4) \tilde{K}_{0,\C}(44'22').
\label{eq:ladder}
\end{multline}
%----------
In order to obtain the applicable form to Eqs.~\eqref{eq:Pi_RAK2} and \eqref{eq:Susceptibility}, we set $\tau'_{1/2} = \tau_{1/2}$ with replacing $\alpha'_1 \to \bar{\alpha}_1$, $\alpha_2 \to \bar{\alpha}_2$, $\alpha'_2 \to \alpha_2$, $\k_1 \to \k_1 + \q/2$, $\k'_1 \to \k_1 - \q/2$, $\k_2 \to \k_2 - \q/2$ and $\k'_2 \to \k_2 + \q/2$.
Eq.~\eqref{eq:ladder} then reads as
%----------
% Equation
%----------
\begin{widetext}
\begin{multline}
\tilde{K}_{0,\q,\C}(12) = - K_{0,\q,\C}(12) 
+ \int_{\C} \dd 3 \int_{\C} \dd 3'
\delta_{\k_1 \k_3} 
G_{\alpha'_3 \bar{\alpha}_1, \k_1-\q/2, \C}(\tau_3 \tau_1) 
G_{\alpha_1 \alpha_3, \k_1+\q/2, \C}(\tau_1 \tau_3) \\
\times \delta_{\C}(\tau_3 \tau'_3) \ii U'_{\k_3 - \k_3'} 
\tilde{K}_{0,\C}(\tau'_3 \tau_2; \alpha_3 \alpha'_3 \bar{\alpha}_2 \alpha_2; \k'_3+\q/2, \k'_3-\q/2, \k_2-\q/2,\k_2+\q/2),
\label{eq:BSE-temp1}
\end{multline}
%----------
where the definition of $U'_{\C}$ has been used (Tables~\ref{tableIII} and \ref{tableIV}) and $K_{0,\q,\C}(12)$ is defined as
\begin{align}
K_{0,\q,\C}(12) = K_{0,\alpha_1 \alpha_2,\q,\C}(\tau_1 \tau_2; \k_1 \k_2 ) 
\equiv \delta_{\k_1 \k_2} G_{\bar{\alpha}_2 \bar{\alpha}_1, \k_1-\q/2, \C}(\tau_2 \tau_1) G_{\alpha_1 \alpha_2, \k_1+\q/2, \C}(\tau_1 \tau_2).
\label{eq:K0}
\end{align}
To proceed further, we only take $\alpha'_3 = \bar{\alpha}_3$ into account in Eq.~\eqref{eq:BSE-temp1} for simplicity.
As a result, we obtain
%----------
% Equation
%----------
\begin{align}
\tilde{K}_{0,\q,\C}(12) = - K_{0,\q,\C}(12) 
+ \int_{\C} \dd 3 \int_{\C} \dd 3'
K_{0,\q,\C}(13) \left\{ \delta_{\alpha_3 \alpha'_3}  \delta_{\C}(\tau_3 \tau'_3) \ii U'_{\k'_3 - \k_3} \right\} \tilde{K}_{0,\q,\C}(3'2).
\label{eq:BSE-temp2}
\end{align}
%----------
In the RAK basis, the Wigner transformation of Eq.~\eqref{eq:BSE-temp2} gives the $4 \times 4$ matrix form [Eq.~\eqref{eq:RAK1}] as
%----------
% Equation
%----------
\begin{align}
\tilde{K}_{0, \q}(\nu; \k_1 \k_2 ) & = - K_{0,\q} (\nu; \k_1 \k_2) 
+ \sum_{\k_3, \k'_3} K_{0, \q}(\nu; \k_1 \k_3) T_{0}(\nu; \k_3 \k'_3) \tilde{K}_{0,\q}(\nu;\k'_3 \k_2) \nonumber \\
& = - K_{0,\q} (\nu; \k_1 \k_2) - \sum_{\k_3, \k'_3} K_{0, \q}(\nu; \k_1 \k_3) T(\nu; \k_3 \k'_3) K_{0, \q}(\nu; \k'_3 \k_2),
\label{eq:BSE-temp3}
\end{align}
%----------
where $T_{0}(\nu; \k_1 \k_2) \equiv \ii U'_{\k_1 - \k_2} I_{4 \times 4}$ and $I_{n \times n}$ is the identity matrix of size $n$.
In the second line, $T(\nu; \k_1 \k_2)$ corresponds to the so-called T matrix~\cite{Schmitt-Rink80,Haug84,Kwong09} written by
%----------
% Equation
%----------
\begin{align}
T(\nu; \k_1 \k_2) = T_{0}(\nu; \k_1 \k_2) + \sum_{\k_3, \k'_3} T_{0}(\nu; \k_1 \k_3) K_{0,\q} (\nu; \k_3 \k'_3) T(\nu; \k'_3 \k_2).
\label{eq:Tmatrix}
\end{align}
%----------
We note that an approximate solution can be obtained for Eq.~\eqref{eq:Tmatrix} if $T_{0}(\nu;\k_1 \k_2)$ depends weakly on the frequency and momentum~\cite{Schmitt-Rink80}
%----------
% Equation
%----------
\begin{align}
T(\nu;\k_1 \k_2) \simeq \left[ I_{4 \times 4} - \sum_{\k_3} T_{0}(\nu;\k_1 \k_3) K_{0,\q}(\nu;\k_3 \k_3) \right]^{-1} T_{0}(\nu;\k_1 \k_2),
\label{eq:Slowly-varying Approx}
\end{align}
%----------
because $T_{0}(\nu;\k'_3 \k_2)$ can be approximated by $T_{0}(\nu;\k_1 \k_2)$ in Eq.~\eqref{eq:Tmatrix}.
This approximation becomes exact for the contact potential $U'_{\q} = U$.
By inserting Eq.~\eqref{eq:Slowly-varying Approx} into Eq.\eqref{eq:BSE-temp3}, we finally obtain 
%----------
% Equation
%----------
\begin{align}
\tilde{K}_{0,\q}(\nu) \equiv
\sum_{\k_1, \k_2} \tilde{K}_{0,\q}(\nu; \k_1 \k_2) 
= - \sum_{\k_1} K_{0,\q}(\nu;\k_1 \k_1) 
\left[ I_{4 \times 4} - \sum_{\k_3} T_{0}(\nu; \k_1 \k_3) \right]^{-1},
\label{eq:Solution}
\end{align}
%----------
which is again in the $4 \times 4$ matrix form.
Here, $K_{0,\q}(\nu;\k_1 \k_1)$ is given from Eq.~\eqref{eq:K0} by
%----------
% Equation
%----------
\begin{subequations}
\label{alleq:K0}
\begin{align}
&K^{\R/\A}_{0,\alpha_1 \alpha_2, \q}(\nu;\k_1 \k_2) = 
 \frac{\delta_{\k_1 \k_2}}{2}
\left[ G^{\K}_{\bar{\alpha}_2 \bar{\alpha}_1, \k_1 - \q/2} * G^{\R/\A}_{\alpha_1 \alpha_2, \k_1 + \q/2}
+G^{\A/\R}_{\bar{\alpha}_2 \bar{\alpha}_1, \k_1 - \q/2}  * G^{\K}_{\alpha_1 \alpha_2, \k_1 + \q/2} \right], \\
&K^{\K}_{0,\alpha_1 \alpha_2, \q}(\nu;\k_1 \k_2) =
 \frac{\delta_{\k_1 \k_2}}{2}
\left[G^{\K}_{\bar{\alpha}_2 \bar{\alpha}_1, \k_1 - \q/2} * G^{\K}_{\alpha_1 \alpha_2, \k_1 + \q/2} \right. \nonumber \\
&\left. \qquad\qquad\qquad\qquad\qquad\qquad\qquad +G^{\R}_{\bar{\alpha}_2 \bar{\alpha}_1, \k_1 - \q/2} * G^{\A}_{\alpha_1 \alpha_2, \k_1 + \q/2}
+G^{\A}_{\bar{\alpha}_2 \bar{\alpha}_1, \k_1 - \q/2} * G^{\R}_{\alpha_1 \alpha_2, \k_1 + \q/2} \right],
\end{align}
\end{subequations}
\end{widetext}
%----------
when we write $[f*g](\nu) \equiv \int \frac{\dd \nu'}{2 \pi} f(\nu' - \nu) g(\nu')$.
$G^{\mathcal{Z}}_{\alpha_1 \alpha_2, \k}(\nu)$ in Eq.~\eqref{alleq:K0} is obtained by using the steady-state solution; see also Appendix~\ref{app:single-particle}.
As a result, $\varPi^{\mathcal{Z}}_{\q}(\nu)$ [Eq.~\eqref{eq:Pi_RAK2}] can be calculated through Eqs.~\eqref{eq:Solution} and \eqref{alleq:K0}.
The formulation for the emission spectrum is thus completed.

In contrast, for the gain-absorption spectrum, we need the fully dressed two-particle GF [Eq.~\eqref{eq:Susceptibility}] that can be calculated through Eq.~\eqref{eq:BSE1} once Eq.~\eqref{eq:Solution} is obtained.
The procedure is quite similar to the way to derive Eq.~\eqref{eq:BSE-temp3} from Eq.~\eqref{eq:ladder}.
We therefore do not repeat the derivation but show only the result as
%----------
% Equation
%----------
\begin{align}
[K^{\R}_{\q}(\nu)]^{-1} = [\tilde{K}^{\R}_{0,\q}(\nu)]^{-1} - \varXi_{\q}(\nu),
\label{eq:KR}
\end{align}
%----------
where $K^{\R}_{\q}(\nu) \equiv \sum_{\k_1 \k_2} K^{\R}_{\q}(\nu; \k_1 \k_2)$ is the $2 \times 2$ matrix in the Nambu space.
$\varXi_{\q}(\nu)$ is defined as
%----------
% Equation
%----------
\begin{align}
\varXi_{\alpha_1 \alpha_2, \q}(\nu) \equiv g_{\alpha_1} g_{\bar{\alpha}_2} \tilde{D}^{\R}_{0,\alpha_1 \alpha_2,\q}(\nu).
\end{align}
%----------
Here, $g_1 = g^{*}$ and $g_2 = g$ as in the caption of Table~\ref{tableIII} and $\tilde{D}^{\R}_{0,\q}(\nu)$ is the retarded part of the partially dressed photon GF,
%----------
% Equation
%----------
\begin{align}
\tilde{D}^{\R}_{0,\q}(\nu) = 
\begin{pmatrix}
\nu - \xi_{\ph,\q} + \ii \kappa & 0\\
0 & - \nu - \xi_{\ph,\q} - \ii \kappa
\end{pmatrix}^{-1},
\label{eq:partial}
\end{align}
%----------
which is equivalent to Eq.~\eqref{eq:DRA} in the limit of $\varPi^{\R}_{\alpha_1 \alpha_2, \q}(\nu) \to 0$.
$K^{\R}_{\q}(\nu)$ is thus obtained from Eq.~\eqref{eq:KR} by using $\tilde{K}^{\R}_{0,\q}(\nu)$ [Eq.~\eqref{eq:Solution}] and $\tilde{D}^{\R}_{0,\q}(\nu)$ [Eq.~\eqref{eq:partial}].
As a consequence, the gain-absorption spectrum can be calculated by inserting the result into Eq.~\eqref{eq:Susceptibility}, where the same notation has been applied to $K^{\mathcal{Z}}_{\alpha_1 \alpha_2, \q}(t_1t_2; \k_1 \k_2)$ as introduced for $\tilde{K}^{\mathcal{Z}}_{0,\alpha_1 \alpha_2, \q}(t_1t_2; \k_1 \k_2)$ just below Eq.~\eqref{eq:Pi_RAK2}.

%=======================================
% Several remarks
%=======================================
\subsection{Several remarks on the causality} \label{subapp:Remarks}
In Section~\ref{sec:Spectral} and Appendices~\ref{subapp:Evaluation of photon GF} and \ref{subapp:Evaluation of two-particle GF}, we have described our formalism to study the emission spectrum and the gain-absorption spectrum.
We have then discussed their properties with several numerical results.
However, we have to mention that $D^{\mathcal{Z}}_{\alpha_1 \alpha_2,\q}(\nu)$ sometimes has the pole(s) in the upper half of the complex $\nu$ plane and the causality of the photon GFs is violated even though such a situation has been avoided in the presented results.
This means that the formalism for the photon GFs has at least one problem because the causality of the GFs must be satisfied in general.
We therefore discuss several possibilities to cause the violation of the causality, here.

For this purpose, let us summarize the general procedures that have been used in our approach to evaluate the photon GFs, which are divided into four steps as follows;
\begin{enumerate}
\item It is assumed that the system is spatially homogeneous and there is a steady state.
\item Only certain kinds of diagrams are taken into account as approximations, and then, the steady-state formalism is derived.
\item Numerical solutions are determined, based on the steady-state formalism.
We note that there are situations where we can find more than one set of solutions including unstable states.
For example, $a_{0} = p_{\k} = 0$ is always one of such solutions.
\item The photon GFs are evaluated by using the obtained steady-state solutions.
However, note that further approximations are used to obtain the photon GFs.
The contribution from the Hartree term is neglected in Eq.~\eqref{eq:Ic} and only $\alpha'_3 = \bar{\alpha}_3$ is taken into account in Eq.~\eqref{eq:BSE-temp1} for simplicity.
\end{enumerate}
In this context, artificial results can be caused by taking the assumptions at step 1, by using the approximations at step 2, by selecting one set of solutions at step 3 and by applying the approximations for the photon GFs at step 4.
The photon GFs will satisfy the causality if all steps work well.
Conversely, at least one of the steps has a problem when the causality is violated. 

In some situations, the violated causality indicates the instability of the solution~\cite{Szymanska07} but we stress that this is not always the case.
The evaluation of the causality depends on the approximations at step 4 even when the same steady-state solutions are used.
In other words, the violated causality for the photon GFs does {\em not} directly mean that the steady-state solution is unstable.
In fact, in some situations, we can find that the time-evolution of the system is settled in the steady-state solution even though the causality of the photon GFs is violated.
One major reason might be that the photon GFs do not play any role to determine the steady-state solution due to the approximation $\varLambda_{\C}(12;3) \simeq 0$ [Eq.~\eqref{eq:Vertex_Approx}; Figure~\ref{fig20:Scheme}].
In this context, $\varLambda_{\C}(12;3) \simeq \bar{g}_{\C}(12;3)$ [Eq.~\eqref{eq:LambdaC_Eq}; Figure~\ref{fig19:VertexEquation}] is left as future work, in which the photon GFs will be determined self-consistently.
This will also be essential when studying the effect of the spontaneous emission.

%======================================================
% Single particle GF in the steady state
%======================================================
\section{Single-particle GF in the steady state}\label{app:single-particle}
We here derive the single-particle GF under the steady-state assumption, which is required in the calculation of $K^{\mathcal{Z}}_{0,\q}(\nu; \k_1 \k_2)$ [Eq.~\eqref{alleq:K0}], for example.
In a similar manner to Eqs.~\eqref{eq:DC_Inv}--\eqref{alleq:photonGFs}, the Dyson equation of Eq.~\eqref{eq:Dyson} reads
%----------
% Equation
%----------
\begin{subequations}
\label{alleq:GRAK}
\begin{align}
&[G^{\R/\A}_{\k} (\nu)]^{-1} = G^{-1}_{0,\k}(\nu) - \varSigma^{\R/\A}_{\k}(\nu), \\
&G^{\K}_{\k} (\nu) = [G^{\R}_{\k} (\nu)]^{-1} \varSigma^{\K}_{\k}(\nu) [G^{\A}_{\k} (\nu)]^{-1},
\end{align}
\end{subequations}
%---------
where $G^{-1}_{0,\k}(\nu)$ corresponds to Eq.~\eqref{eq:EOM_GR_Inverse1} in the limit of $\partial_t = 0$.
By using the self-energy of Eq.~\eqref{eq:Self-energy} with $\k_1 = \k_2 = \k$, Eq.~\eqref{alleq:GRAK} yields
%----------
% Equation
%----------
\begin{subequations}
\label{alleq:GRAK2}
\begin{align}
&G^{\R/\A}_{\k}(\nu) =
\begin{pmatrix}
\nu - \tilde{\xi}_{\e,\k} \pm \ii \gamma & \varDelta_{\k} \\
\varDelta^{*}_{\k} & \nu + \tilde{\xi}_{\h,\k} \pm \ii \gamma
\end{pmatrix}^{-1},
\label{eq:GRA}\\
&G^{\K}_{\k}(\nu) = - \ii 2 \gamma G^{\R}_{\k}(\nu)
\begin{pmatrix}
1-2f^{\B}_{\e}(\nu) & 0 \\
0 & 2f^{\B}_{\h}(-\nu)-1
\end{pmatrix}
G^{\A}_{\k}(\nu),
\end{align}
\end{subequations}
%---------
in the \mbox{e-h} picture of Table~\ref{tableI}.
This means that the single-particle GF $G^{\mathcal{Z}}_{\alpha_1 \alpha_2, \k}(\nu)$ can be obtained when the steady-state solutions of $a_{0}$, $p_{\k}$, $n_{\e/\h,\k}$ and $\mu$ are calculated through Eqs.~\eqref{alleq:MSBEs}--\eqref{alleq: GR_Inv2-Sigma2} with Eq.~\eqref{eq:EOM_GR}.

%======================================================
%   References and Notes
%======================================================
%\bibliography{reference}

\end{document}